\documentclass[12pt]{iopart}
\usepackage{graphicx}

\begin{document}

\title{The physics of chromatin}
\author{Helmut Schiessel}
\address{Max-Planck-Institut f\"{u}r Polymerforschung, Theory Group, P.O.Box 3148,
D-55021 Mainz, Germany}

\begin{abstract}
Recent progress has been made in the understanding of the physical
properties of chromatin -- the dense complex of DNA and histone
proteins that occupies the nuclei of plant and animal cells. Here
I will focus on the two lowest levels of the hierarchy of DNA
folding into the chromatin complex: ({\it i}) the nucleosome, the
chromatin repeating unit consisting of a globular aggregate of
eight histone proteins with the DNA wrapped around: its
overcharging, the DNA unwrapping transition, the ''sliding'' of
the octamer along the DNA. ({\it ii}) The 30nm chromatin fiber,
the necklace-like structure of nucleosomes connected via linker
DNA: its geometry, its mechanical properties under stretching and
its response to changing ionic conditions. I will stress that
chromatin combines two seemingly contradictory features: (1) high
compaction of DNA within the nuclear envelope and at the same time
(2) accessibility to genes, promoter regions and gene regulatory
sequences.
\end{abstract}

\tableofcontents

\maketitle

\section{Introduction}

Higher developed organisms face the problem to store and retrieve
a huge amount of genetic information -- and this in each cell
separately. For instance, the human genom corresponds to 3 billion
base pairs (bp) of the DNA double helix, two copies of which make
up two meters of DNA chains that have to be stored within the tiny
micron-sized nucleus of each cell \cite{alberts94}. These two
meters are composed of 46 shorter DNA pieces, each of which, if
not condensed, would form a swollen coil of roughly 100$\mu m$
diameter -- clearly much too large to fit into the nucleus. On the
other hand, the densely packed genom would form a ball of just
$\sim 2$ $\mu\mbox{m}$ diameter due to the huge aspect ratio of
contour length -- 2 meters -- versus diameter -- 20 \AA\ -- of the
DNA material. Hence, the DNA indeed fits into the nucleus but a
suitable compaction mechanism is required.

DNA is a highly charged macromolecule carrying two negative
elementary charges per 3.4 \AA. In the presence of multivalent
counterions DNA condenses into dense -- often toroidal --
aggregates~\cite{bloomfield96}, resembling the DNA packaged in
virial capsids \cite{olson01,cerritelli97,kindt01}. In viruses the
DNA has just to be stored whereas such a simple way of DNA
compaction cannot work for the long DNA chains in eucaryotic cells
(cells of fungi, plants and animals) where many portions of the
DNA have to be accessible to a large number of proteins (gene
regulatory proteins, transcription factors, RNA polymerases etc.).
Therefore the substrate that these proteins interact with must be
much more versatile to allow access to certain regions of the DNA
and to hide (i.e. to silence) other parts. That way each cell can
regulate the expression of its genes separately according to its
state in the cell cycle, the amount of nutrients present etc.
Furthermore, the differentiation of cells into the various types
that make up a multicellular organism relies to a large extent on
the way the DNA -- which is identical in all the cells -- is
packaged.

The substrate that combines all these features is chromatin, a
complex of DNA and so-called {\it histone proteins}. In 1974 it
has been realized that the fundamental unit of chromatin is the
{\it nucleosome} \cite{kornberg74,kornberg99}: roughly 200 bp of
DNA are associated with one globular octameric aggregate of eight
histone proteins consisting of two molecules each of the four core
histones H2A, H2B, H3 and H4. A stretch of 147 bp DNA is wrapped
in a $1$-$\,$and-$\,3/4$ left-handed superhelical turn around the
octamer and is connected via a stretch of linker DNA to the next
such protein spool. Each octamer together with the wrapped DNA
forms a {\it nucleosome core particle} with a radius of $\sim 5$
nm and a height of $\sim 6$ nm which carries a large negative
electric charge \cite{khrapunov97,raspaud99}.

While the structure of individual core particles is now documented
in great detail mainly on the basis of high-resolution X-ray
analyses \cite{luger97,davey02}, much less is known about the
higher-order structures to which they give rise. When the fiber is
swollen -- as this is the case for low ionic strength -- it has
the appearance of ''beads-on-a-string.''\ It is sometimes referred
to as the ''10-nm fiber'' since its ''beads''\ have $\sim 10$ nm
diameter \cite{thoma79}. With increasing salt concentration,
heading towards physiological conditions (roughly 100 mM), the
fiber becomes denser and thicker, attaining a diameter of $\sim
30$ nm \cite{widom86}. Longstanding controversy surrounds the
structure of this so-called 30{\it -nm fiber}
\cite{vanholde89,widom89,vanholde95,vanholde96}. In the {\it
solenoid models} \cite{thoma79,finch76,widom85}\ it is assumed
that the chain of nucleosomes forms a helical structure with the
linker DNA bent in between whereas the {\it zig-zag} or {\it
crossed linker models}
\cite{woodcock93,horowitz94,leuba94,bednar98,schiessel01b}\
postulate straight linkers that connect nucleosomes that are
located on opposite sites of the fiber. The higher-order folding
of the 30-nm fiber into structures on scales up to microns is yet
to be elucidated. In Fig.~1 I sketch the steps of the DNA folding
starting with a DNA chain of length $\sim 1$ cm and $\sim 10^{6}$
octamers and ending up at the highly condensed {\it chromosome}.
This highly condensed structure occurs before cell division and
contains the chain and its copy neatly packaged for the
distribution into the two daughter cells. The size of the
chromosome is $\sim 10000$ times smaller than the contour length
of the original chain.

\begin{figure}
\begin{center}
\includegraphics*[width=12cm]{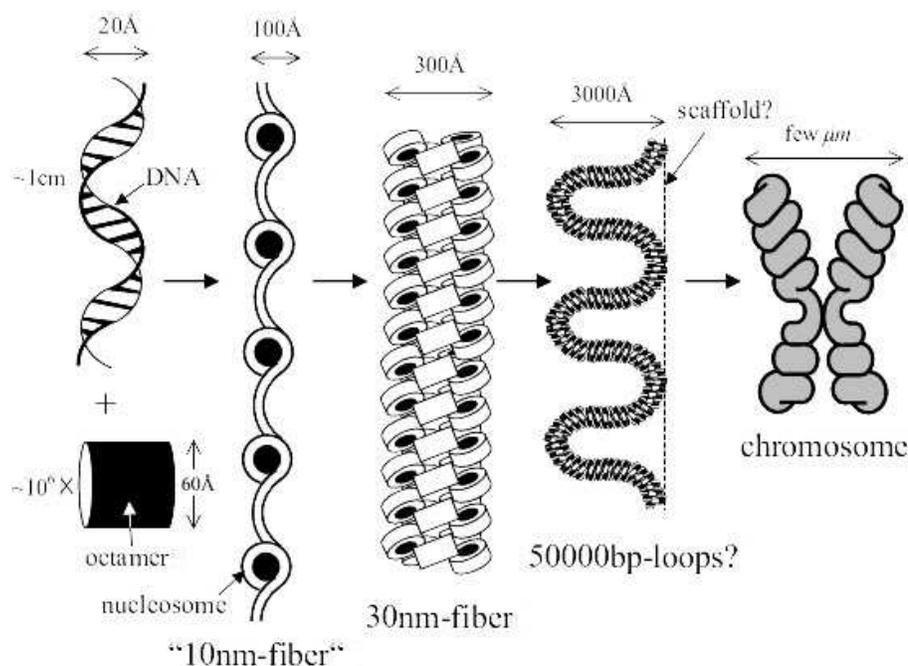}
\end{center}
\caption{Steps of the DNA compaction into chromatin. The DNA
molecule of length $\sim 1$ cm is compacted with the help of
$10^6$ histone octamers leading to 10000-fold reduction of its
original length (see text for details)}
\end{figure}

It is instructive to draw a comparison between the structure and
function of chromatin to that of a daily-life example: the
library. As the nucleus stores the long one-dimensional string of
basepairs, so does the library contain the huge one-dimensional
string of letters, the text written down in all of its books. A
book like Ref.~\cite{alberts94} contains $\sim 10$ km text, a
library with 10000 books stores roughly 100000 km text! How can
the user find and retrieve the little piece of information of
interest? The way this is handled is that the text is folded in a
hierarchical fashion in lines, pages, books and shelves. This
makes it is relatively easy -- with the help of a few markers --
to find the corresponding text passage. Furthermore, all the text
is stored in a dense fashion but the book of interest can be taken
out of the shelve and opened at the appropriate page without
perturbing the rest of the library. Apparently, the result of this
hierarchical structure is a relative high efficiency in storing a
huge amount of information in a relatively small space and, at the
same time, having a high accessibility to it.

The similarities between hierarchies in the library and in
chromatin are pretty obvious. What is less obvious and in many
respects still an open question is how the dense chromatin
structure can be opened locally to allow access to its genes. As
mentioned above for one nucleosomal repeat length, typically 200
bp, 147 bp are wrapped around the octamer, i.e., roughly 75
percent of the DNA chain is tightly associated to the histone
aggregates. It is known that many essential proteins that interact
with DNA do not have access to DNA when it is wrapped (reviewed in
Ref.~\cite{workman98}). Moreover, even the unwrapped sections --
the linker DNA -- are somewhat buried inside the dense 30-nm
fiber. Therefore, it is necessary for the cell to have mechanisms
at hand to open -- unfold -- the fiber and then, somehow, to
unwrap the DNA from the protein spools or to temporarily remove
them from the DNA piece of interest.

This leads to the problem of how the chromatin structure changes
its shape with time. As this structure involves length scales of
many orders in magnitude (from \AA ngstroms to microns) so do its
dynamical processes take place in a wide range of time scales,
beginning with fluctuations of the nucleosome structure in the
micro- to millisecond timescale \cite{widom98} up to large scale
variations in the condensation degree of the chromosome that
follows the cell cycle with a typical period of hours to days
\cite{alberts94}.

It has been shown through competitive protein binding to
nucleosomal DNA \cite{polach95,polach96}\ that thermal
fluctuations might lead to partial unwrapping of the DNA from the
nucleosome. This mechanism provides intermittent access to
nucleosomal DNA. Not only unwrapping but even ''{\it sliding}'' of
nucleosomes along DNA seems to be facilitated by thermal
fluctuations as it has been demonstrated in well-defined {\it in
vitro} experiments \cite{pennings91,meersseman92,pennings94}.
Whereas this kind of repositioning dynamics is quite slow -- even
at elevated temperatures the timescale is of the order of minutes
to hours -- nucleosome repositioning appears also to be of great
importance {\it in vivo} where it is aided by {\it
chromatin-remodeling complexes}, large multi-protein
complexes that use energy by burning ATP (reviewed in Refs.~\cite%
{kornberg99,varga-weisz98,guschin99,peterson00,flaus01}). These complexes
might catalyze and direct the displacement of nucleosomes out of regions
where direct access to DNA has to be granted (like promoter regions of
transcriptionally active genes).

Based on a range of experiments it has also been speculated that RNA
polymerase, the protein complex that transcribes -- copies -- genes, can
itself act ''through'' nucleosomes without having to disrupt the structure
completely \cite%
{studitsky94,studitsky95,felsenfeld96,studitsky97,felsenfeld00}.
An appealing picture is the idea that the polymerase gets around
the nucleosome in a loop. Such a mechanism is especially important
since genes are typically distributed over a DNA portion of the
length of 100's to 10000's of bp which means that there are a few
up to 1000's of nucleosomes with which a polymerase has somehow to
deal with during transcription.

Concerning the unfolding of the 30-nm fiber, I mentioned already above that
a decrease in the ionic strength leads to a swelling of the fiber.
Obviously, {\it in vivo }such a global swelling is not possible within the
tiny space available within the nucleus. Hence only the unfolding of local
regions should be expected as the above given analogy of a library suggests.
Experimental observations seem to indicate that the swelling degree can
indeed be tuned locally by the acetylation of the lysine-rich (i.e. cationic) {\it %
tails} of the eight core histones that appear to be long flexible
polyelectrolyte chains \cite{luger98} that extend out of the
globular part. Furthermore, transcriptionally active chromatin
portions show a depletion in the {\it linker histone} H1, a
cationic proteins that is believed to act close to the entry-exit
region of the DNA at the nucleosome. As long as H1 is present the
fiber is relatively dense and the individual nucleosomes are inert
against thermal fluctuations (no unwrapping, sliding or
transcription through the octamer). If H1 is missing the chromatin
fiber appears to be much more open and the nucleosomes become
''transparent'' and mobile due to thermal fluctuations
\cite{pennings94}.

Undoubtedly chromatin lies in the heart of many essential
biological processes ranging from gene expression to cell
division. Most of these processes are controlled by a huge amount
of specific proteins. Their investigation clearly belongs to the
realm of biologists and is beyond the scope of a physicist.
Nevertheless, many insights gained in this field were achieved
through {\it in vitro} experiments under relative well-controlled
conditions, in some cases essentially only involving DNA and
histone proteins. Furthermore, new physical methods like
micromanipulation experiments allowed to gain access to certain
physical properties of chromatin. The purpose of this paper is to
discuss some of these results and, especially, to review the
physical theories they gave rise to.

The interest of physicists into single nucleosomes was mainly
sparked by the above mentioned fact that the core particle carries
a large negative net charge. This is due to the fact that much
more negatively charged DNA is wrapped around the cationic octamer
than necessary for its neutralization. Beginning around 1998 a
considerable activity started
among several research groups to explain on which physical facts {\it %
overcharging} is based (recently reviewed in \cite{grosberg02}), a
phenomenon not only occurring in chromatin but also in DNA-lipid complexes,
multilayer adsorption of polyelectrolytes etc. To gain insight into this the
nucleosome was translated into different types of toy models, usually
consisting of one charged chain and an oppositely charged sphere and the
amount of wrapping was calculated \cite%
{park99,mateescu99,gurovitch99,netz99,kunze00,nguyen01,nguyen01b,nguyen01c,schiessel01,nguyen01d,gelbart01,kunze02,schiessel02b}
(cf. also the related studies \cite%
{odijk80,pincus84,vongoeler94,wallin96,wallin96b,wallin97,wallin98,haronska98,jonsson01,chodanowski01,sakaue01,messina02,messina02b,jonsson02,akinchina02,dzubiella02}%
). It turned out that overcharging is a fairly robust phenomenon
that occurs even if certain mechanisms are neglected (like
counterion release upon
adsorption). It was also shown that multi-sphere-complexation can lead to {\it %
undercharged} systems \cite{nguyen01b,schiessel01}.

The relevance of these toy models for ''real'' nucleosomes might
be questionable, especially since the binding sites of the DNA to
the octamer are relatively specific \cite{luger97,davey02} and
since under physiological conditions (100 mM) the screening length
is fairly short -- 10 nm, i.e., half the diameter of the DNA
double helix. Nevertheless, these models might give some insight
into the unwrapping transition that takes place when the DNA which
is fairly rigid on the nucleosome length scale unwraps from the
nucleosome due to a decrease of the adsorption energy. This can be
achieved by a change in the salt concentration and has indeed been
observed experimentally \cite{khrapunov97,yager89}. A simple
approach to this problem comparing adsorption energies between a
''sticky'' spool and a semiflexible chain and its bending energy
has been given in Ref.~\cite{marky91} and led to the prediction of
an {\it unwrapping transition} in an all-or-nothing fashion, i.e.,
the cylinder is fully wrapped by the chain or not wrapped at all.
A more complex picture has been found in Refs.~\cite%
{netz99,kunze00} where the unwrapping has been calculated for
complexes with short chains (''nucleosome core particles''). Here
the chain showed different degrees of wrapping (dependent on the
ionic strength) which seems to agree fairly well with the
corresponding experimental observations on core particles
\cite{khrapunov97,yager89}. It was shown in
Refs.~\cite{schiessel02b,schiessel00} that besides the wrapped and
unwrapped structures there is for longer chains the realm of open
{\it multi-loop} complexes (so-called {\it rosettes}) that have
now also been observed in computer simulations \cite{akinchina02}.

The nucleosome repositioning, mentioned above, is another single-nucleosome
problem that has been investigated theoretically \cite%
{schiessel01c,kulic02,kulic02b}. It is suggested that the
nucleosome ''sliding'' is based on the formation of loops at the
ends of the nucleosome which then diffuse as defects around the
spool, leading to the repositioning. This process somewhat
resembles polymer reptation in the confining tube of the
surrounding medium (a polymer network or melt), cf.
Ref.~\cite{degennes71,degennes79,doi86}. A different ''channel''
for repositioning might be a cork screw motion of the DNA helix
around the octamer induced by twist defects which has now also
been studied theoretically \cite{kulic02b}.

On the level of the 30-nm fiber recent progress has been made in
the visualization of these fibers via electron cyromicroscopy
\cite{bednar98}. The micrographs reveal for lower ionic strength
structures that resemble the crossed linker model mentioned above.
However, for increasing ionic strength the fibers become so dense
that their structure still remains obscure. An alternative
approach was achieved via the micromanipulation of single 30-nm
fibers \cite{cui00,bennink01,brower02}. The stretching of the
fibers showed interesting mechanical properties, namely a very low
stretching modulus for small tension, a force plateau around 5 pN
and, at much higher tensions, saw-tooth-type patterns.

These experimental results led to a revival of interest in 30-nm
fiber models.\ Theoretical studies
\cite{schiessel01b,ben-haim01,victor02} as well as computer
simulations \cite{katritch00,wedemann02} attempted to explain the
mechanical properties of the fiber. What all these models have in
common is that they assume straight linkers in accordance with the
experimental observations, at least found for lower ionic strength
\cite{bednar98}. The low stretching modulus of the fibers is then
attributed to the bending and twisting of the linker DNA which is
induced by the externally applied tension. Indeed, all the models
find a relatively good agreement with the experimental data. The
$5pN$ force plateau is usually interpreted as a reversible
condensation-decondensation transition of the fiber which can be
attributed to a small attractive interaction between the nucleosomes \cite%
{schiessel01b,katritch00}. This is also in good agreement with recent
studies on single nucleosome core particles where such a weak attraction has
been observed and attributed to a tail-bridging effect \cite%
{mangenot02,mangenot02b}. Finally, the sawtooth-pattern which leads to a
non-reversible lenghtening of the fiber probably reflects the
''evaporation'' of histones from the DNA \cite{brower02,marko97}.

The geometry of the crossed-linker models suggest that the density of
nucleosomes in the fiber depends to a large extend on the entry-exit angle
of the DNA at the nucleosomes. It is known that in the presence of the
linker histone the entering and exiting strands are forced together in a
''stem'' region \cite{bednar98}. A recent study \cite{schiessel02}
investigated how the electrostatic repulsion between the two strands
dictates this entry-exit angle. Especially, it was shown how this angle can
be controlled {\it in vitro }via a change in the salt concentration. It also
has been speculated that via biochemical mechanisms that control the charges
in the entry-exit region (like the acetylation of certain histone tails) the
cell can locally induce a swelling of the fiber \cite{vanholde96}.

Before discussing in the following all the above mentioned issues
concerning the physics of chromatin, I note that there are also
important studies using a bottom-down approach by studying the
physical properties of whole chromosomes that have been extracted
from nuclei preparing for cell division. Via micropipette
manipulation of these mitotic chromosomes it was demonstrated that
they are extremely deformable by an externally applied tension
\cite{houchmandzadeh97,poirier00,poirier01} and that a change in
ionic strength induces a hypercondensation or decondensation of
the chromosome, respectively \cite{poirier02}. Meiotic and mitotic
chromosomes were compared to simple polymer systems like brushes
and gels \cite{marko97b}. A problem with this bottom-down approach
is still the lack of knowledge of the chromosome structure at this
level and of the proteins that cause them. I will therefore
dispense with giving a discussion on this subject.

This review is organized as follows. In the next section I give a
discussion of single nucleosome problems. After providing some
experimental facts on the structure of the nucleosome (Section
2.1), I discuss simple model systems (Section 2.2), the unwrapping
transition (Section 2.3) and the nucleosome repositioning (Section
2.4). Section 3 features the next level of folding, the 30-nm
fiber. I briefly review some of the proposed models (Section 3.1)
and then give a systematic account on the crossed-linker model
(Sections 3.2 and 3.3) and its mechanical response to stretching,
bending and twisting (Section 3.4). Then fiber swelling (Section
3.5) is discussed. Section 4 gives a conclusion and outlook.

\section{Single nucleosome}

\subsection{Experimental facts on the core particle}

The structure of the nucleosome core particle is known in
exquisite detail from X-ray crystallography: the octamer in
absence of the DNA was resolved at 3.1 \AA\ resolution
\cite{arents91} and the crystal structure of the complete core
particle at 2.8 \AA\ resolution \cite{luger97} and recently at 1.9
\AA\ \cite{davey02}. The octamer is composed of two molecules each
of the four core histone proteins H2A, H2B, H3 and H4. Each
histone protein has in common a similar central domain composed of
three $\alpha $-helices connected by two loops. These central
domains associate pairwise in form of a characteristic
''handshake''\ motif which leads to crescent-shaped heterodimers,
namely H3 with H4 and H2A with H2B\footnote{At physiological
conditions stable oligomeric aggregates of the core histones are
the H3-H4 tetramer (an aggregate of two H3 and two H4 proteins)
and the H2A-H2B dimer. The octamer is stable if it is associated
with DNA or at higher ionic strengths.}. These four
''histone-fold''\ dimers are put together in such a way that they
form a cylinder with $\sim 65$ \AA\ diameter and $\sim $60 \AA\
height. With grooves, ridges and relatively specific binding sites
they define the wrapping path of the DNA, a left-handed helical
ramp of 1 and $3/4$ turns, 147 bp length and a $\sim 28$ \AA\
pitch. In fact, the dimers themselves are placed in the octamer in
such a way that they follow this solenoid: They form a
tetrapartite, left-handed superhelix,
a spiral of the four heterodimers (H2A-H2B)$^{1}$, (H3-H4)$^{1}$, (H3-H4)$%
^{2}$ and (H2A-H2B)$^{2}$. This aggregate has a two-fold axis of
symmetry (the dyad axis) that goes through the
(H3-H4)$_{2}$-tetramer apex and is perpendicular to the superhelix
axis. A schematic view of the nucleosome core particle is given in
Fig.~2.

\begin{figure}
\begin{center}
\includegraphics*[width=6cm]{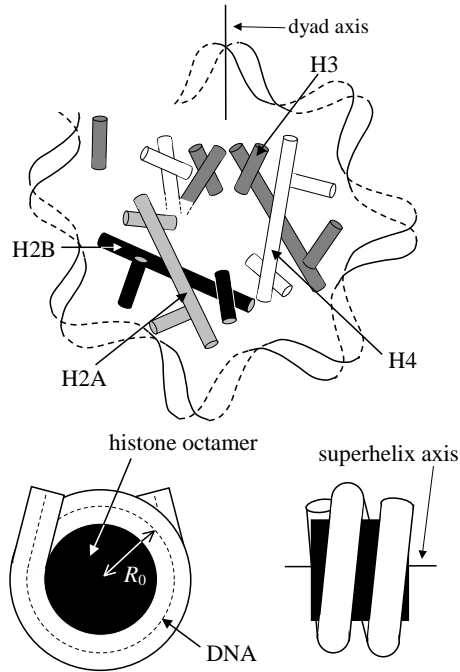}
\end{center}\caption{Schematic views of the nucleosome core particle. At the
top picture the upper half of the 8 core histones and the
nucleosomal DNA are depicted. At the bottom a simplified model is
displayed where the octamer is replaced by a cylinder and the DNA
by a wormlike chain. Also indicated are the dyad axis and the DNA
superhelix axis.}
\end{figure}

There are fourteen regions where the wrapped DNA contacts the octamer
surface, documented in great detail in Refs.~\cite{luger97} and \cite%
{davey02}. These regions are located where the minor grooves of
the right-handed DNA double helix face inwards towards the surface
of the octamer. Each crescent-shaped heterodimer has three contact
points, two at its tips and one in the middle, altogether making
up 12 of the 14 contacts. Furthermore, at each end of the wrapped
section (the termini of the superhelix) there is a helical
extension of the nearby H3 histone making contact to a minor
groove. At each contact region there are several direct hydrogen
bonds between histone proteins and the DNA sugar-phosphate
backbone \cite{luger97} as well as bridging water molecules
\cite{davey02}. Furthermore, there is also always a (cationic)
arginine side chain extending into the DNA minor groove. The free
energies of binding at each sticking point is different which can
be concluded from the fact that for each binding site there is a
different number of hydrogen bonds located at different positions;
this is also reflected in the fluctuations of the DNA phosphate
groups in the nucleosome crystal \cite{luger97}. However, a
reliable quantitative estimate of these energies is still missing.

An indirect method to estimate the adsorption energies at the sticking
points is based on studies of competitive protein binding to nucleosomal DNA
\cite{polach95,polach96,anderson00}. Many proteins are not able to bind to
DNA when it is wrapped on the histone spool due to steric hindrance from the
octamer surface. However, thermal fluctuations temporarily expose portions
of the nucleosomal DNA via the unwrapping from either end of the superhelix.
It was demonstrated that sites which are cut by certain restriction enzymes
showed -- compared to naked DNA -- an increased resistance to digestion by
these enzymes when they are associated to the octamer. Furthermore, the
farer these sites are apart from the termini of the superhelix the less
frequent they become exposed for cutting. The rate for digestion is reduced
roughly by a factor $10^{-2}$ for sites at the superhelical termini and $%
\sim 10^{-4}-10^{-5}$ for sites close to the center of the nucleosomal DNA
portion (i.e., close to the dyad axis). From these findings one can estimate
that the adsorption energy per sticking point is of order $\sim 1.5-2k_{B}T$
.

It is important to note that the $1.5-2k_{B}T$ do not represent the {\it pure%
} adsorption energy but instead the {\it net} gain in energy which
is left after the DNA has been bent around the octamer to make
contact to the sticking point. A rough estimate of this
deformation energy can be given by describing the DNA as a
semiflexible chain with a persistence length $l_{P}$ of $\sim $500
\AA\ \cite{hagerman88}. Then the elastic energy \cite{harries66}
required to bend the 127 bp of DNA around the octamer (10 bp at
each terminus are essentially straight \cite{luger97}) is given by
\begin{equation}
E_{elastic}/k_{B}T=l_{P}l/2R_{0}^{2}  \label{bend}
\end{equation}
Here $l$ is the bent part of the wrapped DNA, $\sim 127\times 3.4$
\AA $=431$ \AA\ and $R_{0}$ is the radius of curvature of the
centerline of the wrapped DNA (cf. Fig. 2) which is roughly 43
\AA\ \cite{luger97}. Hence, the bending energy is of order
$58k_{B}T$. This number, however, has to be taken with caution.
First of all it is not clear if Eq.~\ref{bend} is still a good
approximation for such strong curvatures. Then it is known that
the DNA does not bend uniformly around the octamer \cite{luger97}
and finally the DNA might show modified elastic properties due to
its contacts with the octamer. Nevertheless, using this number one
is led to the conclusion that the
bending energy per ten basepairs, i.e., per sticking site, is of order $%
60k_{B}T/14\sim 4k_{B}T$.

Together with the observation that the net gain per sticking point is $\sim
2k_{B}T$ this means that the pure adsorption energy is on average $\sim
6k_{B}T$ per binding site. Note that the huge pure adsorption energy of $%
\sim 6k_{B}T\times 14\sim 85k_{B}T$ per nucleosome is cancelled to a large
extend by the $\sim 58k_{B}T$ from the DNA bending, a fact that has
important consequences for the unwrapping transition discussed in Section
2.3 and especially, for the nucleosome repositioning reviewed in Section 2.4.

Of great importance are also flexible, irregular tail regions of
the core histones which make up $\sim $28\% of their sequences
\cite{luger98}. Each histone has a highly positively charged,
flexible tail (which is the N-end of the polypeptide chain
\cite{alberts94}) that extends out of the nucleosome structure.
Some of them exit between the two turns of the wrapped DNA, others
on the top or bottom of the octameric cylinder. These N-tails are
extremely basic due to a high amount of lysine and arginine
aminoacids (aa's). They are sites of posttranslational
modification and are crucial for chromatin regulation. Especially,
the tails have a strong influence on the structure of the 30nm
chromatin fiber as I will discuss in more detail in Section 3.
X-ray scattering data on core particles \cite{mangenot02} suggest
that the tails are adsorbed on the complex for small ionic
strengths and extended at high salt concentrations (cf. Fig. 8 in
that paper). If the octamer is associated with a longer DNA piece
the N-tails desorb at higher ionic strength.

Finally, let me mention the amount of charges found on the
nucleosome core particle \cite{khrapunov97}. The histone octamer
contains 220 basic side chains (arginine and lysine). From these
are about 103 located in the flexible histone tails mentioned
above. The rest, 117 residues, are in the globular part of the
octamer, of which 31 are exposed to the solvent, the rest being
involved in intra- and interprotein ionic interactions. On the
other hand, one has 147 bp of DNA wrapped around the octamer, each
contributing two phosphate groups. Hence there are 294 negative
charges from the DNA versus 220 positive charges of the octamer
(or even less, 134 -- not counting the charges buried inside the
octamer), i.e., the nucleosomal complex is overcharged by the DNA.
At first sight this is a surprising fact and indeed led to the
development of simplified toy models containing charged spheres
and oppositely charged chains which I will discuss in the
following section.

\subsection{Polyelectrolyte--charged sphere complexes as model systems for
the nucleosome}

The complexation of (semi-)flexible polyelectrolytes and oppositely charged
macroions is an important ingredient in biological processes. For instance,
the non-specific part of the interaction between DNA and proteins is
governed by electrostatics \cite{strauss94}. In fact, the nucleosome is an
example of this kind of interaction. A number of experimental \cite%
{braunlin82,mascotti90,raedler97,koltover99} and theoretical studies \cite%
{park99,record78,bruinsma98,harries98,sens00} have demonstrated that the
complexation of {\it highly} charged macroions (e.g. DNA) is governed by an
unusual electrostatics mechanism: {\it counterion release}. The free energy
of complexation is then dominated by the entropy increase of the released
counterions that had been condensed before complexation. This electrostatic
contribution to the free energy has to compete with the energy cost of
deforming one or both macromolecules to bring them in close contact.

In the next section I discuss how the counterion release leads to
an overcharged sphere-chain complex which might be considered as a
simplified model system of the nucleosome. I also consider the
case of a polyelectrolyte chain placed in a solution of oppositely
charged spheres (Section 2.2.2). Both sections follow closely the
treatments given in Refs.~\cite{park99,schiessel01}. Afterwards I
review studies where it was found that overcharging occurs also in
weakly charged systems due to ''standard'' electrostatics (Section
2.2.3). Section 2.2.4 is devoted to physiological conditions
(strong screening) and also to the question whether such toy
models can ''explain'' the large net charge of nucleosomes.

\subsubsection{Single-sphere complex (highly charged case)}

Consider a single sphere of radius $R_{0}$ with its charge $eZ$
homogeneously smeared out over the surface (the ''octamer'') and a
flexible rod with a charge per unit length $-e/b$, persistence
length $l_{P}$, contour length $L\gg R_{0}$ and radius $r$ (the
''DNA chain''). Both are placed in a salt solution characterized
by a Bjerrum length $l_{B}\equiv e^{2}/\epsilon k_{B}T$ ($\epsilon
$: dielectric constant of the solvent; in water $\epsilon =80$ and
$l_{B}=7$\AA\ at room temperature) and a Debye screening length
$\kappa ^{-1}=\left( 8\pi c_{s}l_{B}\right) ^{-1/2}$; furthermore
the solvent is treated as a continuum. Clearly, such a model
neglects all the intricate features of the nucleosome discussed in
Section 2.1. It is, however, indispensable to start from such a
simple model system to identify general features that occur in
polyelectrolyte-macroion complexes.

I will focus in this section on salt concentrations $c_{s}$ that are
sufficiently small such that $\kappa ^{-1}$ is large compared to the sphere
radius, $\kappa R_{0}\ll 1$. The persistence length is assumed large
compared with $R_{0}$. The chain is here highly charged which means that the
so-called {\it Manning parameter} $\xi \equiv l_{B}/b$ is required to be
much larger than one. In this case $\left( 1-\xi ^{-1}\right) L/b\simeq L/b$
counterions are condensed on the chain reducing the effective line charge
density to the value $-e/l_{B}$ \cite{oosawa71,manning78}. The entropic cost
to ''confine'' a counterion close to the chain is $\Omega k_{B}T$ with $%
\Omega =2\ln \left( 4\xi \kappa ^{-1}/r\right) $ \cite{schiessel01,rouzina96}%
. This leads to the following entropic electrostatic charging free energy of
the isolated chain in the salt solution:
\begin{equation}
\frac{F_{chain}\left( L\right) }{k_{B}T}\simeq \frac{\Omega L}{b}
\label{fcc}
\end{equation}
On the other hand the corresponding electrostatic charging free energy of
the spherical macroion of charge $Z$ is given by
\begin{equation}
\frac{F_{sphere}\left( Z\right) }{k_{B}T}\simeq \left\{
\begin{array}{ll}
\frac{l_{B}Z^{2}}{2R_{0}} & \mbox{for}\;\left| Z\right| <Z_{max} \\
\left| Z\right| \tilde{\Omega}\left( Z\right) & \mbox{for}\;\left| Z\right|
\gg Z_{max}%
\end{array}
\right.  \label{fcs}
\end{equation}
where $\tilde{\Omega}\left( Z\right) =2\ln \left( \left| Z\right|
l_{B}\kappa ^{-1}/R_{0}^{2}\right) $ \cite{oosawa71} and $Z_{max}
\approx \tilde{\Omega} R_{0}/l_{B}$ (see below). For weakly
charged spheres ($\left| Z\right| <Z_{max}$) $F_{sphere}$ is the
usual electrostatic charging energy. In the highly charged case,
$\left| Z\right| \gg Z_{max}$, most of the counterions are
localized close to the sphere with an entropic
cost $\tilde{\Omega}\left( Z\right) k_{B}T$ per counterion leading to Eq.~%
\ref{fcs} for $\left| Z\right| \gg Z_{max}$. Only the small fraction $%
Z_{max}/Z$ of counterions is still free leading to an effective
sphere charge $Z_{max}$. The value of $Z_{max}$ follows from the
balance of electrostatic
charging energy $l_{B}Z_{max}^{2}/2R_{0}$ and counterion entropy $-\tilde{%
\Omega}\left( Z\right) Z_{max}$ \cite{alexander84}.

The total free energy of the sphere-chain complex can be determined as
follows. Assume that a length $l$ of the chain has been wrapped around the
sphere. Divide the sphere-chain complex in two parts: the sphere with the
wrapped part of length $l$ of the chain and the remaining chain of length $%
L-l$. The first part -- which I will refer to as the ''complex'' -- carries
a net charge $Z\left( l\right) =Z-l/b$. The electrostatic free energy $%
F_{compl}\left( l\right) $ of the complex is then estimated to be equal to $%
F_{sphere}\left( Z\left( l\right) \right) $ (neglecting higher-order
multipole contributions). There is a special length $l_{iso}=bZ$, the {\it %
isoelectric wrapping length}, at which $Z\left( l_{iso}\right) =0$. The
usual principle of charge neutrality would lead one to expect that the total
free energy is minimized at this point.

The total free energy of the sphere-chain complex is approximately given by
the following terms \cite{schiessel01}:
\begin{equation}
F_{1}\left( l\right) =F_{compl}\left( l\right) +F_{chain}\left( L-l\right)
+F_{compl-chain}\left( l\right) +E_{elastic}\left( l\right)  \label{ftotal}
\end{equation}
The first two terms have already been discussed. The third term is the
electrostatic free energy of the interaction between the complex and the
remainder of the chain which is of the order
\begin{equation}
\frac{F_{compl-chain}\left( l\right) }{k_{B}T}\simeq Z^{*}\left( l\right)
\ln \left( \kappa R_{0}\right)  \label{fsc}
\end{equation}
where $Z^{*}\left( l\right) $ is the {\it effective} charge of the complex
(the smaller value of $Z\left( l\right) $ and $Z_{max}$). The final term in
Eq.~\ref{ftotal} describes the elastic energy of the wrapped portion of the
chain that has a typical curvature $1/R_{0}$ and is given already above, Eq.~%
\ref{bend}.

Following Ref.~\cite{schiessel01} the two cases $\left\vert
Z\left( l\right) \right\vert <Z_{max}$ and $\left\vert Z\left(
l\right) \right\vert >Z_{max}$
have to be treated separately. The first case applies for wrapping lengths $%
l $ between $l_{min}=l_{iso}-bZ_{max}$ and $l_{max}=l_{iso}+bZ_{max}$. The
free energy \ref{ftotal} takes then the form
\begin{equation}
\frac{F_{1}\left( l\right) }{k_{B}T}\simeq \frac{l_{B}}{2R_{0}}\left( Z-%
\frac{l}{b}\right) ^{2}+\frac{Cl}{b}+const.  \label{fl2}
\end{equation}
where $C=l_{P}b/2R_{0}^{2}-\ln \left( \kappa R_{0}\right) -\Omega $. For the
second case, when $\left\vert Z\left( l\right) \right\vert >Z_{max}$, one
finds
\begin{equation}
\frac{F_{1}\left( l\right) }{k_{B}T}\simeq \frac{B^{\mp }l}{b}+const
\label{fl3}
\end{equation}
with $B^{\mp }=l_{P}b/2R_{0}^{2}-\Omega \mp \tilde{\Omega}$. The
\textquotedblright $-$\textquotedblright\ sign refers to the case $l<l_{min}$
(i.e. $Z\left( l\right) >Z_{max}$) when for every segment $b$ of adsorbed
length a negative counterion of the sphere and a positive counterion of the
chain are released while the \textquotedblright $+$\textquotedblright\ sign
refers to the case $l>l_{max}$ (equivalently, $Z\left( l\right) <-Z_{max}$)
when for every adsorbed segment a positive counterion is transferred from
the chain to the sphere leading to a change $k_{B}\left( \Omega -\tilde{%
\Omega}\right) $ of its entropy. The three different cases are depicted in
Fig.~3.

\begin{figure}
\begin{center}
\includegraphics*[width=6cm]{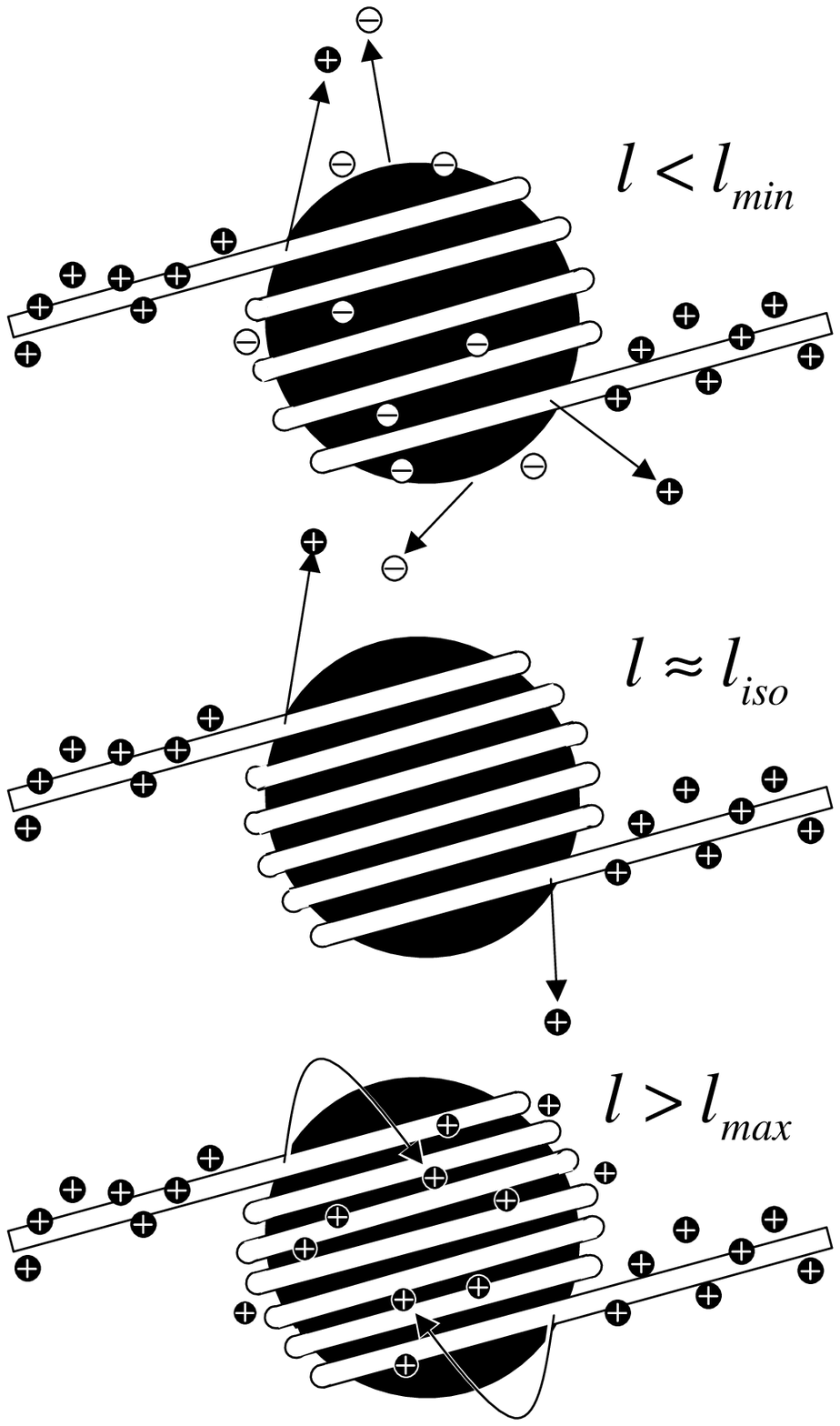}
\end{center}
\caption{Schematic view of the complex between a highly charged
chain and an oppositely charged sphere. Depicted are three
scenarios. From top to bottom: For short wrapping lengths
complexation is driven by release of counterions from sphere and
chain (top picture). For intermediate values (around the
isoelectric wrapping length) all counterions of the sphere have
been released, further complexation leads, however, still to
release of counterions from the chain. For even larger wrapping
lengths there is no further counterion release.}
\end{figure}

Using Eqs.~\ref{fl2} and \ref{fl3} one can describe the
complexation as a function of chain stiffness. For large $l_{P}$,
$B^{-}>0$ and there is no wrapping; the free energy is minimized
for $l=0$. There is, however, still the possibility of more open
complexes with many point contacts between the sphere and the
chain which we shall discuss in Section 2.3. As $l_{P}$ is
reduced, $B^{-}$ changes sign
which marks the onset of wrapping. For $B^{-}<0$ and $C>0$ the minimum of $%
F_{1}\left( l\right) $ lies in between $l_{min}$ and $l_{iso}$. According to
Eq.~\ref{fl2} the position of the free energy minimum $l^{*}$ is given by
\begin{equation}
l^{*}=l_{iso}-CR_{0}/\xi  \label{ls}
\end{equation}
a result first given by Park, Bruinsma and Gelbart \cite{park99}. Further
reduction of $l_{P}$ leads to increasing $l^{*}$ until the complex reaches
the isoelectric point at $C=0$. For smaller $l_{P}$, $C<0$ and according to
Eq.~\ref{ls} $l^{*}>l_{iso}$ so that the complex is overcharged.
Consequently, for a fully flexible chain with $l_{P}=0$, the complex is {\it %
always} overcharged. The critical persistence length below which complexes
are overcharged is $l_{P}=2\left( \Omega +\ln \left( \kappa R_{0}\right)
\right) R_{0}^{2}/b$.

It can be seen clearly from this line of arguments that it is the
release of counterions from the chain that drives the
overcharging. What opposes this effect is the charging energy of
the complex, the repulsion between the chain and the overcharged
complex and, most importantly, the bending stiffness of the chain.

\subsubsection{Multi-sphere complex (highly charged case)}

In the previous section it was discussed how the release of
counterions from the chain upon adsorption causes overcharging.
Even though the focus of this chapter is on single-nucleosome
properties it is instructive to consider next the case of a highly
charged chain placed in a solution of oppositely, highly charged
spherical macroions. This case has been investigated by myself,
Bruinsma and Gelbart \cite{schiessel01}. The solution is
represented as a reservoir with concentration $c_{m}$ of
uncomplexed spheres. The chemical potential is the sum of the
usual ideal solution term and the electrostatic free energy of a
spherical macroion with charge $Z\gg Z_{max}$ (cf. Eq.~\ref{fcs}):
\begin{equation}
\frac{\mu _{sphere}}{k_{B}T}=\ln \left( c_{m}R_{0}^{3}\right) +Z\tilde{\Omega%
}\simeq Z\tilde{\Omega}  \label{ms}
\end{equation}
The number of spheres that complex with the chain are determined
by requiring this chemical potential to equal that of the
complexed spheres. In Ref.~\cite{schiessel01} we assumed a
beads-on-a-string configuration, with a mean spacing $D$ between
spheres. The Euclidean distance $S$ between the beginning and the
end of this configuration is related to the number of complexed
spheres via $N=S/D$. The wrapping length $l$ per sphere follows
from $S$ and $L$ to be $L\simeq Nl+S-2NR_{0}$. The Gibbs free
energy of the bead-on-a-string configuration is given by
\begin{equation}
G\left( N,l\right) =NF_{1}\left( l\right) +F_{int}\left( N,l\right) -\mu
_{sphere}N  \label{g1}
\end{equation}
where $F_{1}$ is the above given single-sphere complex free energy, Eq.~\ref%
{ftotal}, and $F_{int}$ is the interaction between the complexed spheres.
For sphere-sphere spacings $D\left( N,l\right) =S\left( N,l\right) /N$ with $%
2R_{0}<D<\kappa ^{-1}$, the complexed spheres feel a mutual electrostatic
repulsion approximately given by (for $\left| Z\left( l\right) \right|
<Z_{max}$)
\begin{equation}
\frac{F_{int}\left( N,l\right) }{k_{B}T}\simeq \Lambda N\frac{%
l_{B}Z^{2}\left( l\right) }{D\left( N,l\right) }  \label{fint}
\end{equation}
with $\Lambda \simeq 2\ln \left( \kappa ^{-1}N/L\right) $. For $D\approx
2R_{0}$ adjacent spheres interact via a strong excluded volume interaction,
for $D\gg \kappa ^{-1}$ the electrostatic interaction is screened.

$G\left( N,l\right) $ has to be minimized with respect to both $N$
and $l$. We showed in Ref.~\cite{schiessel01} that due to the
large chemical potential of the spheres (last term in
Eq.~\ref{g1}) it is energetically favorable to keep adding spheres
to the chain up to the point when $D\approx 2R_{0}$. At this point
the hard-core repulsion terminates complexation and the chain is
completely ''decorated'' with spheres. It follows then that the
number $N$ of spheres and the wrapping length $l$ per sphere are
related via $N\simeq L/l$, i.e., essentially the whole chain is in
the wrapped state. This argument holds for any
$l_{min}<l<l_{max}$. Because of this relation (namely $N=L/l$) the
Gibbs free energy depends only the number $N$ of complexed
spheres:
\begin{equation}
\frac{G\left( N\right) }{k_{B}T}\simeq N\left\{ \left( \Lambda +1\right)
\frac{l_{B}}{2R_{0}b^{2}}\left( l_{iso}-L/N\right) ^{2}-\frac{\mu _{sphere}}{%
k_{B}T}\right\} +const.  \label{gtot}
\end{equation}

Clearly, the first term of Eq.~\ref{gtot} favors the isoelectric
configuration $L/N=l_{iso}$. However, because of the second term, we can
lower the free energy further by increasing $N$ beyond $L/l_{iso}$. This is
not a small effect since $\mu _{sphere}/k_{B}T$ is of order $Z\gg Z_{max}$
while the first term of Eq.~\ref{gtot}, the capacitive energy, is of order $%
\left( l_{B}/R_{0}\right) Z_{max}^{2}\approx Z_{max}$ (since $Z_{max}\approx
R_{0}/l_{B}$). The spheres in the many-sphere complex are thus {\it %
undercharged}. The optimal wrapping length is as follows
\begin{equation}
l\simeq l_{iso}\left( 1-\frac{\tilde{\Omega}}{\Lambda +1}\frac{R_{0}/l_{B}}{Z%
}\right)  \label{lon}
\end{equation}

Physically, this effect can be illustrated by first setting $L/N=l_{iso}$.
In that case the complex is isoelectric. Now add one more sphere. By equally
redistributing the chain length between the $N+1$ spheres, one has an
individual wrapping length $l=L/\left( N+1\right) $ close to the isoelectric
one. Therefore the previously condensed counterions of the added sphere are
released and increase their entropy. By adding more and more spheres --
while reducing $l=L/N$ -- more and more counterions are liberated.

In both cases, the single-sphere case of the previous section and
the multi-sphere case discussed here, it is the counterion release
that is the driving force which brings oppositely, highly charged
macroions together. In the first case the release of the cations
of the chain is responsible for bringing more monomers to the
complex than necessary for its neutralization; in the second case
it is the release of the anions of the spheres that attracts more
spheres to the chain than ''optimal'' and the spheres are
undercharged. Comparing the similarities between the two cases it
might be more appropriate to say that in the latter case the
spheres overcharge the chain.

Finally, there is also the possibility to have a solution of
chains and spheres in a certain stoichiometric ratio such that
there are $N<L/l^{*}$ (with $l^{*}$ being the single-sphere
wrapping length, Eq.~\ref{ls}). Then essentially all spheres will
complex with the chains (due to the large contribution from the
counterions to the chemical potential, Eq.~\ref{ms}). Since in
this case there is enough chain available, each sphere will be
overcharged by the chain. Indeed, using Eq.~\ref{g1} with $\mu
_{sphere}=0$ we found in Ref.~\cite{schiessel01}
\begin{equation}
l^{*}\simeq l_{iso}-C\frac{R_{0}}{\xi }\left( 1-\frac{2\Lambda R_{0}N}{L}%
\right)  \label{ls2}
\end{equation}
as the optimal wrapping length. This is the single chain wrapping length,
Eq.~\ref{ls}, with a slightly reduced deviation from the isoelectric point
due to the electrostatic repulsion between the complexed spheres, Eq.~\ref%
{fint}.

\subsubsection{Weakly charged case}

The first theoretical models on sphere-chain complexation were presented in
1999, each of which used quite a different approach to this problem \cite%
{park99,mateescu99,gurovitch99,netz99}. Park, Bruinsma and Gelbart \cite%
{park99} considered a semiflexible and highly charged chain and showed that
counterion release leads to overcharging, as discussed in Section 2.2.1, cf.
Eq.~\ref{ls}. The other studies \cite{mateescu99,gurovitch99,netz99}
considered weakly charged chains and arrived at the conclusion that also in
this case overcharging should be a common phenomenon.

Gurovitch and Sens \cite{gurovitch99} studied a point like central
charge (the ''sphere'') and a connected chain of charges (the
flexible chain). On the basis of a variational approach
(self-consistent field theory using an analogy to quantum theory
\cite{degennes71}) they came to the conclusion that the chain
collapses on the central charge even if the total charge of the
resulting complex becomes ''overcharged.'' The critical polymer
charge up to which this collapse occurs is 15/6 times the central
charge. In a subsequent discussion \cite{golestanian99} it became
clear that this number has to be taken with caution and that other
effects, like the formation of tails, loops etc. were not included
in the class of trial function which were used in that study.

Mateescu, Jeppesen and Pincus \cite{mateescu99} used a purely
geometrical approach in order to calculate the zero-temperature
state of a complex of a sphere and a perfectly flexible chain in
the absence of any small ions (no salt, no counterions). They
divided the chain into two regions, one straight tail (or two
tails on opposite sites of the complex) and a spherical shell
around the macroion. The only approximation in that study was to
uniformly smear out the monomer charges within the spherical
shell. Starting from a point-like sphere and then gradually
increasing its radius $R_{0}$ they found the following typical
scenario (cf. Fig.~1 in that paper): For very small $R_{0}$ the
complex is slightly undercharged and shows two tails. With
increasing $R_{0}$ more and more chain wraps around the sphere
leading to an overcharging of the complex. For sufficiently large
sphere radius the whole chain is adsorbed. Before this point is
reached there are -- for sufficiently long chains -- two jump-like
transitions: one from the two-tail to the one-tail configuration
and then one from the one-tail case to the completely wrapped
state.

Finally, Netz and Joanny \cite{netz99} considered the complexation between a
semiflexible chain and a sphere. For simplicity, they considered a
two-dimensional geometry and calculated -- using a perturbative approach --
the length of the wrapped section and the shape of the two-tails for
different salt concentrations. That study focuses on the wrapping transition
and its discussion (together with that of subsequent studies, Ref.~\cite%
{kunze00,kunze02}) will be relegated to Section 2.3.

The four studies mentioned above agreed in that respect that
overcharging should be a robust phenomenon occurring in these
systems but there was still a transparent argument missing that
would clarify the nature of the underlying mechanism that leads to
overcharging. Nguyen and Shklovskii \cite{nguyen01} bridged this
gap by showing that correlations between the charged monomers
induced by the repulsion between the turns of the wrapped chain
can be considered as the basis for this effect. They studied again
a fully flexible chain and neglected the entropy of the chain
configurations. The chain is assumed to be in the one-tail
configuration with the tail radially extending from the sphere.
Then -- as it is the case in Ref.~\cite{mateescu99} -- the energy
of the chain-sphere complex is completely given by the
electrostatic interactions between the different parts:
\begin{eqnarray}
\frac{E}{k_{B}T}&\simeq& \frac{l_{B}\left( l-l_{iso}\right) ^{2}}{2R_{0}b^{2}}+%
\frac{l_{B}l}{b^{2}}\ln \left( \frac{\Delta }{r}\right)
+\frac{l_{B}\left( L-l\right) }{b^{2}}\ln \left(
\frac{L-l}{r}\right) \nonumber \\
&&+\frac{l_{B}\left( l-l_{iso}\right) }{b^{2}}\ln \left(
\frac{L-l+R_{0}}{R_{0}}\right) \label{nguyen1}
\end{eqnarray}
I use here the same symbols as in the previous sections (cf.
beginning of Section 2.2.1; $l_{iso}=bZ$ denotes again the
isoelectric wrapping length). The first term in Eq.~\ref{nguyen1}
is the charging energy of the complex (the sphere plus the wrapped
chain of length $l$), the second term is the self energy of the
wrapped chain portion and will be discussed in detail below. The
third term is the self-energy of the tail of length $L-l$, and the
forth term accounts for the interaction between the complex and
the tail.

I discuss now the second term in Eq.~\ref{nguyen1} following the
arguments given in Ref.~\cite{nguyen01}. The length $\Delta $
denotes the typical distance between neighboring turns of the
wrapped chain, i.e. $\Delta \approx R_{0}^{2}/l$. Consider an
isoelectric complex, $l=l_{iso}$, and assume that $\Delta \ll
R_{0}$ (multiple turns). Pick an arbitrary charged monomer on the
wrapped chain. It ''feels''\ the presence of other neighboring
charged monomers up to a typical distance $\Delta $ beyond which
the chain charges are screened by the oppositely charged
background of the sphere. Hence the wrapped portion of the chain
can be ''divided'' into fractions of length $\Delta $ that behave
essentially like rods of that length and of radius $r$ having a
self energy $\sim l_{B}\left( \Delta /b^{2}\right) \ln \left(
\Delta /r\right) $. One has $l/\Delta $ such portions leading
indeed to $l_{B}\left( l/b^{2}\right) \ln \left( \Delta /r\right)
$. Another interpretation of this term can be given as follows
(again following Ref.~\cite{nguyen01}): One can consider the
formation of
the complex as a two-step process. First one brings in sections of length $%
R_{0}$ and places them on the sphere in random positions and
orientations. This leads to a self energy $\sim l_{B}\left(
R_{0}/b^{2}\right) \ln \left( R_{0}/r\right) $ whereas the
interaction of each segment with the random background charge can
be neglected. Then, as the second step, one reorients and shifts
these pieces on the ball in order to minimize their mutual
electrostatic repulsion, i.e., one forms something like an
equidistant coil with distance $\Delta $ between the turns. Now
there is an additional contribution stemming from the attraction
between each chain piece of length $R_{0}$ with a stripe on the
sphere of length $R_{0}$ and width $\Delta $ leading to a gain in
the electrostatic energy, scaling as $-l_{B}\left(
R_{0}/b^{2}\right) \ln \left( R_{0}/\Delta \right) $. This
contribution from all these $R_{0}$-sections ($l/R_{0}$ pieces
leading to $-l_{B}\left(
l/b^{2}\right) \ln \left( R_{0}/\Delta \right) $) constitutes the {\it %
correlation energy} of the wrapped chain. Together with the self energy of
these pieces ($l/R_{0}$ times $l_{B}\left( R_{0}/b^{2}\right) \ln \left(
R_{0}/r\right) $) the correlation leads to $l_{B}\left( l/b^{2}\right) \ln
\left( \Delta /r\right) $ which is indeed the second term of Eq.~\ref%
{nguyen1}.

What is now the prediction of Eq.~\ref{nguyen1}? Minimization of $E$ with
respect to $l$ leads to the following condition \cite{nguyen01}
\begin{equation}
\left( l-l_{iso}\right) \left( \frac{1}{R_{0}}-\frac{1}{L-l+R_{0}}\right)
=\ln \left( \frac{l}{R_{0}}\frac{L-l}{L-l+R_{0}}\right) +2\simeq \ln \left(
\frac{l_{iso}}{R_{0}}\right)  \label{nguyen2}
\end{equation}
On the right hand side the argument of the logarithm was
simplified assuming $L-l\gg R_{0}$ (long tail), $l\gg R_{0}$ (many
turns of the wrapped chain) and $l\approx l_{iso}$.
Eq.~\ref{nguyen2} can be interpreted as follows: The left hand
side describes the cost (if $l>l_{iso}$) of bringing in a chain
segment from the tip of the tail to the surface, the simplified
term on the right hand side is the gain in correlation energy (cf.
Eq.~\ref{nguyen1}). This leads to the following optimal wrapping
length:
\begin{equation}
l^{*}\simeq l_{iso}+R_{0}\ln \left( \frac{l_{iso}}{R_{0}}\right)
\label{nguyen3}
\end{equation}
which demonstrates that the correlations induce indeed an
overcharging of the complex.

Note that Eq. \ref{nguyen3} gives the asymptotic value of $l^{*}$
for long chains, $L\gg l^{*}$, the case where most of the monomers
are located in the tail. As discussed in detail in
Ref.~\cite{nguyen01} one encounters a discontinuous transition
from the one-tail configuration to the completely collapsed state
when one
decreases $L$ to such a value that the length of the tail is just of order $%
R_{0}$ (up to a logarithmic factor). This collapse is similar to
the collapse discussed in Ref.~\cite{gurovitch99} but occurs for
much shorter
chains (namely for chains of the length $l_{iso}+R_{0}\ln \left( ...\right) $%
). The authors also considered the two-tail configuration and
showed that it it is formed, again in a discontinuous fashion, for
very large chains of length $L> l_{iso}^{2}/R_{0}$. This might be
-- contrary to the claim in Ref.~\cite{nguyen01} -- in qualitative
agreement with the prediction of Ref.~\cite{mateescu99} who found
-- for sufficiently large spheres -- with increasing $L$ two
discontinuous jumps, from the collapsed to the one-tail
configuration and from the one-tail to the two-tail state (cf.
Fig.~1 in that paper).

It is also worth mentioning that this theory can be easily extended to
semiflexible chains. One has to add Eq.~\ref{bend} to the free energy \ref%
{nguyen1} and finds (following the steps that led to Eq.~\ref{nguyen3}) for
the optimal wrapping length
\begin{equation}
l^{*}\simeq l_{iso}+R_{0}\ln \left( \frac{l_{iso}}{R_{0}}\right) -\frac{%
l_{P}b^{2}}{2R_{0}l_{B}}  \label{wrap}
\end{equation}
i.e., the mechanical resistance of the chain against bending decreases the
wrapped amount.

In Refs.~\cite{nguyen01b} and \cite{nguyen01c} Nguyen and
Shklovskii also considered the many-sphere case for weakly charged
components. Similar to the case considered in the previous section
they found in the case of an abundance of spheres undercharged
complexes (a phenomenon which they call ''polyelectrolyte charge
inversion'' as opposed to ''sphere charge inversion''). It was
shown that in this case the imbalance is mainly caused by the
reduction of the self-energy of each complexed sphere (cf.
Ref.~\cite{nguyen01b} for details).

\vspace{0in}Nguyen and Shklovskii argue that the correlation
effect is the basis of all the phenomena discussed in Section 2.2.
The approximation given in Ref.~\cite{mateescu99}, for instance,
is to homogeneously smear out the charges of the wrapped chain and
in this respect it overestimates the
gain in electrostatic energy upon complexation, i.e., the second term in Eq.~%
\ref{nguyen1} is neglected. They call this approximation the
''metallization approach'' \cite{nguyen01d}, an approximation that
obviously holds for sufficiently tight wrapping only. Also in that
reference they argue that the overcharging via counterion release
as discussed in Refs.~\cite{park99} (cf. Section 2.2.1) is
ultimately based on the correlation effect. Their argument is as
follows: Consider an isoelectric single-chain complex and assume
that the chain is wrapped in a random fashion around the sphere.
Then the electrical field close to the complex is essentially
vanishing. If more chain would be wrapped around the complex no
counterions would be released. It is only the fact that the chain
will be adsorbed in an orderly fashion due to its self-repulsion
that each section is surrounded by a correlation hole that leads
to counterion release even beyond the isoelectric point.

Sphere-chain complexes have also been considered in several computer
simulations \cite%
{wallin96,wallin96b,wallin97,wallin98,haronska98,jonsson01,chodanowski01,sakaue01,messina02,messina02b,jonsson02,akinchina02,dzubiella02}%
. Wallin and Linse \cite{wallin96}\ studied the effect of chain flexibility
on the geometry of a complex of a single sphere with a polyelectrolyte -- we
will come back to this problem in Section 2.3. The same authors also varied
the line charge density of the polyelectrolyte \cite{wallin96b}\ and the
radius of the sphere \cite{wallin97}; finally they considered the case when
there are many chains present \cite{wallin98}. Chodanowski and Stoll \cite%
{chodanowski01} considered the complexation of a flexible chain on
a sphere (assuming Debye-H\"{u}ckel interaction) and found good
agreement with Ref.~\cite{nguyen01} concerning overcharging and
the discontinuous transition to the one-tail configuration for
longer chains. The case of multisphere adsorption was studied by
Jonsson and Linse, having flexible \cite{jonsson01} or
semiflexible \cite{jonsson02} chains and taking explicitly into
account the counterions of the spheres and the chain. Their
findings show the same qualitative features as discussed in this
section above and in Section 2.2.2. A recent study by Akinchina
and Linse \cite{akinchina02} focused systematically on the role of
chain flexibility on the structure of the complex; the results
will be discussed below in Section 2.3. Messina, Holm and Kremer
\cite{messina02,messina02b} demonstrated that in the case of
strong electrostatic coupling (large values of $l_{B}$) it is even
possible that a polyelectrolyte chain forms a complex with a
sphere that carries a charge of the {\it same} sign -- a process
which is made possible by correlation effects making use of the
neutralizing counterions. Most recently Dzubiella, Moreira and
Pincus \cite{dzubiella02} studied the polarizibility of overall
neutral chain-sphere complexes in electrical fields as well as the
interaction between two such complexes.

\subsubsection{Physiological conditions}

Up to here I discussed only sphere-chain models for the case of
weak screening, $\kappa R_{0}<1$. At physiological condition,
however, the screening length is roughly 10 nm and hence ten times
smaller than the
overall diameter of the nucleosome. This section is devoted to this case ($%
\kappa R_{0}\gg 1$).

I will mainly focus here on weakly charged chains and spheres
where linear Debye-H\"{u}ckel theory can be applied. For strong
screening, $\kappa R_{0}\gg 1$ the potential $\phi _{sphere}$
close to the ball looks essentially like that
of a charged plane with charge density $Z/\left( 4\pi R_{0}^{2}\right) $%
: $e\phi _{sphere}\left( h\right) /k_{B}T\simeq \left(
l_{B}Z/2\kappa R_{0}^{2}\right) e^{-\kappa h}$ ($h$: height above
surface). Neighboring turns of the adsorbed chain have locally the
geometry of (weakly) charged rods for which it has been predicted
that they form a lamellar phase \cite{netz99b,schiessel01d}. The
lamellar spacing $\Delta $ follows from the competition between
the chain-sphere attraction and the chain-chain repulsion. The
chain-sphere attraction leads to the following adsorption energy
per area:
\begin{equation}
\frac{f_{chain-sphere}}{k_{B}T}\simeq -\frac{l_{B}Z}{2\kappa
R_{0}^{2}b\Delta }  \label{ads}
\end{equation}
assuming that the chain is so thin that its adsorbed charged monomers feel
an unscreened attraction to the surface, $\kappa r\ll 1$. To calculate the
rod-rod repulsion one starts from the potential around a single rod: $e\phi
_{chain}\left( R\right) /k_{B}T=-2l_{B}b^{-1}K_{0}\left( \kappa R\right) $ ($%
R$: radial distance from rod axis); $K_{0}$ denotes the modified
Bessel
function that has the asymptotics $K_{0}\left( x\right) \simeq -\ln x$ for $%
x\ll 1$ and $K_{0}\left( x\right) \simeq \left( \pi /2x\right)
^{1/2}\exp \left( -x\right) $ for $x\gg 1$. This leads to the
following free energy density of the chain-chain repulsion:
\begin{equation}
\frac{f_{chain}}{k_{B}T}=\frac{2l_{B}}{b^{2}\Delta }\sum_{k=1}^{\infty
}K_{0}\left( k\kappa \Delta \right)  \label{dh4}
\end{equation}
To proceed further one might consider two limiting cases. If the lamellar
spacing is much smaller than the screening length, $\kappa \Delta \ll 1$,
the sum in Eq.~\ref{dh4} can be replaced by an integral \cite%
{netz99b,schiessel01d}:
\begin{equation}
\frac{f_{chain}}{k_{B}T}\approx \frac{2l_{B}}{b^{2}\Delta }\int_{0}^{\infty
}K_{0}\left( k\kappa \Delta \right) dk=\frac{\pi l_{B}}{b^{2}\kappa \Delta
^{2}}  \label{dh5}
\end{equation}
The free energy density $f=f_{chain-sphere}+f_{chain}$ is then
minimized for the {\it isoelectric} lamellar spacing: $\Delta
=4\pi R_{0}^{2}/\left( bZ\right) $ leading to the wrapping length
$l\simeq 4\pi R_{0}^{2}/\Delta =bZ=l_{iso}$. Note, however, that
going to the continuous limit means to smear out the charges,
neglecting the correlation energy discussed after
Eq.~\ref{nguyen1}. This is similar to the approximation used by
Mateescu, Jeppesen and Pincus \cite{mateescu99} who studied the
unscreened case (discussed above in Section 2.2.3). There they
considered, however, the self-energy of the tails, a contribution
driving more chain monomers to the sphere and hence leading to
overcharging. This overcharging was overestimated since the
correlation effects (included in
the theory of Nguyen and Shklovskii \cite{nguyen01}, cf. also Eq.~\ref%
{nguyen1}) were washed out. Here, on the other hand, for strong
screening the self-energy of a chain section remains the same,
whether it is adsorbed or not (on a length scale $\kappa ^{-1}$ it
always looks straight). Hence it is here appropriate not to
include the tail contribution.

This might lead one to expect that there is no overcharging for
the case of strong screening. However, as mentioned above, there
is an approximation involved when going from Eq.~\ref{dh4} to
\ref{dh5}. This approximation is only good for $\Delta \simeq r$.
A more careful calculation leads also here to the prediction of
overcharging. To see this one has to realize that $\int
K_{0}\left( k\kappa \Delta \right) dk-\sum\nolimits_{k}K_{0}\left(
k\kappa \Delta \right) \approx \int_{0}^{1}K_{0}\left( \kappa
\Delta \right) dk\simeq -\ln \left( \kappa \Delta \right) $.
Taking this into account one can replace Eq.~\ref{dh5} by
\begin{equation}
\frac{f_{chain}}{k_{B}T}\simeq \frac{\pi l_{B}}{b^{2}\kappa \Delta ^{2}}%
\left( 1+\frac{2}{\pi }\kappa \Delta \ln \left( \kappa \Delta \right) \right)
\label{dh6}
\end{equation}
Minimizing the free energy density with this additional
contribution (coming from correlation effects) leads to a slightly
smaller lamellar spacing
\begin{equation}
\Delta \approx \frac{b^{-1}}{\frac{Z}{4\pi R_{0}^{2}}+\frac{\kappa }{\pi b}%
\ln \left( \frac{Zb}{4\pi R_{0}^{2}\kappa }\right) }  \label{lam}
\end{equation}
and to a wrapping length that is larger than the isoelectric one
(overcharging):
\begin{equation}
l^{*}=\frac{4\pi R_{0}^{2}}{\Delta }\approx l_{iso}+4R_{0}^{2}\kappa \ln
\left( \frac{l_{iso}}{4\pi R_{0}^{2}\kappa }\right)  \label{l2}
\end{equation}
Lowering the ionic strength leads to smaller $\kappa $-values and
hence to a reduction of the degree of overcharging. When $\kappa
^{-1}\approx R_{0}$ one recovers Eq.~\ref{nguyen3}, the result
presented by Nguyen and Shklovskii \cite{nguyen01} for the case of
weak screening.

Netz and Joanny \cite{netz99} also considered the case of spheres
with an even smaller charge density $Z/4\pi R_{0}^{2}$ where
$\Delta
>\kappa ^{-1}$. In that case only the interactions with the two next
neighboring turns count. From Eq.~\ref{dh4} follows
\begin{equation}
\frac{f_{chain}}{k_{B}T}=\frac{\sqrt{2\pi }l_{B}}{b^{2}\kappa ^{1/2}\Delta
^{3/2}}e^{-\kappa \Delta }  \label{dh7}
\end{equation}
Minimizing $f=f_{chain-sphere}+f_{chain}$ leads then approximately to
\begin{equation}
\Delta \simeq \kappa ^{-1}\left( \ln \left( \frac{R_{0}^{2}\kappa }{bZ}%
\right) +1\right)  \label{lam2}
\end{equation}
i.e., the spacing is of the order of the screening length (up to logarithmic
corrections). The overcharging can then become very large. Clearly the term
''overcharging'' becomes quite questionable when there is such a strong
screening that charges in the complex interact only very locally over length
scales of order $\kappa ^{-1}\ll R_{0}$.

So far the bending energy was not accounted for, i.e., the chain was assumed
to be perfectly flexible. Bending leads to an additional energy (per area): $%
f_{bend}\simeq l_{P}/\left( 2R_{0}^{2}\Delta \right) $. This contribution
scales with the lamellar spacing as $1/\Delta $ as also the chain-sphere
attraction, Eq.~\ref{ads}, does. One can therefore interpret the bending to
renormalize the sphere charge to a smaller value $\widetilde{Z}%
=Z-l_{P}b\kappa /l_{B}$. In fact, $l_{iso}$ in Eq.~\ref{l2} has to
be replaced by $l_{iso}-l_{P}b^{2}\kappa /l_{B}$.

When going to highly charged systems one encounters nonlinear
screening (counterion condensation). This case has been
extensively discussed by Nguyen, Grosberg and Shklovskii
\cite{nguyen00}. They showed that the case $\Delta >\kappa ^{-1}$
corresponds essentially to the above described case of strongly
screened lamellas, Eq.~\ref{lam2}, but with an effective rod line
charge density $1/l_{B}$ (instead of the bare value $1/b$). There
is nearly no counterion release in this case. The other limit
$\Delta \ll \kappa ^{-1}$
is highly non-linear and quite complicated. For details I refer the reader to Ref.~\cite%
{nguyen00} (section VI in that paper).

Concluding, what can be learned from the sphere-chain systems with regard to
the nucleosome structure? It is clear that many of the assumptions entering
the models do not agree with the details of the nucleosome conformation.
Most of the studies discussed in Sections 2.2.1 to 2.2.3 assume a weak
screening which does not correspond to physiological conditions. Many
studies neglect the bending energy which is a major energetic penalty for
DNA wrapping around the octamer. Hydrogen bonds, solvent effects are not
included. Most importantly, these studies do not account for the fact that
the wrapping path is clearly prescribed by more or less specific binding
patches on the octamer surface (cf. Section 2.1). And the octamer is, of
course, not a sphere. Even when going to the strong screening case,
discussed above, the assumed lamellar arrangement of the wrapped chain has
only a vague resemblance to the 1 and 3/4 turns of wrapped nucleosomal DNA.
Here one might at least say that the $\sim 28$\AA\ pitch of the superhelical
ramp is in rough agreement with the prediction $\Delta \approx \kappa ^{-1}$
given in Eq.~\ref{lam2}.

Nevertheless, the improved understanding of the chain-sphere
complexes that has been achieved in the recent years is in my
opinion very helpful. Many of the investigated model systems
resemble closely complexes between
macroions (colloids \cite{ganachaud97,sukhorukov98}, dendrimers \cite%
{kabanov00}, charged micelles \cite{mcquigg92} etc.) and synthetic polymers
which is of technological relevance as a means of modifying macroion
solution behavior. But also for the nucleosome the sphere-chain systems help
to understand its behavior under changing ionic conditions. Especially when
the ionic strength is lowered the electrostatics becomes long ranged. In
this case many local details become overruled by the electrostatics. The
behavior of the nucleosome at low and high ionic strength will be the
subject of the next section.

\subsection{ Unwrapping transition}

The nucleosomal complex is only stable at intermediate salt
concentrations. In Section 2.3.1 I will give an overview over
experimental results that shed some light on the instabilities of
the nucleosome core particle at low and high ionic strengths. I
also report on how these instabilities can be understood in the
framework of a sphere-chain model (for {\it short} chain length).
Then I provide a thorough discussion of the instabilities of
sphere-chain complexes at high ionic strength in Section 2.3.2 and
at low ionic strength in Section 2.3.3 -- both, for chains of {\it
arbitrary} chain length.

\subsubsection{Instabilities of the nucleosome core particle at low and at
high ionic strength}

Yager, McMurray and van Holde \cite{yager89} characterized the
stability of the nucleosome core particle as a function of the
salt concentration (NaCl) and the concentration of core particles
(measured via the 260 nm absorbance). Using a variety of
experimental methods (velocity sedimentation, gel exclusion
chromatography and gel electrophoresis) they arrived at the
following main conclusions (cf. also the schematic phase diagram,
Fig.~4). For not too low concentrations core particles are stable
for ionic strengths ranging from 2 mM to 750 mM (called region 1
in that paper; this includes physiological relevant salt
concentrations $\sim 100$ mM). For slightly higher salt
concentrations (region 3 \cite{yager89}) or low concentration of
core particles (region 2) the DNA is partially dissociated; an
equilibrium between histones, free DNA and core particles is
observed. Especially, in region 3 there is also the occurrence of
(H2A-H2B)-depleted particles. At salt concentrations beyond 1.5 M
the core particle is completely dissociated into histone oligomers
(the (H2A-H2B)-dimer and the (H3-H4)$_{2}$-tetramer) and free DNA
(region 4). On the other end, for very low salt concentration $<1$
mM (called region 1E in \cite{yager89}) one finds an ''expanded''
state of the nucleosome.

\begin{figure}
\begin{center}
\includegraphics*[width=10cm]{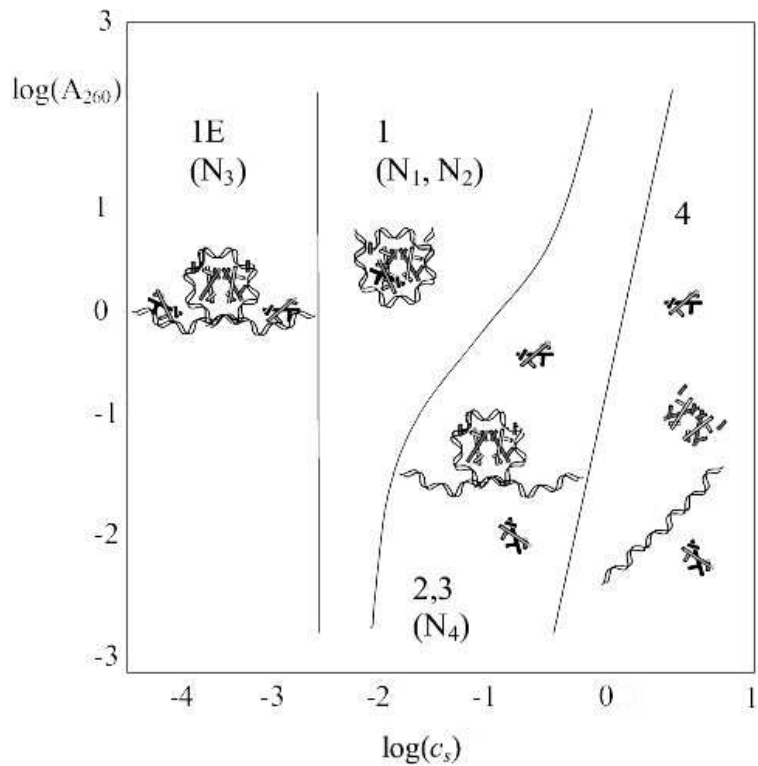}
\end{center}
\caption{The effects of salt on the conformation and stability of
nucleosome core particles as a function of the salt concentration
$c_s$ and the DNA concentration measured by the 260 nm adsorbance
(adapted from Ref.~\cite{yager89}). The notations of
Ref.~\cite{yager89} for the different states are given as well as
the ones of Ref.~\cite{khrapunov97} (in brackets).}
\end{figure}

Using a different experimental approach (measurement of the
fluorescence of the aa tyromisine in the histone proteins)
Khrapunov et al.~\cite{khrapunov97} came to similar conclusions
(cf. Fig.~4): For ionic strengths between 5 and 600 mM the core
particle is intact but one finds different degrees of contact
between the core histones (the resulting two forms are called
N$_{1}$ and N$_{2}$ in \cite{khrapunov97}). At larger ionic
strength ($\approx 1.2$ M) the terminal regions of the DNA unwrap
and the (H2A-H2B)-dimers are
dissociated (N$_{4}$) and at an even larger value ($\approx 1.5$ M) the (H3-H4)%
$_{2}$-tetramer leaves the DNA. Finally, at low salt concentations
one encounters an open state (called the N$_{3}$-form): the dimers
break their contact with the tetramer and the DNA termini unwrap
(with the dimers still attached to them).

The key features of the behavior of core particle DNA (neglecting
the substructure of the octamer) are indeed recovered in the
framework of the sphere-chain models. Most clearly this has been
demonstrated in a study by Kunze and Netz \cite{kunze00,kunze02}
(cf. also \cite{netz99}). They considered the complexation of a
charged, semiflexible chain with an oppositely charged sphere. All
charges in their system interact via a
standard Debye-H\"{u}ckel potential. The optimal DNA configuration ${\bf r}%
\left( s\right) $ ($0\leq s\leq L$) follows then from the minimization of
the free energy functional
\begin{eqnarray}
\frac{{\cal F}\left\{ {\bf r}\left( s\right) \right\} }{k_{B}T}&=&\frac{l_{P}}{%
2}\int_{0}^{L}ds\left( \frac{1}{R\left( s\right) }\right) ^{2}-\frac{l_{B}Z}{%
b\left( 1+\kappa R_{0}\right) }\int_{0}^{L}\frac{e^{-\kappa \left(
\left| {\bf r}\left( s\right) \right| -R_{0}\right) }}{\left| {\bf
r}\left( s\right) \right| } \nonumber \\
&&+\frac{l_{B}}{b^{2}}\int_{0}^{L}ds\int_{s}^{L}ds^{\prime }%
\frac{e^{-\kappa \left( \left| {\bf r}\left( s\right) \right| -\left| {\bf r}%
\left( s^{\prime }\right) \right| \right) }}{\left| {\bf r}\left( s\right) -%
{\bf r}\left( s^{\prime }\right) \right| }  \label{dh}
\end{eqnarray}
The first term on the right hand side is the bending energy of the
chain where $R\left( s\right) $ denotes its curvature at point $s$
along the contour. The second and third term account for the
electrostatic attraction between monomers and the sphere and the
monomer-monomer repulsion, respectively. The symbols are the same
as used above (cf. Section 2.2.1). Kunze and Netz chose the
parameters ($l_{P}$, $L$, $b$ and $R_{0}$) such as to mimic the
values of the core particle and varied $Z$ and $\kappa ^{-1}$ as
''free''\ parameters. The optimal shape ${\bf r}\left( s\right) $
was found by numerical minimization of Eq.~\ref{dh} and
characterized by two order parameters, a rotational and a
torsional one. They found the following overall picture: For
reasonable\ values of the sphere charge $Z$ one finds for
vanishing ionic strength ($\kappa ^{-1}\rightarrow \infty $) an
open, planar configuration where only a small fraction of the
chain is wrapped whereas the two tails (of equal length) are
extended into roughly opposite directions. This is reminiscent of
the open structures reported in the
experimental studies (region 1E in \cite{yager89}, N$_{3}$%
-form in \cite{khrapunov97}). Upon addition of salt the structure
stays first extended but looses at some point its rotational
geometry (in form of a transition from a two- to a one-tail
configuration) and then at even higher ionic strength the chain
goes from a planar to a non-planar configuration. It begins then
to wrap more and more and -- at some point -- regains its
rotational symmetry (the wrapping path resembles then some kind of
''tennisball seam pattern''). At that point already (which is well
below physiological ionic strengths) the chain is almost
completely wrapped. It stays in this wrapped state up to very high
salt concentrations. Only then the chain unwraps in a
discontinuous fashion because the chain-sphere attraction is
sufficiently screened. Also these features of the complex (the
wrapped compact state in a wide range around physiological
conditions and
the unwrapping at high salt content) reflect findings in Refs.~\cite%
{khrapunov97} and \cite{yager89}. The behavior of such complexes
for chains of {\it arbitrary} length is the subject of the next
two sections.

\subsubsection{The rosette state at high ionic strength}

For physiological conditions (or for even higher salt
concentrations) the electrostatic interaction between the DNA
chain and the octamer can be considered as short-ranged. It is
then usually sufficient to assume some attractive short-range
interaction between the chain and the octamer with a range of
interaction $\kappa ^{-1}\ll R_{0}$. In this spirit Marky and Manning \cite%
{marky91} considered the wrapping of a semiflexible chain around a
cylinder. They came up with the picture of a simple ''all or
none'' unwrapping transition. Denote with $\lambda $ the
adsorption energy per length of the chain on the cylinder (in
units of $k_{B}T$). Then for each additional wrapped length
$\Delta l$, one gains $\Delta E_{ads}/k_{B}T=-\lambda \Delta l$.
On the other hand, in order to wrap the chain it has to be bent
with a curvature $R_{0}^{-1}$ with $R_{0}$ being the radius of the
octamer; this leads to an energetic cost $\Delta
E_{elastic}/k_{B}T=l_{P}\Delta l/2R_{0}^{2}$, cf. Eq.~\ref{bend}.
From this follows that if the adhesion energy $\lambda $ is larger
than
\begin{equation}
\lambda _{c}=\frac{l_{P}}{2R_{0}^{2}}  \label{lambdac}
\end{equation}
then more and more chain will wrap around the histone spool (up to a point
when there are no adsorption sites available anymore). On the hand, for $%
\lambda <\lambda _{c}$, the chain unwraps completely. This
unwrapping transition can also be induced by increasing the
persistence length of the chain, cf. Eq.~\ref{lambdac}.

This, however, is not the complete picture. Already the numerical
study by Wallin and Linse \cite{wallin96} indicated that with
increasing chain stiffness one encounters a {\it gradual} change
of the conformations of the complexed chain towards more extended
structures. In Ref.~\cite{schiessel00} myself, Rudnick, Bruinsma
and Gelbart showed in a systematic analytical study that there is
indeed a wide range of parameters in which more open, multi-leafed
(''rosette'') states occur in this system. The results of this
model study together with some additional material will be
presented in the current section. Following that reference I will
first discuss the ground-state configurations of the sphere-chain
complex and then account for thermal fluctuations. Then I will
present the general phase diagram that
includes wrapped ($\lambda >\lambda _{c}$) and open ($\lambda <\lambda _{c}$%
) structures.

In Ref.~\cite{schiessel00} we started from the popular {\it
worm-like chain model} (WLC) which provides a good description of
the mechanical properties of DNA (for reviews cf.
Refs.~\cite{frank97,schlick95}). The chain molecule is represented
by a semiflexible tube of radius $r$ characterized by two elastic
moduli, the bending and torsional stiffnesses. The elastic energy
of a WLC of length $L$ can be expressed as
\begin{equation}
E_{elastic}=\frac{1}{2}\int_{0}^{L}ds\left[ A\left( \frac{1}{R\left(
s\right) }\right) ^{2}+C\left( \frac{d\Theta }{ds}\right) ^{2}\right]
\label{WLC}
\end{equation}
Here $A$ is the bending stiffness and $1/R\left( s\right) $ the curvature of
the chain at point $s$ along its contour. The stiffness is usually expressed
as $A=k_{B}Tl_{P}$ where $l_{P}$ is the orientational persistence length of
the chain -- as given above in Section 2.1, cf. Eq.~\ref{bend}. The
torsional angle of the chain is $\Theta $ and the torsional stiffness is $C$%
, for DNA $C\simeq k_{B}T\times 750$\AA\ \cite{hagerman88}. In addition to
this bending contribution there is a short-ranged attraction between the
chain and the sphere (or cylinder) with a range of interaction $\delta \ll r$%
.

As mentioned above for strong attraction $\lambda >\lambda _{c}$
the chain is wrapped around the sphere. On the other hand, if
$\lambda <\lambda _{c}$, the WLC can only make point contacts to
the sphere. The energy of a contact point (in units of $k_{B}T$)
\begin{equation}
\mu \simeq \lambda \sqrt{R_{0}\delta }  \label{mu}
\end{equation}
follows from the length $\sqrt{R_{0}\delta }$ of chain portion
around that point contact that is located within the distance
$\delta $ from the sphere. In Ref.~\cite{schiessel02b} I gave the
quantities $\lambda $ and $\delta $ in terms of strongly screened
electrostatics. In that case $\delta =\kappa
^{-1}$. The adsorption energy per length can be estimated from the Debye-H%
\"{u}ckel electrostatic potential close to the surface (cf. beginning of
Section 2.2.4). From Eq.~\ref{ads} follows
\begin{equation}
\lambda =\frac{l_{B}Z}{2\kappa R_{0}^{2}b}  \label{lamda}
\end{equation}
and hence the unwrapping into the rosette takes place when $\lambda $
reaches the critical value given by Eq.~\ref{lambdac}, i.e., when the
persistence length
\begin{equation}
l_{P}=\frac{l_{B}Z}{\kappa b}  \label{unwrap}
\end{equation}
is reached. That this is the upper bound for $l_{P}$ for having a stable
wrapped complex has been predicted by Netz and Joanny \cite{netz99} (cf.
Eq.~(35) in that paper).

For $\lambda >\lambda _{c}$ the optimal number $M^{\ast }$ of point contacts
as well as the preferred configuration of the chain is obtained by a
minimization of the energy
\begin{equation}
E=E_{elastic}-k_{B}T\mu M  \label{e1}
\end{equation}
with $E_{elastic}$ given by Eq.~\ref{WLC}. In
Ref.~\cite{schiessel02b} we searched first for the minima of the
elastic energy (zero temperature conformations), neglecting
thermal fluctuations that were included in a second step (see
below).

The search can be performed systematically by applying the {\it Kirchhoff analogy} \cite%
{love44,nizette99} which relates stationary points of the WLC energy, Eq.~%
\ref{WLC}, to the well-studied classical mechanics problem of the
trajectory of a supported, symmetric spinning top in a gravity
field. It can be easily
shown that the action of the spinning top has precisely the same form as Eq.~%
\ref{WLC} with the time playing the role of the arc length, the
orientation of the top corresponding to the tangent vector of the
WLC, the gravitational force being a tension acting on the
rod\footnote{The tension comes here from the \textquotedblright
sticky\textquotedblright\ sphere that induces the rosette
structure discussed below. Later, in Section 2.4.2, I will give an
example where it becomes more obvious how the rod tension formally
enters the Hamiltionian in form of a Lagrange multiplier $T$
-- leading to a term that resembles the potential energy of the spinning top.%
} etc. This analogy has been repeatedly applied to DNA related
problems during the last 20 years
(e.g. see \cite%
{benham77,benham79,lebret79,lebret84,shi94,fain97,swigdon98,fain99,coleman00,zandi01}%
). For a nice visual review on the spinning top-elastic rod
analogy the reader is referred to Ref.~\cite{nizette99}.

In Ref.~\cite{schiessel00} we presented solutions of the
corresponding Euler-Lagrange equation that combine the following
features: ({\it i}) The WLC closes on itself\footnote{This
assumption is only for technical reasons. As discussed below, the
behavior for an open chain is essentially identical. The solutions
for closed WLC can be characterized in terms of the topologically
conserved linking number. Loops with a non-zero linking number
show a configuration that combines twist and spatial distortion
(known as \textquotedblright writhing\textquotedblright ).}; ({\it
ii}) it is possible to inscribe a
sphere of radius $R_{0}$ inside the WLC that touches the WLC at $M$ points; (%
{\it iii}) there is no self-intersection of the WLC chain with
itself if it is surrounded by a tube of radius $r$; and ({\it iv})
the solution is stable against small perturbations. The resulting
rosette-type configurations can be characterized by the number of
loops $M$. Fig.~5 shows such a rosette (computed numerically); as
indicated in the figure it is indeed possible to inscribe a sphere
(or a cylinder) in the central hole of the rosette. For each $M$,
we adjusted the linking number of the loop to minimize the elastic
energy. By varying the degree of spatial distortion (characterized
by the so-called "writhe") a family of solutions was obtained, for
given $M$, with different hole diameters. Solutions with the
maximum amount of writhe have the smallest central hole diameter
as well as the lowest elastic energy. The inset of Fig.~5 shows
the elastic energy of a loop of length $L$, in dimensionless
units, for the rosette state, as a function of the degree of
writhe.

\begin{figure}
\begin{center}
\includegraphics*[width=9cm]{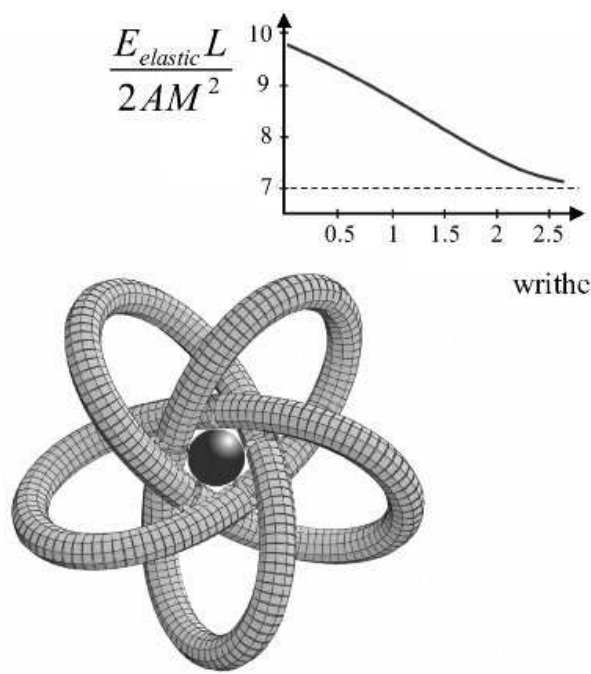}
\end{center}
\caption{Five-leafed rosette. This configuration corresponds to
the minimal energy solution of the Euler-Lagrange equation with a
maximum amount of writhe and the smallest central hole (cf. text
for details)}
\end{figure}

These solutions are actually saddle points of the WLC energy,
i.e., there is a finite subset of infinitesimal distortions that
lower the elastic energy of the WLC; the rest raises the energy or
leaves it unchanged. The role of the sphere is to stabilize this
saddlepoint and turn it into a real maximum (for a detailed
discussion cf. Ref.~\cite{fain99}). The energy of a minimum-hole
rosette depends on the overall chain length $L$ and the number of
point contacts $M$ (= number of leaves) as
\begin{equation}
\frac{E_{min}\left( M\right) }{k_{B}T}\simeq \frac{2\chi l_{P}M^{2}}{L}-\mu M
\label{emin}
\end{equation}
with $\chi =7.02$ (see inset of Fig.~5). This result can be
understood making use of results of the earlier work by Yamakawa
and Stockmayer (YS) \cite{yamakawa72} who showed that a loop of
length $l$, formed by imposing common endpoints on a WLC strand,
assumes the form of a lemniscate-shaped
leaf with an 81-degrees apex angle. The elastic bending energy of a leaf is $%
e\left( l\right) =2\chi A/l$ and $E_{min}\left( M\right) $ given above just
equals $Me\left( L/M\right) $ plus the adhesion energy. We verified
numerically that the leafs of the rosette indeed have apex angles close to
81 degrees. The energy $E_{min}\left( M\right) $ exhibits a minimum as a
function of the number $M$ of rosette leaves for
\begin{equation}
M^{*}=\frac{\mu L}{2\chi l_{P}}  \label{ns}
\end{equation}
We thereby obtained an open, multileafed structure controlled by the
adhesion energy and the persistence length that competes with the wrapped
state. Packing considerations imply an upper limit for the number of loops
of order $M_{max}\approx \left( R_{0}/r\right) ^{3/2}$; this is the maximal
number of contacts, each excluding an area $\approx r\sqrt{R_{0}r}$, that
can be closely packed on the surface of the sphere.

Note that the above results do not depend critically on the assumption that
the chain forms a closed loop. In fact, for a chain with open ends we expect
the same number of rosette leaves, each again having approximately the shape
of a YS-loop with an $81{{}^{\circ }}$ apex angle. This means that here
neighboring leaves will have a relative orientation of $\sim 180{{}^{\circ }}%
-81{{}^{\circ }}=99{{}^{\circ }}$. In addition, the leaves will be
slightly twisted with respect to each other (like propeller
blades) to account for the mutually excluded volume. I give an
overview over the results in a \textquotedblright phase
diagram\textquotedblright\ (ground state configurations) in Fig.~6
(adapted from Ref.~\cite{schiessel02b}; an alternative
presentation with a different choice of axes is presented in
Ref.~\cite{schiessel00}).

\begin{figure}
\begin{center}
\includegraphics*[width=7cm]{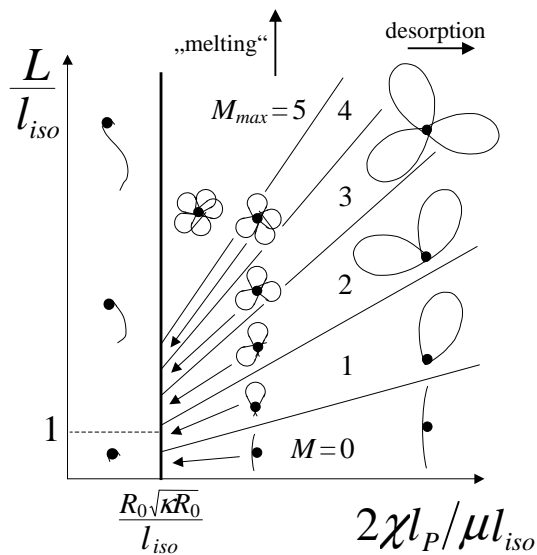}
\end{center}
\caption{The sphere-chain complex in the case of short-ranged
attraction (for instance, at high ionic strength). Depicted is the
diagram of states as a function of the total length $L$ of the
chain and its persistence length $l_P$ divided by the point
contact energy $\mu$ (both axes in units of $l_{iso}$). The thick
vertical line indicates the sharp unwrapping transition from the
wrapped to the rosette-type complexes.}
\end{figure}

We next studied in Ref.~\cite{schiessel00} the stability of the
rosette against {\it thermal fluctuations}. We started from a
single, large loop of length $L$ and constructed the rosette step
by step, by attaching to the sphere lemniscate-shaped leafs of
variable length of the kind examined by YS \cite{yamakawa72}. The
finite-temperature free energy cost $F\left( l_{leaf},\phi \right)
$ of introducing a single leaf of length $l_{leaf}$ and apex angle
$\phi $ into a large strand was computed by YS. Using
path-integral methods they found \cite{yamakawa72}
\begin{equation}
\frac{f\left( l_{leaf},\phi \right) }{k_{B}T}\simeq \left\{
\begin{array}{ll}
2\chi \frac{l_{P}}{l_{leaf}}+\ln \frac{l_{leaf}}{l_{P}}+W\left( \phi \right)
+... & \mbox{for}\;l_{leaf}\ll l_{P} \\
\frac{3}{2}\ln \frac{l_{leaf}}{l_{P}}+... & \mbox{for}\;l_{leaf}\gg l_{P}%
\end{array}
\right.  \label{flf}
\end{equation}
The function $W\left( \phi \right) $ has a minimum when the apex angle of
the leaf is approximately $81{{}^\circ}$. The logarithmic contribution to $%
f\left( l_{leaf},\phi \right) $ -- associated with the
configurational entropy of the loop neglecting excluded volume
interaction ($\theta $ solvent) -- imposes a free energy penalty
for {\it large} leaves, $l_{leaf}\gg l_{P}$ (corresponding to the
entropy of a closed random walk in 3 dimensions). The enthalpic
$1/l_{leaf}$ contribution imposes an energy penalty for {\it
small} leaves. Hence, for given $\phi $, $f\left( l_{leaf},\phi
\right) $ as a function of $l_{leaf}$ has a shallow minimum, its
value being close to that of the persistence length $l_{P}$. The
total free energy cost $F_{M}\left(
\left\{ l_{i}\right\} \right) $ of introducing into a large loop an $M$%
-leafed rosette for a fixed distribution $\left\{ l_{i}\right\} $ of leaf
lengths is then given by

\begin{equation}
\frac{F_{M}\left( \left\{ l_{i}\right\} \right) }{k_{B}T}\simeq
\sum\limits_{i=1}^{M}\frac{f\left( l_{i}\right) }{k_{B}T}-\mu M  \label{fn}
\end{equation}
Here
\begin{equation}
\frac{f\left( l\right) }{k_{B}T}=\frac{2\chi l_{P}}{l}+\frac{3}{2}\ln \left(
\frac{l}{l_{P}}\right)  \label{fli}
\end{equation}
constitutes an interpolation formula between the large and small
$l$ limits of $f\left( l,81{{}^{\circ }}\right) $ as given by
Eq.~\ref{flf}. The partition function $Z_{M}$ for an $M$-leafed
rosette follows by integration over all possible leaf
distributions
\begin{equation}
Z_{M}=\int_{0}^{\infty }\prod_{i=1}^{M}\frac{dl_{i}}{r}\exp \left[ -\frac{1}{%
k_{B}T}\left( \sum\limits_{i=1}^{M}f\left( l_{i}\right)
+P\sum\limits_{i=1}^{M}l_{i}\right) \right] =\left( Z_{1}\right) ^{M}
\label{gp}
\end{equation}
Assuming $\mu \gg 1$ (strong sticking points) one can assume the number of
leaves always to be maximal, $M=M_{max}$ (the index is dropped here and in
the following for simplicity of notation). In Eq.~\ref{gp} a Lagrangian
multiplier $P$ is introduced in order to satisfy the constraint $%
\sum_{i=1}^{M}l_{i}=L$. Physically, $P$ is the overall tension of the loops
induced by their adhesion to the sphere. The free energy follows to be
\begin{equation}
G\left( P\right) =-k_{B}T\ln Z_{M}=Mk_{B}T\left\{ 2\sqrt{\frac{2\chi l_{P}P}{%
k_{B}T}}+\ln \left( \frac{r/l_{P}}{\sqrt{\pi /2\chi }}\right) \right\}
\label{gp2}
\end{equation}

It is interesting to note that $G\left( P\right) $ is
mathematically identical to the free energy of a one-dimensional
many-body system of $M$ particles under a ''pressure'' $P$
confined to a circular track of length $L$. The particles are
interacting via a ''nearest-neighbor pair-potential'' $f\left(
l\right) $ while $k_{B}T\mu $ is the ''chemical potential'' of the
particles. According to Eq.~\ref{fli}, the effective pair
potential $f\left( l\right) $ is concave (i.e., $d^{2}f/dl^{2}<0$)
for inter-particle spacings exceeding a spinodal threshold spacing
of order $l_{P}$. Experience with mean-field theory of many-body
systems suggests that one should expect phase-decomposition if the
average spacing between particles exceeds a spinodal threshold.

Using $L=dG/dP$ one finds the non-linear ''tension-extension'' curve $%
P\left( L\right) =2\chi l_{P}k_{B}T\left( M/L\right) ^{2}$. Using Eqs.~\ref%
{gp} and \ref{gp2}, it is straightforward to compute the first and (reduced)
second moments of the leaf size distribution:
\begin{equation}
\left\langle l_{leaf}\right\rangle =-k_{B}T\frac{d}{dP}\ln Z_{1}=\frac{L}{M}
\label{ll}
\end{equation}
and
\begin{equation}
\frac{\sqrt{\left\langle \left( l_{leaf}-\left\langle l_{leaf}\right\rangle
\right) ^{2}\right\rangle }}{\left\langle l_{leaf}\right\rangle }=\frac{1}{2}%
\sqrt{\frac{L}{\chi Ml_{P}}}  \label{dl}
\end{equation}
Thus one encounters no phase-coexistence: the leaf size grows with chain
length in the same manner as the ''$T=0$'' solution. This does not mean,
however, that the rosette structure is not altered when the mean leaf size
exceeds $l_{P}$, because the reduced second moment starts to exceed one at
that point (actually, at $\left\langle l_{leaf}\right\rangle $ exceeding $%
4\chi l_{P}$). One can therefore identify $L/M\approx l_{P}$ as the onset
point of heterogeneity of the leaf size distribution; the orderly, symmetric
rosette is starting to ''melt''. It must be emphasized though that $G\left(
P\right) $ is analytic and that there is no true thermodynamic singularity.

\begin{figure}
\begin{center}
\includegraphics*[width=8cm]{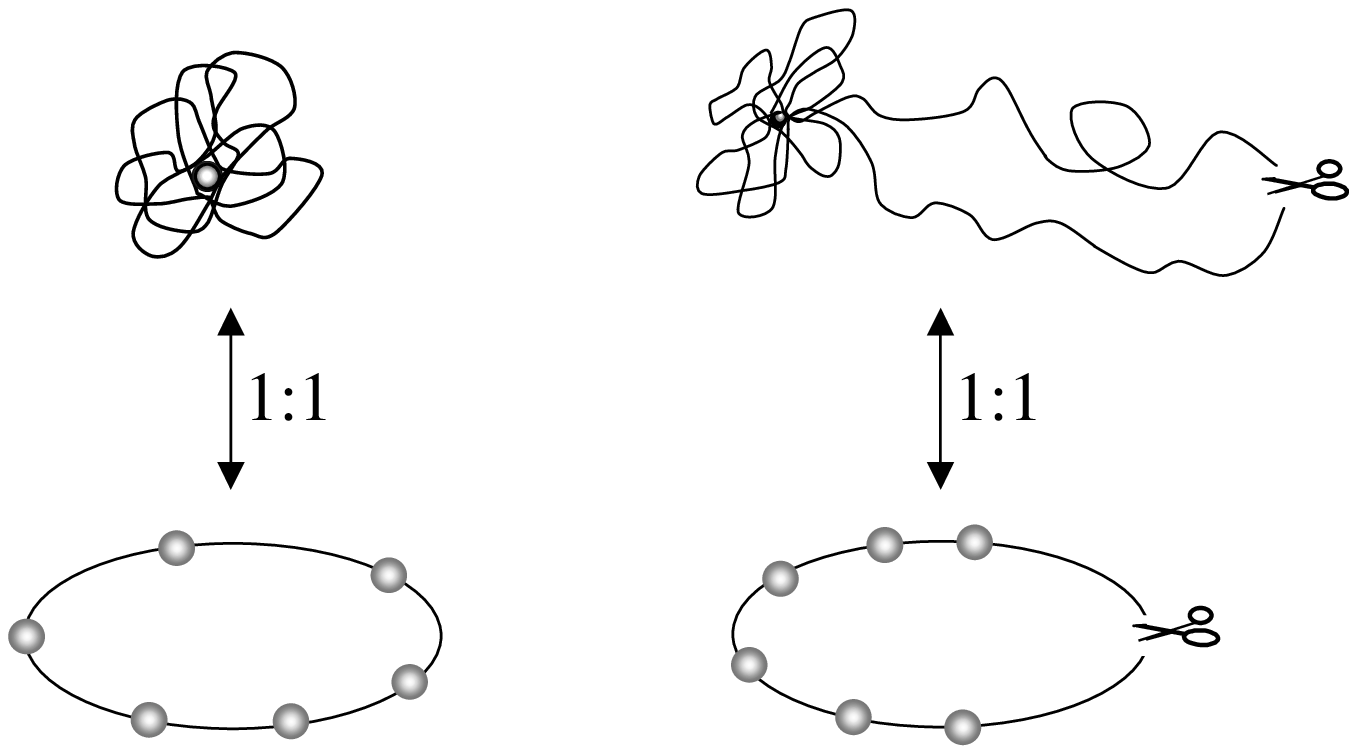}
\end{center}
\caption{Molten rosettes of closed (left) and open chains (right).
These structures are extremely fragile: cutting one loop leads to
two long tails at the expense of the other loops. Shown is also
the corresponding problem of particles on a one-dimensional closed
(or open) track that interact via a nearest-neighbor pair
potential.}
\end{figure}

This heterogeneous rosette state is very fragile. For instance,
the loop size distribution will change drastically if one
considers an {\it open} chain, i.e., if one allows for two free
ends. Consider a chain of length $L$ with $M$ leaves of length
$l_{i}$ ($i=1,...,M$) and the two chain ends of length $l_{0}$ and
$l_{M+1}$. One requires $\sum\nolimits_{i=0}^{M+1}l_{i}=L$. The
calculation of the partition function goes along similar lines as
above. One has just to multiply $Z_{M}$ in Eq.~\ref{gp} with $%
\int_{0}^{\infty }dl_{0}dl_{M+1}\exp \left( -P\left( l_{0}+l_{M+1}\right)
/k_{B}T\right) /r^{2}$ which accounts for the free ends. This leads to $%
G\left( P\right) $, Eq.~\ref{gp2}, with an additional additive term $%
2k_{B}T\ln \left( rP/k_{B}T\right) $. The \textquotedblright
pressure\textquotedblright\ has to be chosen such that $L=dG/dP=M\sqrt{2\chi
l_{P}k_{B}T/P}+2k_{B}T/P$, i.e.
\begin{equation}
\sqrt{P}=\sqrt{\frac{2k_{B}T}{L}+\frac{\chi k_{B}Tl_{P}M^{2}}{2L^{2}}}+\frac{%
M}{L}\sqrt{\frac{\chi k_{B}Tl_{P}}{2}}  \label{rp}
\end{equation}
The average loop size is now given by
\begin{equation}
\left\langle l_{leaf}\right\rangle =-k_{B}T\frac{d}{dP}\ln Z_{1}=\sqrt{\frac{%
2\chi k_{B}Tl_{P}}{P}}\simeq \left\{
\begin{array}{ll}
\frac{L}{M} & \mbox{for}\;\frac{L/M}{l_{P}}\ll M \\
\sqrt{\chi l_{P}L} & \mbox{for}\;\frac{L/M}{l_{P}}\gg M%
\end{array}
\right.  \label{lsize}
\end{equation}
The second moment of the loop size distribution obeys
\begin{equation}
\frac{\sqrt{\left\langle \left( l_{leaf}-\left\langle l_{leaf}\right\rangle
\right) ^{2}\right\rangle }}{\left\langle l_{leaf}\right\rangle }=\left(
\frac{k_{B}T}{8\chi l_{P}P}\right) ^{1/4}\simeq \left\{
\begin{array}{ll}
\frac{1}{2}\sqrt{\frac{L/M}{\chi l_{P}}} & \mbox{for}\;\frac{L/M}{l_{P}}\ll M
\\
\frac{1}{2}\left( \frac{L}{\chi l_{P}}\right) ^{1/4} & \mbox{for}\;\frac{L/M%
}{l_{P}}\gg M%
\end{array}
\right.  \label{deltal}
\end{equation}
Hence one recovers the result for the closed-loop case but only for $L/M\ll
Ml_{P}$. In the opposite case, $L/M\gg Ml_{P}$, one finds a different
scaling $\left\langle l_{leaf}\right\rangle \propto \sqrt{l_{P}L}$ (Eq.~\ref%
{lsize}) which shows that the mean leaf size is now small compared to $L/M$.
Most of the chain is part of the two free ends that emerge from the rosette;
the average length of a free end is $\left\langle l_{0}\right\rangle
=k_{B}T/P$ which leads indeed to $L/2-M\sqrt{\chi l_{P}L}/2$. A schematic
view of molten rosettes formed from closed and open chains and the
corresponding analogy of a 1D gas of particles is shown in Fig.~7.

In Appendix A I present some new results that extend the above calculations
to the case of rosette formation in $d$ dimensional space. This will shed
some light on the nature of the \textquotedblright phase
coexistence\textquotedblright\ within molten rosette structures. In the next
section (2.3.3) I shall show that rosette structures do also occur at low
ionic strength. In that section I will also contrast the unwrapping
transitions into the rosette at low and high ionic strength. Furthermore, I
will speculate if rosette structures could occur in DNA-histone complexes.

\subsubsection{The rosette state at low ionic strength}

The rosette configuration discussed in the last section is a way
to bring at least a small fraction of the chain in close contact
to the ball (in the form of point contacts). The majority of the
monomers resides in the loops that do not ''feel'' the presence of
the sphere but are needed to connect the point contacts via small
curvature sections. At first sight one might thus expect the
rosettes to be a special feature for chain-sphere complexes with a
short ranged attraction.

This is, however, not true. Rosettes are quite robust and occur
also in systems with a much larger range of interaction. Recently
This became clear in a Monte Carlo study by Akinchina and Linse
\cite{akinchina02}. They considered the complexation of a
semiflexible charged chain with an oppositely charged ball that
carries the same absolute charge as the chain (isoelectric
complex). No small ions were present so that the charged monomers
were attracted to the sphere via a long-ranged $1/r$-interaction.
The authors simulated systems with chains of different persistence
lengths and linear charge densities as well as spheres with
different radii. Depending on the choice of parameters they
encountered a multitude of structures -- ranging from collapsed
structures with a ''tennisball seam pattern'' or solenoid
arrangement of the wrapped chain \cite{kunze00,kunze02} to open
multi-leafed structures very much resembling the ones discussed in
the previous section. The rosette structures occur for stiffer
chains on smaller spheres. The example configurations in
Ref.~\cite{akinchina02} clearly show some rosette structures with
one, two and three leaves (cf. Fig.~1, system II in that paper).
That these are representative example configuration can most
clearly be seen in the adsorption probability of monomers as a
function of the monomer index (Fig.~3 in \cite{akinchina02}).

To understand better why rosettes occur also in the long-ranged
case I developed a scaling theory for this system in
Ref.~\cite{schiessel02b} that I will outline in the following.
Consider first sufficiently short chains $L=bN\leq bZ=l_{iso}$
where the chain charge is smaller than (or equals) the sphere
charge. The energy of the rosette with $M$ leaves is then
approximately given by
\begin{equation}
\frac{E_{rosette}}{k_{B}T}\simeq \frac{l_{P}}{L}M^{2}-\frac{l_{B}Z}{b}M
\label{rosette1}
\end{equation}
The first term is the bending energy of $M$ leaves of length $L/M$
and typical curvature $\sim M/L$. This has, of course, the same
form as the elastic contribution to the energy of the rosette at
strong screening, first term of Eq.~\ref{emin} (up to a numerical
constant that I do not consider here). The second term is the
attraction between the ball charge $Z$ and the chain charge $L/b$
over the typical distance $L/M$. Remarkably this term shows the
same scaling with $M$ as the second term of Eq.~\ref{emin} that
describes the energy of the point contacts! One has just to
identify the point contact energy $\mu $ (for the strong screening
case) with the leaf-sphere attraction
\begin{equation}
\mu =\frac{l_{B}Z}{b}  \label{rosette2}
\end{equation}
for the unscreened case. The optimal leaf number is thus again (cf. Eq.~\ref%
{ns})
\begin{equation}
M^{\ast }\simeq \frac{\mu L}{l_{P}}  \label{rosette3}
\end{equation}
and the leaf size is
\begin{equation}
l_{leaf}\simeq \frac{L}{M^{\ast }}\simeq \frac{l_{P}}{\mu }
\label{rosette3b}
\end{equation}
The rosette state competes with the wrapped state that was already discussed
in Section 2.2. The rosette state is expected to transform {\it continuously}
into the wrapped structure when $L/M^{\ast }\simeq R_{0}$; then the leaves
become so small that they touch with their contour the surface of the
sphere. Indeed, by setting $M=L/R_{0}$ in Eq.~\ref{rosette1} one finds
\begin{equation}
\frac{E_{wrap}}{k_{B}T}\simeq \frac{l_{P}L}{R_{0}^{2}}-\frac{l_{B}ZL}{bR_{0}}%
=\left( \frac{l_{P}}{R_{0}^{2}}-\frac{l_{B}Z}{bR_{0}}\right) L
\label{rosette4}
\end{equation}
which can be considered as the free energy of the wrapped state:
The first term is the bending energy, Eq.~\ref{bend}, and the
second accounts for the electrostatic chain-sphere attraction. All
other electrostatic contributions (as written down in
Eq.~\ref{nguyen1}) are much smaller and do not occur on this level
of approximation.

On the right hand side of Eq.~\ref{rosette4} I arranged the terms in such
way that one can deduce directly an unwrapping transition at $%
l_{P}/R_{0}^{2}=l_{B}Z/bR_{0}$, i.e., at
\begin{equation}
l_{P}=\mu R_{0}  \label{rosette5}
\end{equation}
At first sight one might expect that at that point the chain
unwraps in a strongly discontinuous fashion similar to the cases
discussed above (the highly charged case, Eq.~\ref{fl3}, and the
short-range case discussed at the beginning of Section 2.3.2).
However, this \textquotedblright unwrapping
point\textquotedblright\ corresponds just to the point $L/M^{\ast
}\simeq R_{0}$ when loops form on the sphere. This leads to a
smooth transition as pointed out before Eq.~\ref{rosette4}. That
the unwrapping transition occurs rather smoothly at low ionic
strength and sharp at high ionic strength has been predicted by
Netz and Joanny \cite{netz99} even though the authors did not
allow in their study for rosette structures.

To complete the picture one has also to study chains that are longer than
the isoelectric length, $L>l_{iso}$. Then at least three terms are needed to
capture the essential physics of the rosette state:
\begin{equation}
\frac{E_{rosette}}{k_{B}T}\simeq \frac{l_{P}}{l}M^{2}-\frac{l_{B}Z}{b}M+%
\frac{l_{B}l}{b^{2}}M  \label{rosette6}
\end{equation}
Here the monomers are ''allowed'' to distribute between the
rosette of length $l$ and a tail of length $L-l$. The first two
terms are the same as above, Eq.~\ref{rosette1}, the last term
describes the self-repulsion of the monomers that constitute the
rosette (additional logarithmic terms accounting for the
self-energy of the tail and its interaction with the rosette, cf.
Eq.~\ref{nguyen1}, are smaller and neglected here). Minimization
with respect to $l$ leads to the optimal rosette length
\begin{equation}
l^{\ast }\simeq b\sqrt{\frac{l_{P}M}{l_{B}}}  \label{rosette7}
\end{equation}
The free energy \ref{rosette6} with the optimal wrapping length $l^{\ast }$
is minimized for the following number of leaves:
\begin{equation}
M^{\ast }=\frac{\mu l_{iso}}{l_{P}}  \label{rosette8}
\end{equation}
Hence -- on this level of approximation -- $l^{\ast }\simeq b\sqrt{%
l_{P}M^{\ast }/l_{B}}\simeq l_{iso}$, i.e., the rosette monomers just
compensate the central ball charge; the rest of the monomers extends away
from the rosette in a tail of length $L-l_{iso}$. Each leaf is of size
\begin{equation}
l_{leaf}\simeq l_{P}/\mu  \label{rosette9}
\end{equation}
The rosette disappears at $l_{iso}/M^{\ast }=R_{0}$, i.e., when Eq.~\ref%
{rosette5} is fulfilled. It is then replaced by a wrapped chain of
length $l_{iso}$ (plus additional correction terms such as the one
given in Eq.~\ref{wrap}) and a tail of length $L-l_{iso}$.

\begin{figure}
\begin{center}
\includegraphics*[width=7cm]{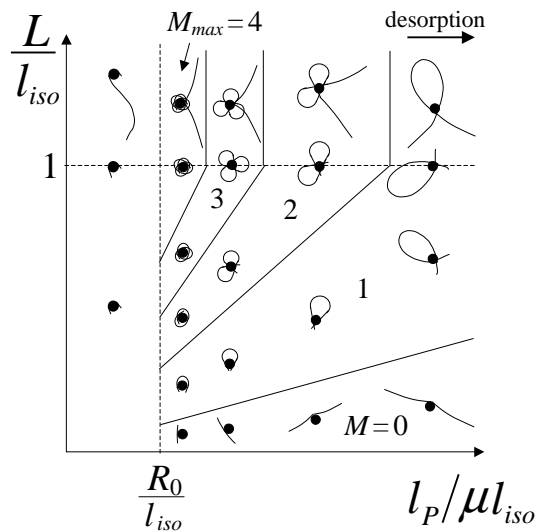}
\end{center}
\caption{The sphere-chain complex at low ionic strength. The axes
are chosen as in Fig.~6 with $\mu$ being now the leaf-sphere
attraction, Eq.~\ref{rosette2}. The unwrapping transition is
smooth in this case.}
\end{figure}

In Fig.~8 I depict the complete diagram of the sphere-stiff chain complexes
to be expected in the long-range case. I again plot $L$ {\it vs.} $%
x=l_{P}/\mu $ (in units of $l_{iso}$), which is for rosettes just
the leaf size, Eq.~\ref{rosette3b}. When one starts in this
diagram at a large $x$-value and goes towards smaller values (with
some arbitrarily fixed value $L<l_{iso}$) then all leaves shrink
and more and more leaves can form. At $x=R_{0}$ the maximal number
of leaves (for that given value of $L$) is reached and at the
same time the leaves disappear simultaneously in a continuous fashion. For $%
x<R_{0}$ the chain wraps around the sphere. For $L>l_{iso}$ the excess
charges are accommodated in tails and all rosettes have the same lengh $%
l_{iso}$. The borderlines between different rosette ground states are then
independent of the total length of the chain and thus appear as vertical
lines. Desorption occurs when the free energy, Eq.~\ref{rosette1} or \ref%
{rosette6}, equals the thermal energy $k_{B}T$. This point is
reached when
\begin{equation}
\frac{l_{P}}{\mu }\simeq \left\{
\begin{array}{ll}
\sqrt{l_{P}L} & \mbox{for}\;L\leq l_{iso} \\
\sqrt{l_{P}l_{iso}} & \mbox{for}\;L>l_{iso}%
\end{array}
\right.  \label{rosette10}
\end{equation}
An arrow in Fig.~8 indicates the direction in which desorption takes place.

\begin{figure}
\begin{center}
\includegraphics*[width=8cm]{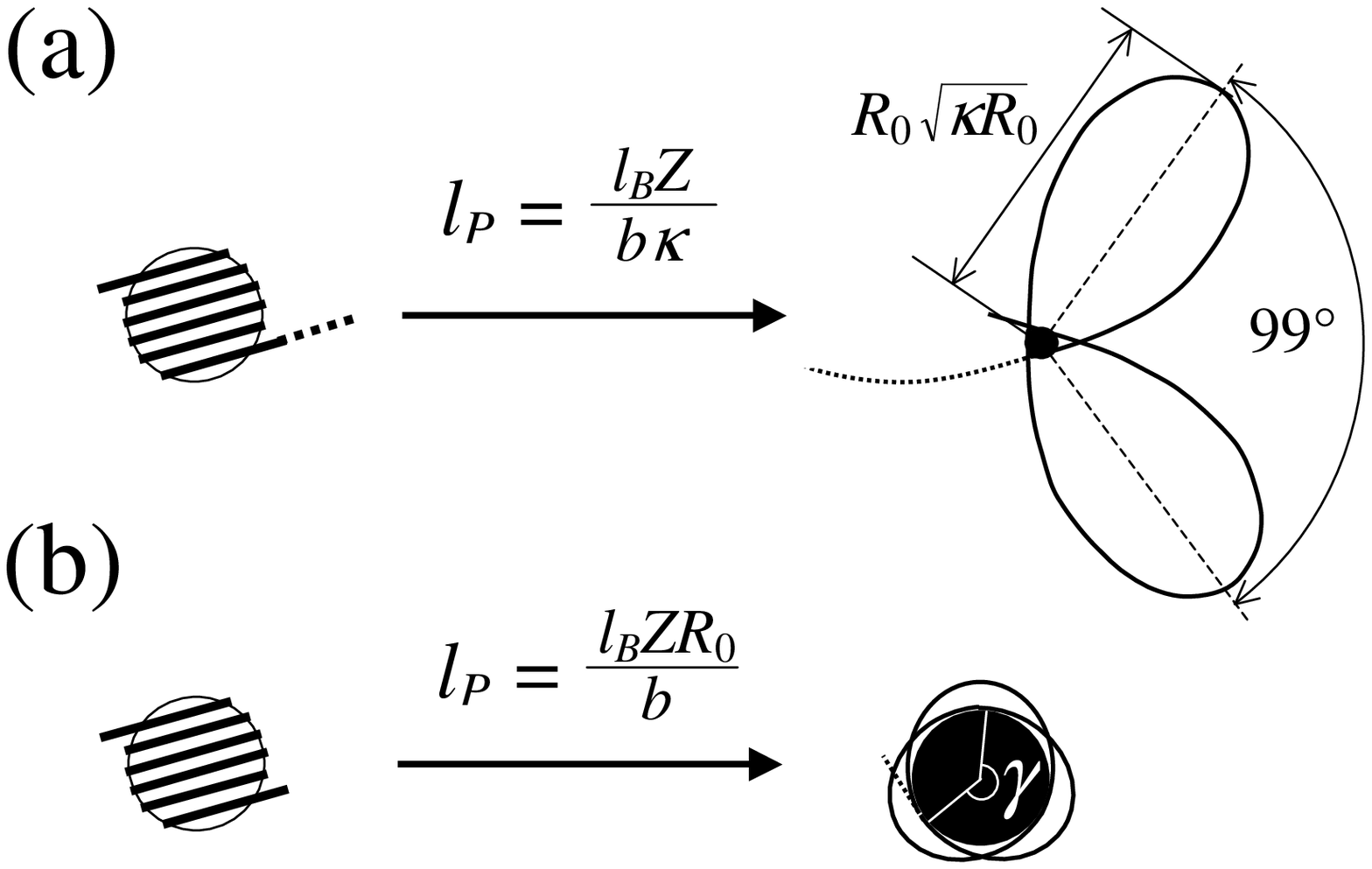}
\end{center}
\caption{The unwrapping transition for (a) short-ranged and (b)
long-ranged interaction. In the former case the chain unwraps
discontinuously into large-leafed Yamakawa-Stockmeyer loops, in
the latter case the transition is continuous.}
\end{figure}

Before comparing these results with the Monte Carlo simulations by
Akinchina and Linse \cite{akinchina02} and with the properties of
the nucleosomal complex it is instructive to take a closer look at
the {\it unwrapping transition} and to contrast the short- and the
long-ranged case. The former case, $\kappa R_{0}\gg 1$, is
depicted in Fig.~9(a). As discussed in Section 2.3.2 the
unwrapping transition is expected to occur at $\lambda _{c}\simeq
l_{P}/2R_{0}^{2}$ (Eq.~\ref{lambdac}) which leads to
Eq.~\ref{unwrap}. At this point the structure jumps in a strongly
discontinuous fashion into a {\it large}-leafed rosette with
leaves of size $l_{leaf}=L/M^{\ast }$. Using Eqs.~\ref{mu} and
\ref{ns} one finds indeed
\begin{equation}
l_{leaf}\simeq R_{0}\sqrt{\kappa R_{0}}\gg R_{0}  \label{critleaf}
\end{equation}
As discussed in the paragraph after Eq.~\ref{ns} neighboring leaves have a
relative orientation of $\sim 99{{}^{\circ }}$ which I also indicated in
Fig.~9(a).

The unwrapping at low ionic strength is depicted in Fig.~9(b) and
goes as follows. When the chain becomes so stiff that
$l_{P}/R_{0}^{2}>l_{B}Z/\left(
bR_{0}\right) $ the wrapped state is not stable anymore (cf. Eq.~\ref%
{rosette4}). At that time many small leaves ($M^{\ast }=L/R_{0}$ ones) form
simultaneously in a continuous fashion. Their size scales as $l_{leaf}\simeq
R_{0}$, the precise prefactor being not accessible to our scaling argument.
The typical opening angle $\gamma $ of the loop at the point of its
formation scales as $\left( L/R_{0}\right) /M^{\ast }\approx 1$, again with
an unknown numerical value. A multi-leafed configuration slightly above the
unwrapping point is depicted in Fig.~9(b).

Additional insight can be gained by generalizing the attractive force
between a given chain charge and the sphere by a power law $-AZ/r^{\alpha }$
with an arbitrary exponent $\alpha >0$. An integer value $\alpha =d-2$ with $%
d=3,4,...$ can be interpreted as a charged chain that adsorbs on a
$d$-dimensional ball in a $d$-dimensional space. The electrostatic
term for the rosette in Eq.~\ref{rosette1} takes then the form
$-AZN^{\alpha }/\left( bL^{\alpha
-1}\right) $ and the one for the wrapped state scales as $%
-AZL/bR_{0}^{\alpha }$. Unwrapping takes place at $l_{P}^{*}\simeq AZ/\left(
bR_{0}^{\alpha -2}\right) $. At this critical value the energy of the rosette $%
E_{rosette}\left( l_{P}=l_{P}^{*}\right) $ has (as a function of
$M$) a minimum at $M^{*}\simeq L/R_{0}$ for $\alpha <2$ ($d<4$),
suggesting a rather smooth unwrapping transition similar to the
one depicted in Fig.~9(b). This
minimum turns into a maximum at $\alpha =2$ ($d=4$). For larger values of $%
\alpha $ we find $M^{*}=0$, i.e., the unwrapping transition is
sharp, similar to the short-ranged case discussed in the previous
section.

Let me also shortly come back to the highly charged case reviewed
in Section 2.2.1. In that case the dominant contribution to the
complexation energy is the release of counterions which is a
rather short-ranged interaction and consequently the unwrapping
transition should be expected to be discontinuous -- even at low
ionic strength. To determine the unwrapping point one has to use
Eq.~\ref{fl3} (with the ''$-$''\ sign) from which follows that
unwrapping into the rosette occurs at $l_{P}\simeq \left( \Omega
+\widetilde{\Omega }\right) R_{0}^{2}/b$. The point contact energy
is of order $\left( \Omega +\widetilde{\Omega }\right)
\sqrt{R_{0}\lambda _{GC}}/b$ where the so-called Guoy-Chapman
length $\lambda _{GC}\simeq 1/\left( \sigma l_{B}\right) $ is the
thickness of layer of condensed counterions around the
sphere ($\lambda _{GC}\ll R_{0}$ for strong counterion condensation \cite%
{schiessel01}). The leaf size at the unwrapping point is then
given by
\begin{equation}
l_{leaf}\simeq \sqrt{\frac{R_{0}}{\lambda _{GC}}}R_{0}\gg R_{0}
\label{lleaf2}
\end{equation}
which indeed indicates a sharp unwrapping transition for highly
charged systems.

We compare now the results of this section to the Monte Carlo
simulations by Akinchina and Linse \cite{akinchina02}. As
mentioned above the simulated systems were always at the
isoelectric point, i.e. $L=bZ=l_{iso}$. Furthermore there was no
screening. The simulation results have thus to be compared with
Fig.~8. Four systems have been considered, each having
a fixed set of parameters $b$, $Z$ and $R_{0}$ but with 7 different values of $%
l_{P}$. This means that for each case the systems were located on
the dashed horizontal line at $L=l_{iso}$ in Fig.~8. In one system
(called system II in \cite{akinchina02}) the continuous
development from a wrapped to the rosette configurations has been
seen very clearly. Example configurations are shown in Fig.~1,
system II in that paper. For $l_{P}=7$ \AA\ the chain is wrapped,
at $l_{P}=60$ \AA\ there is already a slight indication of very small loops ($%
N=4$ or $5$, cf. the small oscillations in Fig.~3, systems II, open
squares). The next system depicted has already a much stiffer chain, $%
l_{P}=250$ \AA , and shows very clearly three leaves, then two leaves at $%
l_{P}=500$ \AA\ and one leaf for the stiffest chain, $l_{P}=1000$
\AA . In Fig.~8 I have chosen the parameters such that $N_{\max
}\simeq l_{iso}/R_{0}$
equals 4 so that this corresponds roughly to system II in \cite{akinchina02}%
. To compare with the simulations one has to follow the
$L=l_{iso}$-line in Fig.~8: One starts with wrapped structures for
$x=l_{P}/\mu <R_{0}$ and finds then the continuous evolution of
rosettes when the line $x=R_{0}$ is crossed. The leaves grow at
the expense of their number (first 4, then 3, 2 leaves), just as
it has been observed in the simulations. The other three systems
considered have different sets of parameters $R_{0}$ and $b$. The
observed behavior of these systems is also in good agreement with
the theoretical
predictions. For details I refer the reader to the last section of Ref.~\cite%
{schiessel02b}.

Finally let me speculate up to what extend the results of the
previous and the current section apply to the nucleosome system
under different ionic conditions. Especially let me ask if rosette
structures could in principle occur on DNA-histone complexes.
First consider the core particle (147 bp DNA). As already
discussed in Section 2.2.4 at physiological salt concentrations
($\sim 100$ mM) one has $\kappa ^{-1}\simeq 10$ \AA\ so the
short range case of the previous section applies. However, the estimate for $%
\lambda $ from Eq.~\ref{lamda} is not reliable since $r\simeq 10$ \AA\ %
(partial screening), since the binding sites between DNA and the
histones are quite specific and since linear Debye-H\"{u}ckel
theory is not reliable for such highly charged components. As
mentioned in Section 2.1 $\lambda $ can be derived instead
experimentally from competitive protein binding to nucleosomal DNA
\cite{polach95} to be of order $6k_{B}T$ per sticking point, i.e.,
$\lambda \approx \left( 1/5\right) $ \AA $^{-1}$. This is {\it
roughly} 5 times larger than what one would expect from
Eq.~\ref{lamda} (but note also that contact is made mainly between
DNA minor grooves which are 10 basepairs apart!). In any case,
Eq.~\ref{unwrap} predicts an unwrapping for sufficiently small
values of $\kappa ^{-1}$ but the numbers are not reliable. On the
other hand, the unscreened long-range case of the current section
applies when $c_{s}<1$ mM ($\kappa ^{-1}>100$ \AA ). As mentioned
in Section 2.3.1 the completely wrapped configuration is not
stable at this point anymore. This is not surprising since the
nucleosomal DNA overcharges the protein octamer by at least 74
negative charges (if not by 160 charges since 86 charged residues
are inside the octamer, cf. Section 2.1). Hence it is expected
that a considerable part of the terminal DNA unwraps and is part
of one or two tails. In Fig.~8 this corresponds to the
wrapped chain structures with tail that are found for small values of $%
l_{P}/\mu <R$ and large values of $L>l_{iso}$.

It would be interesting to redo the experiments
\cite{khrapunov97,yager89} mentioned in Section 2.3.1 with
complexes between histone octamers and DNA segments that are
longer than 147 basepairs. For sufficiently large salt
concentrations I expect the formation of DNA rosettes (if there is
no interference with the partial disintegration of the octamer
which could be avoided by introducing covalent linkages between
the core histones). For the other limit (low salt) it might be
appropriate to use the argument for highly charged chains and
spheres as given before Eq.~\ref{lleaf2}. The linear charge
density of DNA is very high (2 phosphate groups per bp, i.e., per
3.4 \AA ). Manning theory \cite{manning78} indeed predicts that
counterion condensation reduces the linear charge density to
$-e/l_{B}$ with $l_{B}=7$ \AA . Also the charge of the histone
octamer is so high that counterion condensation is important. It
was argued above that the unwrapping occurs around $l_{P}=\left(
\Omega +\widetilde{\Omega }\right) R_{0}^{2}/b$ ($\Omega $ and
$\widetilde{\Omega }$ are numbers of order (but larger than) one).
However, since the DNA persistence length is so small that
$l_{P}<R_{0}^{2}/b \approx($50 \AA$)^{2}/1.7$ \AA $\simeq 1500$
\AA\ I expect the wrapped state to be stable at low ionic strength
and there should be no formation of rosettes in this limit.

\subsection{Nucleosome repositioning}

It has been shown that nucleosome repositioning occurs
spontaneously via thermal fluctuations (under certain conditions).
This autonomous repositioning is the subject of the current
section. I first review the relevant experiments (2.4.1) and then
discuss three theoretical models proposed to account for this
effect: bulge diffusion (2.4.2), large loop repositioning (2.4.3)
and twist diffusion (2.4.4). I contrast the three cases in Section
2.4.5 and speculate that similar modes might be catalyzed by
remodeling complexes that use the energy of ATP hydrolysis.
Nucleosome repositioning induced through transcription on short
DNA segments is also briefly discussed in that section.

\subsubsection{Experiments}

An early study of uncatalyzed nucleosome repositioning was
presented by Beard \cite{beard78}. He constructed
''chromatin-DNA-hybrids'' where segments of radioactively labelled
naked DNA were covalently joined to sections of chromatin fibers
(derived from a simian virus). Remarkably the experiments
suggested that nucleosomes spontaneously reposition themselves and
invade the naked DNA. This was shown by several experimental
methods: ({\it i}) the occurrence of a radioactive component in
the $\sim 175$ bp-band of a gel electrophoresis experiment that
separated the products of a nuclease digestion. ({\it ii})
Electron-micrographs showed an increased spacing of the
nucleosomes on the chromatin pieces and ({\it iii}) binding of
radioactive components of the recut hybrids on nitrocellulose
filters (to which naked DNA would not bind under the given
conditions). All these methods indicated that something
''happens'' on the time scale of hours in a 150 mM NaCl solution
at elevated temperatures ($37^{\circ }$C). The methods used gave,
however, not more quantitative information about the nucleosome
migration rate. Furthermore, the conditions were not well-defined
enough to exclude ATP-driven processes. Finally, the histone tails
belonging to the virial DNA were strongly acetylated compared to
the ''normal'' chromatin of its host cell (which -- as shown much
later -- has not a big effects on nucleosome mobility, see below).

In that paper Beard discussed also three different possible modes
of nucleosome repositioning: (a) jumping where the octamer
dissociates {\it completely} from the DNA and complexes at another
position, (b) sliding or rolling in which the octamer moves along
the DNA without dissociation and (c) displacement transfer where a
naked segment of the DNA displaces a nucleosomal DNA section and
takes over. Beard argues that two of the mechanism (jumping and
displacement transfer) imply that in the presence of competing DNA
nucleosomes would be transferred from one DNA chain to another, a
fact that was not observed in his experiment when chromosomes and
radioactively labelled, naked DNA were mixed together. Nucleosomes
could only migrate onto the naked DNA when the chromosome and DNA
pieces were covalently joined. Beard concluded that ''sliding and
rolling modes'' should be responsible for octamer repositioning.

Spadafora, Oudet and Chambon \cite{spadafora79} showed via gel
electrophoresis that nucleosomal rearrangement occurs in fragments
of rat liver chromatin when certain conditions are fulfilled:
Either one needs to go to high ionic strength above $600mM$ NaCl
or the fiber has to be depleted of the linker histone H1; in that
case rearrangement occurred also around physiological conditions.
It was observed that when either of these conditions is fulfilled
{\it and} if temperatures are elevated (again $37^{\circ }$C),
then nucleosomes seem to move closer to each other (in the
time range of hours), away from the natural 200 bp repeat length towards a $%
\sim $140 bp repeat length. That length corresponds to a close
packing of nucleosomes, a fact that the authors assigned to an
internucleosomal attraction. Furthermore, since H1 is dissociated
from the nucleosome at around 600 mM salt content they concluded
that one role of H1 is to prevent nucleosome mobility.

Similar conclusions were also drawn by Watkins and Smerdon
\cite{watkins85} from corresponding experiments on human
chromatin. An interesting additional experiment studied the
exchange of histone proteins between different DNA chains. This
was shown by mixing $^{14}$C labelled chromatin and $^{3}$H
labelled naked DNA. It became clear that at {\it high }ionic
strength (600 mM) there is a formation of (probably intact)
nucleosomes on the competing DNA whereas at physiological
conditions this was not unambiguously detected.
Moreover, the double-radioactive component showed the footprint of a $\sim $%
146 bp repeat length. This might indicate that a combination of
histone protein transfer and subsequent nucleosome ''sliding''
into tight packing took place.

\begin{figure}
\begin{center}
\includegraphics*[width=6cm]{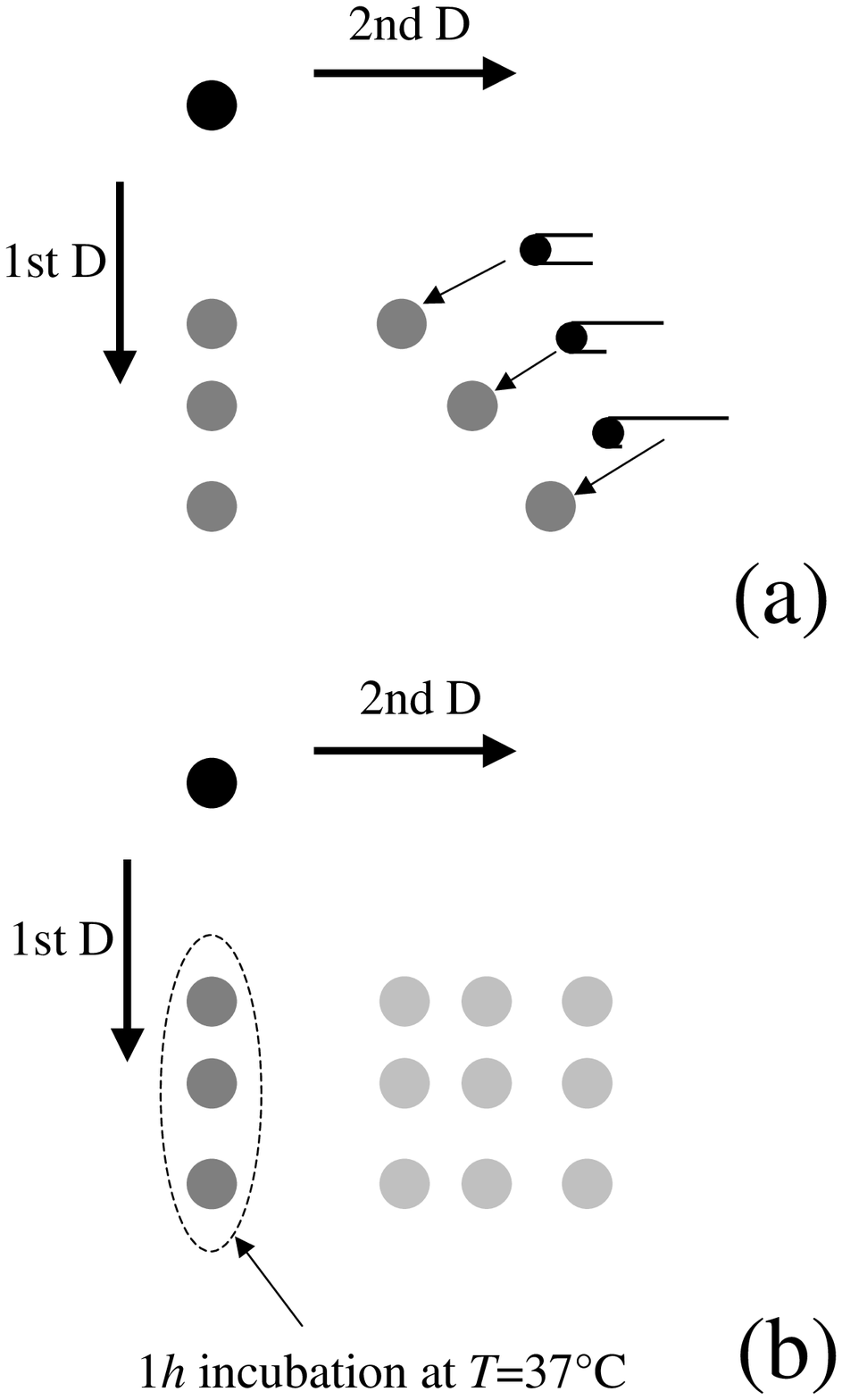}
\end{center}
\caption{Schematic view of the two-dimensional gel electrophoresis
experiment by Pennings et al.~\cite{pennings91} that allowed to
demonstrate autonomous nucleosome repositioning. (a) Under
conditions where no repositioning takes place the final products
line up on a diagonal -- each spot corresponding to a certain
nucleosome position. (b) If the mononucleosomes are incubated the
final products form a square of dots.}
\end{figure}

An important series of experiments on nucleosome repositioning under rather
well-defined conditions was performed by Pennings, Meersseman and Bradbury
\cite{pennings91,meersseman92,pennings94}. Even though the original focus of
this study was to understand better the {\it positioning} of nucleosomes on
special natural sequences that have a high affinity to octamers \cite%
{simpson83}, the authors came up with elegant methods to monitor
the nucleosome {\it re}positioning. It was found \cite{pennings91}
that on tandem repeats of 5s rDNA positioning sequences (each of
length 207 bp) nucleosomes assemble in one dominant position
surrounded by minor positions multiples of 10 bp apart. The most
interesting observation was that there is a dynamic redistribution
between these positions. This was shown by cutting the 207$_{18}$
chromatin into its repeating subunits and then studying the
nucleosome dynamics on such 207 bp fragments. The authors took
advantage of the fact that different nucleosome positions on the
chain give rise to different electrophoretic
mobilities\footnote{One reason is the bend induced by the octamer;
it is known that a bend on a naked DNA fragment affects its
mobility in the gel in very much the same way \cite{wu84}. Another
reason is the inhomogeneity of the charge distribution along the
chain that traps such a polymer in the gel in a U-like
conformation (cf. Fig. 3 in Ref.~\cite{loomans97,schiessel97}).}
and that the motion of the nucleosome along the chain can be
suppressed by subphysiological temperatures or ionic strengths,
and by the presence of Mg$^{2+}$. The 207 bp mononucleosomes were
first separated by an initial dimension of the electrophoresis in
conditions where mobility is suppressed. In this way they obtained
different bands indicating a set of preferred positions. An entire
track from such a gel was then incubated for some period of time
in new conditions where mobility may occur, then changed back to
physiological conditions where mobility is suppressed again, and
run in a {\it second}, equivalent, dimension of gel
electrophoresis. Essentially, the first dimension of
electrophoresis created a nonequilibrium distribution. Depending
on the conditions this distribution relaxed during the subsequent
incubation, which in turn was detected as products moving off the
diagonal in the second dimension of electrophoresis -- cf. Fig.~10
for a schematic depiction of the two-dimensional separation
technique.

In Ref.~\cite{pennings91} it was found that substantial
redistribution took place when the sample was incubated for an 1
hour at $37^{\circ}$C but not at $4^{\circ}$C. The experiments
were done in low ionic conditions in a tris-borate buffer
(0.5$\times $TBE; cf. Ref.~\cite{schwinefus00} for a discussion of
the effects of this buffer on naked DNA). There was a set of
preferred positions, all multiples of 10 bp (the DNA helical
pitch) apart. That means that the nucleosomes had all the same
rotational positioning with respect to the DNA. Another feature
that was observed is that the nucleosomes have a preference for a
positioning at the ends of the DNA fragments, a typical feature
for nucleosomes on short DNA that was recently discussed by Sakaue
et al.~\cite{sakaue01}. The 5s rDNA positioning sequence itself,
however, is located more towards the middle. When gel-separating
the mononucleosomes directly after they have been excised from the
tandemly repeated nucleosomes this position led indeed to the
strongest band. After incubation, however, the end positions
showed the highest probability.

The authors extended their study to head-to-tail dimers of 5S rDNA (207$_{2}$%
) \cite{meersseman92}. In a first dimension of a 2D gel electrophoresis the
mononucleosomes were separated according to their position on the dimer.
This was followed by an incubation at $4{{}{{}^{\circ }}}$ or $37{{}{%
{}^{\circ }}}$C and a subsequent cutting of the dimer into its
monomers. The resulting product was then separated through
electrophoresis in a second dimension. Again for the sample
incubated at elevated temperatures a repositioning of the
nucleosomes was found. Interestingly, however, the study indicated
that the repositioning took place only within a cluster of
positions around each positioning sequence but not between them, a
fact that was shown by radioactive labelling of one half. This
finding indicates that there is no ''long-range'' repositioning at
low ionic strength. Other systems studied in that paper were
fragments of H1-depleted native chromatin and nucleosomes
reconstituted on {\it Alu} repeats; also in these cases a
repositioning was detected as a result of an elevated temperature
incubation. The authors concluded that the repositioning ''may be
visualized as following a corc screw movement
within the superhelical path of the DNA'' \cite%
{meersseman92}. The same authors studied in Ref.~\cite{pennings94} the nucleosome mobility on the 207$%
_{2}$-dimer in the presence of linker histone H1 (or its avian
counterpart H5) and found that the mobility of nucleosomes was
dramatically reduced.

Ura et al.~\cite{ura96,ura97} -- following
Ref.~\cite{meersseman92} -- studied nucleosome mobility on the
207$_{2}$-dimer under varying conditions, namely in the presence
of various chromosomal proteins and in the case when the core
histones are acetylated, respectively. In the former case mobility
was suppressed (depending on the type and concentration of the
chromosomal protein), in the latter case the mobility was not
changed much.

Flaus et al.~\cite{flaus96} developed a different strategy to
determine nucleosome positioning and repositioning. In their
method they used a chemically modified H4 histone that induces --
after addition of some chemical -- a cut on the nucleosomal DNA 2
bp away from the dyad axis. Via gel electrophoresis of the
resulting product they were able to determine the nucleosome
position with base pair resolution. Using this method Flaus and
Richmond \cite{flaus98} studied the nucleosome dynamics on a mouse
mammary tumor sequence which revealed several features of
repositioning more clearly. The longest fragment of this sequence
studied was 438 bp long and had two positioning sequences where
two nucleosomes assembled, each at a unique position. These
positions were also found when mononucleosomes were assembled on
shorter fragments that included only one of the two positioning
sequences. The authors studied the degree of repositioning of the
mononucleosomes on such shorter fragments (namely nucleosome A on
a 242 bp- and nucleosome B on a 219 bp-fragment) as a function of
heating time (ranging from 20 to 80 minutes) and temperature
(ranging from $0^{\circ }$ to $50^{\circ }$C). It was found that
the repositioning rates -- as estimated from the occurrence and
intensity of new bands -- increase strongly with temperature but
also depend on the positioning sequence (and/or length of the
fragment). The difference of repositioning for the two sequences
is remarkable: at 37${{}^{\circ }}$C one has to wait $\sim 90$
minutes for the A242 and more than 30 hours for the B219 for
having half of the material repositioned. Another feature found
was again a preference for end positions (roughly 70 bp from the
dyad axis, similar to the finding in Ref.~\cite{pennings91}). For
nucleosome B which showed a slower repositioning the set of new
positions were all multiples of 10 bp apart (namely at a 20, 30,
40, 50 bp-distance from the starting position), i.e., they all had
the same rotational phase. On the other hand, nucleosome A did not
show such a clear preference for the rotational positioning. It
was argued that these differences reflect specific features of the
underlying base pair sequences involved. Nucleosome B is complexed
with a DNA sequence that has AA/AT/TA/TT dinucleotides that show a
10 bp periodicity inducing a bend on the DNA whereas nucleosome A
is positioned via homonucleotide tracts.

The authors speculate in Ref.~\cite{flaus98} that the preference
of end positions might be caused by one (or several) of the
following mechanisms (1) direct histone interaction with a special
structure at the DNA terminus, (2) relief of the repulsion between
entering and exiting strand and (3) entropy gain by having a long
unbound DNA extension.

Recently Hamiche et al.~\cite{hamiche01} demonstrated that the
(uncatalyzed) nucleosome mobility along DNA depends on the
presence of histone tails. Especially, in the absence of the
N-tail of H2B that passes in between the two turns of the
nucleosomal DNA \cite{luger97} spontaneous repositioning of the
nucleosomes was detected (here {\it during} a gel electrophoresis
in a second direction).

How does repositioning work? The studies seem to indicate that the
octamer is not transferred to competing DNA (at least hardly under
physiological conditions, cf.~\cite{watkins85}). It is also clear
that repositioning should preferably not involve the simultaneous
dissociation of all the 14 bindings sites which would be too
costly to be induced by thermal fluctuations. A mechanism that
requires only the breakage of a few contacts is loop diffusion.
Strictly speaking one has to distinguish here between two
different kinds of loops: small loops or bulges (2.4.2) and big
loops (2.4.3) that show a qualitative different energetics and
lead also to a different picture of the overall nucleosome
repositioning dynamics. Another possibility could be the above
mentioned corkscrew motion as suggested in
Ref.~\cite{meersseman92} (and even before by van Holde and Yager
\cite{vanholde85}). This could be facilitated through the twist
diffusion of small defects that will only require to break one or
two contacts at a time, a mechanism discussed in Section 2.4.4.

\subsubsection{Bulge diffusion}

In Ref.~\cite{schiessel01c} myself, Widom, Bruinsma and Gelbart
argued that the repositioning of nucleosomes without dissociation
from the DNA chain that wraps them might be possible through the
diffusional motion of small intra-nucleosomal loops. This
biological process is analogous to the familiar physical situation
of reptation of ''stored
length'' in polymer chains. Thirty years ago de Gennes \cite%
{degennes71} discussed the motion of a flexible chain trapped in a
gel, modelled by a matrix of fixed point-like obstacles that
cannot be crossed by the polymer. Fig.~11 depicts schematically
the mechanism whereby diffusion
of these ''defects'' of stored length $%
\Delta L$ gives rise to overall translation of the chain. Specifically, when
the loop moves through the monomer at {\it B}, this monomer is displaced by
a distance{\it \ }$\Delta L$. de Gennes wrote down a conservation equation
for this motion of defects along the trapped chain and calculated its
overall mobility, and thereby, in particular, the molecular-weight
dependence of the overall translation diffusion coefficient. In our present
situation the reptation dynamics do not arise from obstacles due to a host
matrix (as in a gel) or to other chains (as in a melt), but rather to loops
associated with unsaturated adsorption of the DNA on the protein complex.
Similar physics arise in the lateral displacements of a linear polymer
adsorbed on a bulk solid surface. Sukhishvili et al.~\cite%
{sukhishvili00,sukhishvili02}, for example, have measured the translational
motion of adsorbed polyethylene glycol (PEG) on functionalized (hydrophobic)
silica, specifically the dependence of its center-of-mass diffusion constant
on molecular weight. They find an unusual scaling behavior -- but one that
can be accounted for by \textquotedblleft slack between sticking
points\textquotedblright , so that lateral motion of the polymer proceeds
via a caterpillar-like diffusion of chain loops. In the case of
intra-nucleosomal loops considered in this section, one is essentially in
the limit of infinite molecular weight, because of the chain length being
large compared to the bead (solid substrate) diameter. Furthermore, one
deals with a lower dimensional problem, since the DNA chain is wrapped
(absorbed) on a 1D path rather than a 2D surface. But the basic features of
loop formation and diffusion, and subsequent motion of the overall chain --
in particular the exclusive role of equilibrium fluctuations in driving
these processes -- are the same in both cases.

\begin{figure}
\begin{center}
\includegraphics*[width=7cm]{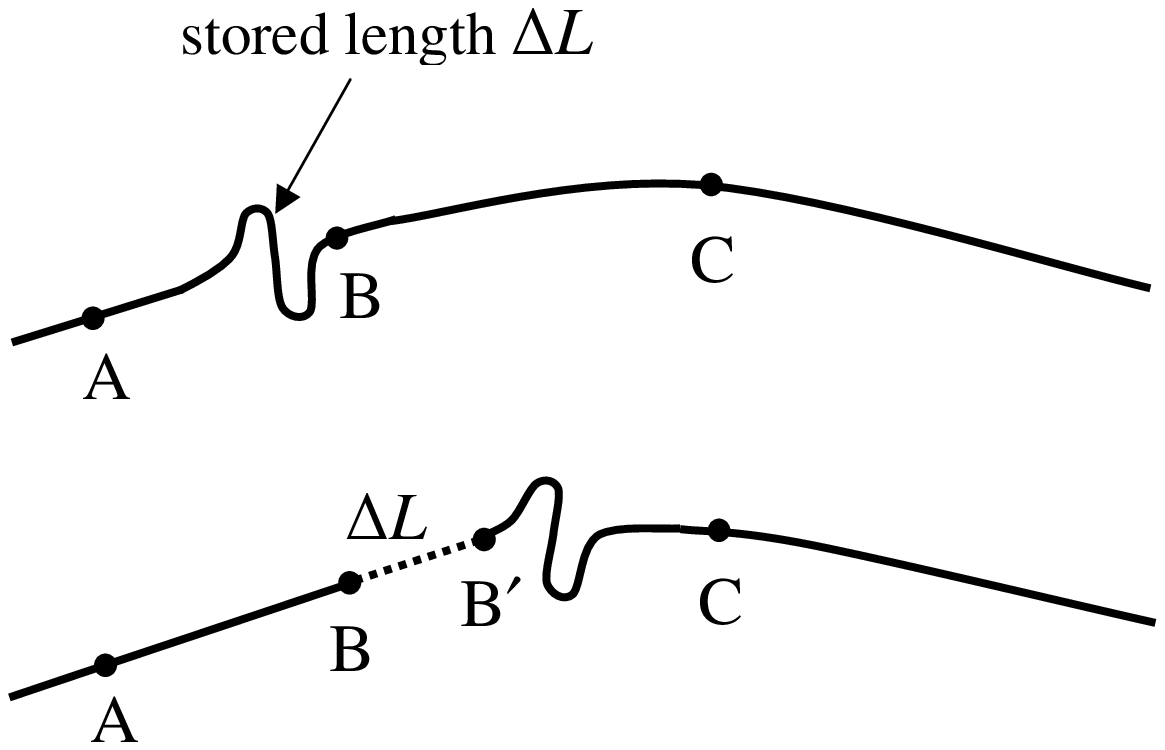}
\end{center}
\caption{According to de Gennes and Edwards translational
diffusion of trapped chains can be envisaged by a reptation
mechanism. When a defect of stored length $\Delta L$ passes
through a monomer at $B$ it is moved by that amount to a new
position $B'$.}
\end{figure}

As shown in earlier studies of competitive protein binding to nucleosomal
DNA \cite{polach95,polach96,anderson00}, thermal fluctuations lead to
lengths of the chain becoming unwrapped at the ends of its adsorbed portion.
If some length of linker is pulled in before the chain re-adsorbs, then an
intra-nucleosomal loop is formed -- see Fig.~12(b). In Ref.~\cite%
{schiessel01c} we calculated first the equilibrium shape and length
distribution of these loops, in terms of the chain bending stiffness $A$,
adsorption energy per unit length $\lambda $, and protein aggregate size $%
R_{0}$ (to be more precise: the radius of curvature of the DNA
centerline). We then considered the diffusion of these loops from
one end of the nucleosome to the other. Finally, treating this
motion as the elementary step in the diffusion of the nucleosome
itself along the wrapping chain, we were able to make estimates of
the nucleosome repositioning rates as a function of $A$, $\lambda
$, $R_{0}$, and solvent viscosity $\eta $.

\begin{figure}
\begin{center}
\includegraphics*[width=4cm]{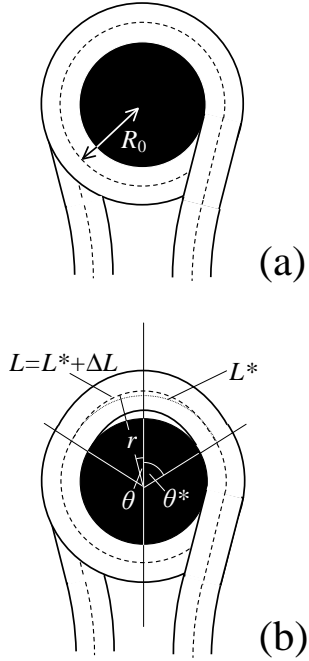}
\end{center}
\caption{(a) Top view of the defect-free nucleosome looking down
the superhelical axis. (b) Geometry of an intra-nucleosomal
bulge.}
\end{figure}

I present here (in more detail than in our letter
\cite{schiessel01c}) the calculation of the equilibrium
statistical mechanical probability associated with the formation
of a {\it small} intra-nucleosomal loop (large loops are
considered in the next section). When no loops are present, a
nucleosome consists of a length $l$ of DNA chain wrapped
continuously around the octamer, see Fig.~12(a). In reality, as
discussed in Section 2.1, the configuration of the adsorbed chain
is a left-handed superhelix (of contour length $l$) spanning the
full height of a  cylinder. One can proceed, however, without
making any explicit assumptions about the shapes of either the
histone octamer or the wrapped DNA.

Consider a fluctuation in which some length of the chain becomes
unwrapped (this can {\it only} happen at the end of the adsorbed
portion of chain) and simultaneously some length, say $\Delta L$,
of linker (i.e., previously unadsorbed chain) is ''pulled in''
before the chain re-adsorbs. The fluctuation has then produced a
loop of contour length
\begin{equation}
L=L^{\ast }+\Delta L  \label{L}
\end{equation}
where $L^{\ast }$ is the ''exposed'' length of nucleosome
associated with the loop; see Fig.~12(b). Note that $\Delta L$
shows a strong preference for values that are multiples of 10 bp
lengths, since this is the periodicity of the DNA helical pitch
(binding sites are located where minor grooves face inwards to the
octamer, cf. Section 2.1); other values require a twisting of the
loop DNA which is energetically costly. For the energy associated
with forming a loop of this kind, one can write
\cite{schiessel01c}:
\begin{equation}
\frac{\Delta U}{k_{B}T}=\frac{E_{elastic}}{k_{B}T}-\left( \frac{l_{P}}{%
2R_{0}^{2}}-\lambda \right) L^{\ast }=\frac{l_{P}}{2}\int_{loop}\frac{ds}{%
R^{2}\left( s\right) }-\left( \frac{l_{P}}{2R_{0}^{2}}-\lambda \right)
L^{\ast }  \label{deltau}
\end{equation}
The first term is the bending energy of the loop with $1/R\left( s\right) $
being the local curvature at distance $s$ along its contour, cf. also Eq.~%
\ref{WLC}. The second term accounts for the length $L^{\ast }$ which has
been adsorbed and bent with curvature $1/R_{0}$ {\it before} loop formation.
The mathematical details of the functional minimization of $\Delta U$, Eq.~%
\ref{deltau}, are presented in Appendix B. There it is shown that
the formation energy of an optimal small loop for given extra
length $\Delta L$ is approximately given by\footnote{In
Ref.~\cite{schiessel01c} we made one further approximation, namely
-- when inserting $\theta ^{*}$ into Eq.~\ref{deltau4} -- we
neglected the second and third term that nearly cancel when one
inserts the typical parameters of the nucleosome. This led us to
Eq.~(2b) in that paper. Here I present the full expression,
Eq.~\ref{deltau3}.}
\begin{equation}
\frac{\Delta U}{k_{B}T}\simeq \frac{6}{5}\left( 20\pi ^{4}l_{P}\lambda
^{5}\right) ^{1/6}\left( R_{0}\Delta L\right) ^{1/3}-\frac{l_{P}}{2R_{0}^{2}}%
\Delta L  \label{deltau3}
\end{equation}
Using Eq.~\ref{deltau3} one finds that the energy to form a loop of minimal
size of $34$ \AA\ (one helical pitch) is roughly 27$k_{B}T$ (assuming $%
R_{0}=43$ \AA , $\lambda =0.176$ \AA $^{-1}$, $l_{P}=500$ \AA );
larger bulges are even more expensive. Here seems to arise a
conceptional problem of the small loop mechanism: the formation
energy of a small loop is of the same order as the complexation
energy of the nucleosome itself (ca. $30k_{B}T$, cf. Section 2.1)!
However, note two points: ({\it i}) The worm-like chain model is
not very reliable for such strong curvatures and might
overestimate the actual bending energy. ({\it ii}) We assumed in
the calculation leading to Eq.~\ref{deltau3}
that $TR_{0}^{2}/A\simeq \pi ^{2}/\theta ^{\ast 2}\gg 1$ (cf. Appendix B). From Eq.~\ref%
{theta} follows, however, that $\theta ^{\ast }\simeq 1.75$ and
hence $\pi ^{2}/\theta ^{\ast 2}\approx 3$. So one is actually in
the crossover region between small and large ''tensions'' $T$.

To account for this fact, the calculation can be redone using a
refined estimation
for $\lambda _{\pm }$, namely $\lambda _{+}=1/\lambda _{-}=\sqrt{2}+\sqrt{%
TR_{0}^{2}/A}$ which is a much simpler expression than Eq.~\ref{lpm} but
still is a rather good approximation; especially it shows the right
asymptotic behavior, Eq.~\ref{lp}. This leads to $\theta ^{*}\approx 1.3$
and $\Delta U\approx 20k_{B}T$ for $\Delta L=10$ bp.

Now the probability distribution for the formation of loops of size $\Delta
L $ is then simply given by the corresponding Boltzmann factor, normalized
such that the maximal number of loops (expected for the unphysical case $%
\Delta U=0$) is the geometrically possible one, i.e. $l/L^{*}$:
\begin{equation}
n_{eq}\left( \Delta L\right) \simeq \frac{l}{L^{*}}e^{-\Delta U/k_{B}T}
\label{neq}
\end{equation}
Even the value $20k_{B}T$ calculated above shows that the
spontaneous formation of a small loop is a {\it very} rare event.
I will check in the following if it occurs sufficiently often so
that it could account for the experimentally observed autonomous
repositioning rates discussed in the previous section. In order to
proceed here the dynamics of the loops needs to be considered and
that of the resulting nucleosome repositioning.

The key idea \cite{schiessel01c} here is that diffusion of the
histone octamer along the DNA is achieved by formation and
annihilation of loops. Let $D$ denote the diffusion constant
relevant to this motion of the ball along the chain, and let $w$
be the rate at which loops are formed (by incorporation of linker
length) $D=w\Delta L^{2}$. These loops ``disappear'' due to their
diffusion ``off''\ the ball, at a rate that is proportional to the
instantaneous number of loops, i.e., at a rate $C_{A}n$.
Accordingly, the overall rate of formation of loops is given by
$w-C_{A}n$, which must vanish at equilibrium, implying
$n_{eq}=w/C_{A}$. Since this number is much smaller than unity, we
are justified in assuming that only one loop at a time needs to be
considered in treating the diffusion of intra-nucleosomal loops.
It follows from the Boltzmann expression for $n_{eq}$ (see
\ref{neq}) that $w$, the rate of loop formation, is given by
\begin{equation}
w\simeq C_{A}\left( l/L^{*}\right) \exp \left( -\Delta U/k_{B}T\right)
\label{rate}
\end{equation}
and $D$ by $w\Delta L^{2}$.

It remains only to evaluate $C_{A}$, characterizing the rate of
diffusion of loops ``off''\ the ball. Let $D^{+}$ denote the
diffusion constant associated with this motion ($D^{+}$
characterizes the diffusion of loops {\it through} a wrapped ball,
as opposed to the coefficient $D$ that describes diffusion of the
{\it ball} along the chain). Since the distance which the loop
must move to leave the ball is $l$, the wrapping length,
one can write $C_{A}^{-1}\simeq l^{2}/D^{+}$%
. From the Stokes-Einstein relation one has furthermore that $%
D^{+}=k_{B}T/\zeta $ where $\zeta \simeq \eta L^{*}$ is the friction
coefficient of the loop, with $\eta $ the effective solution viscosity. $%
L^{*}$, as before, is the exposed length of the octamer associated with the
loop, and hence provides the loop size relevant to its diffusion along the
(1D!) nucleosome path of the chain. This hydrodynamic description is
justified by the fact that loop diffusion requires unbinding of only a
single sticking site, whose binding energy is of order of a few $k_{B}T$.
Combining all of the results from this and the preceding paragraph then
gives
\begin{equation}
D\approx \frac{k_{B}T}{\eta l}\left( \frac{\Delta L}{L^{*}}\right) ^{2}\exp
\left( -\Delta U/kT\right)  \label{diff}
\end{equation}
with $L^{*}$ given by Eq.~\ref{theta} and $\Delta U$ by
Eq.~\ref{deltau3}.

Recalling $\theta ^{*}=1.3$ and $\Delta U=20k_{B}T$ for $\Delta
L\simeq 34$ \AA\ and taking reasonable estimates for $\eta $ (a
centipoise), $R_{0}$(43 \AA ) and $l$ ($500$ \AA ) we find that
$D$ is of order $10^{-16}cm^{2}/s$. Hence typical repositioning
times are in the order of an hour; furthermore there is a strong
dependence on the temperature. A closer comparison of these
theoretical estimates with the experiments (discussed in Section
2.4.1) will be given in the discussion (Section 2.4.5) after I
have also presented the theories for large loops and twist
diffusion.

\subsubsection{Large loop repositioning}

The perturbation calculation presented in Ref.~\cite{schiessel01c}
and reviewed in the previous section allows only to study small
loops that store an amount of excess length of $10$ or $20$ bp. To
describe also large loops a different approach is necessary as it
was presented by Kuli\'{c} and myself in Ref.~\cite{kulic02}. In
that paper we made use of the Euler-Kirchhoff theory for the
static equilibrium of rods which allowed us to describe loops of
{\it any} given excess lengths. The outcome of that paper changed
our view of how repositioning via loop formation should work;
besides the local repositioning based on bulge diffusion there
should also occur a long-range hopping via large loops -- at least
for the case of very low nucleosome line densities as they are
often encountered in {\it in vitro} experiments.

In Ref.~\cite{kulic02} we started again from the Hamiltionian given by Eq.~%
\ref{deltau}. The section of the DNA constituting the loop has a contour
length $L$ and is parametrized by its arc length $s$ ranging from $-L/2$ to $%
L/2$. $L$ is the sum of the exposed length $L^{\ast }=2\theta ^{\ast }R_{0}$
($2\theta ^{\ast }$: opening angle) and the excess length $\Delta L$, see
Eq.~\ref{L}. In order to compute the ground state for a trapped
intranucleosomal loop the total energy \ref{deltau} has to be minimized
under two constraints: ({\it i}) The excess length $\Delta L$ is prescribed
so that the following relation between the opening angle $\theta ^{\ast }$
and the total loop length $L$ has to be fulfilled
\begin{equation}
\Delta L=L-2\theta ^{\ast }R_{0}=const.  \label{Constraint1}
\end{equation}
({\it ii}) At the two ends $s=\pm L/2$ the rod has to be tangential on an
inscribed circle of given radius (representing the nucleosome):
\begin{equation}
R_{0}=\left\vert \frac{y\left( L/2\right) }{-x^{\prime }\left( L/2\right) }%
\right\vert =const.  \label{Constraint2}
\end{equation}
Here $x\left( s\right) $ and $y\left( s\right) $ are the Cartesian
coordinates of the rod axis as a function of the arc-length
parameter $s$ (cf. Fig.~13). The absolute value in the second
constraint needs to be introduced formally for dealing with
crossed rod solutions (which are considered later on) and can be
omitted for simple uncrossed loops.

\begin{figure}
\begin{center}
\includegraphics*[width=8cm]{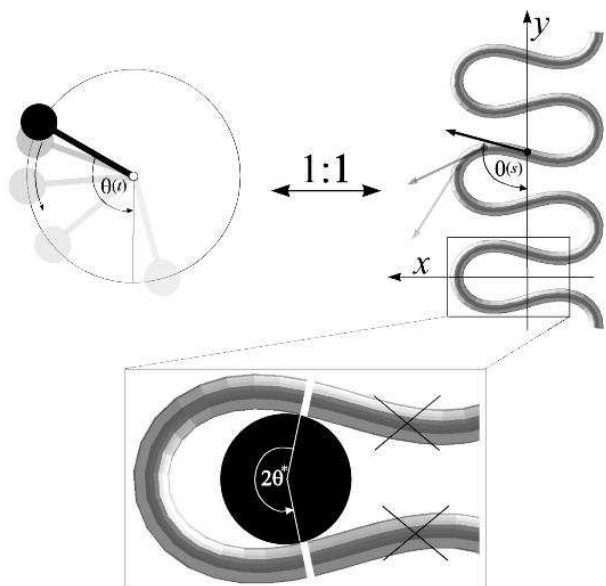}
\end{center}
\caption{The Kirchhoff analogue for the case of a planar pendulum
and a planar semiflexible rod under tension. The inset displays
how to construct a intranucleomal loop by inscribing a circular
disc representing the octamer.}
\end{figure}

For the analytical description of the loop geometry it is
convenient to introduce the angle $\theta =\theta \left( s\right)
$ between the DNA tangent and the $Y$-axis (cf. Fig.~13) that
describes the DNA centerline (note that this is the same angle
$\theta $ as the one introduced in the previous section, cf.
Fig.~12). Integrating the sine (cosine) of $\theta $ over the arc
length parameter $s$ yields the $X$ ($Y$) Cartesian coordinate of
any point along the rod, and the derivative $\theta ^{\prime }$
gives the rod curvature $R^{-1}$. Furthermore the nucleosome
opening angle $\theta ^{\ast }$ is simply related to $\theta $ at
the boundary, namely $\theta ^{\ast }=\theta \left( L/2\right) $
for simple loops and $\theta ^{\ast }=\pi -\theta \left(
L/2\right) $ for crossed loops (see below).

The two constraints Eqs.~\ref{Constraint1} and \ref{Constraint2} can be
rewritten in terms of $\theta $ and then be introduced into the minimization
by two Lagrange multipliers $T_{1}$ and $T_{2}$. We then arrive at the
following functional
\begin{eqnarray}
{\cal F}\left\{ \theta \left( s\right) \right\}
&=&A\int\limits_{0}^{L/2}\left( \theta ^{\prime }\right) ^{2}ds
-\left( \frac{A}{2R_{0}^{2}}-k_{B}T\lambda \right) L^{\ast
}\nonumber \\
&&+T_{1}\left[ L-(\Delta L+L^{\ast })\right] +T_{2}\left[
\int\limits_{0}^{L/2}\cos \theta ds-R_{0}\sin \theta ^{\ast
}\right]  \label{EnergyFunctionalFull}
\end{eqnarray}
Here the first term is the bending energy, the second accounts for
the exposed length $L^{\ast }\equiv 2\theta ^{\ast }R_{0}$ and the
third and
forth term are the imposed length and tangency constraints. Eq.~\ref%
{EnergyFunctionalFull} can be rearranged in the more familiar form
\begin{equation}
\int_{0}^{L/2}\left( A\left( \theta ^{\prime }\right)
^{2}+T_{2}\cos \theta \right) ds+b.t.
\label{EnergyFunctionalSimplified}
\end{equation}
where b.t. denotes boundary terms (depending on $\theta \left(
L/2\right) $ only) that obviously do not contribute to the first
variation inside the relevant $s$ interval. The integral in
Eq.~\ref{EnergyFunctionalSimplified} is analogous to the action
integral of the plane pendulum with $A\left( \theta ^{\prime
}\right) ^{2}$ corresponding to the kinetic and $-T_{2}\cos \theta
$ to the potential energy. The latter analogy is nothing else but
Kirchhoff's kinetic mapping between deformed rods and the spinning
top that contains the present problem as a simple special case
(cf. the paragraph after Eq.~\ref{e1} for a brief discussion of
the Kirchhoff analogy).

Kirchhoff's analogy provides one directly with explicit
expressions for DNA shapes subjected to twist, bending and various
geometric/topological constraints. Here, for the case of planar
untwisted rods, also called the {\it Euler elastica}, where the
corresponding ''spinning top'' reduces to the plane pendulum, the
rod conformations are most generally given by
\begin{equation}
\cos \theta \left( s\right) =1-2m{\rm sn}^{2}\left( \frac{s}{\Lambda }\mid
m\right)  \label{GenSol}
\end{equation}
This can be integrated to obtain the general planar rod shapes in
Cartesian coordinates:

\begin{eqnarray}
x\left( s\right) &=&2\sqrt{m}\Lambda {\rm cn}\left( \frac{s}{\Lambda }\mid
m\right)  \label{CartesianX} \\
y\left( s\right) &=&2\Lambda E\left( \frac{s}{\Lambda }\mid m\right) -s
\label{CartesianY}
\end{eqnarray}
with sn, cn$(.\mid m)$ (and later below dn) denoting the Jacobi elliptic
functions with the parameter $m$ and $E\left( u\mid m\right) $ being the
incomplete elliptic integral of the second kind in its ''practical''
form \cite{abramowitz72}. The two parameters $%
m>0$ and $\Lambda >0$ in Eqs.~\ref{CartesianX} and
\ref{CartesianY} characterize the shape and the scale of the
solution, respectively. These solutions are up to trivial plane
rotations, translations, reflections and shifting of the contour
parameter $s\rightarrow s+s_{0}$ the most general solutions to the
Euler elastica. For different parameters $m$ one obtains different
rod shapes corresponding to different solutions of the plane
pendulum motion \cite{nizette99}. The case $m=0$ describes a
pendulum at rest which corresponds to a straight rod. For $0<m<1$
one has strictly oscillating pendulums corresponding to point
symmetric rod shapes where the turning points of the pendulum have
their counterparts in points of inflection of the rod. For
$m<0.\allowbreak 72$ the rod is free of self intersections like
the one depicted in Fig.~13. For $m$ larger than $0.72$ the rods
show varying complexity with a multitude of self-intersections and
for $m=1$ one has the homoclinic pendulum orbit corresponding to a
rod solution with only one self intersection that becomes
asymptotically straight for $s\rightarrow \pm \infty $. For even
higher values of $m$, i.e., for $m\geq 1$ one has revolving
pendulum orbits corresponding to rods
with self-intersections lacking point symmetry. Finally, the limiting case $%
m\rightarrow \infty $ corresponds to the circular rod shape.

In order to describe a trapped loop one needs to use Eqs.~\ref{CartesianX}
and \ref{CartesianY} imposing the constraints \ref{Constraint1} and \ref%
{Constraint2}. For details of this calculation I refer the reader
to Ref.~\cite{kulic02}. There we present explicit solutions for
the scaling parameter $\Lambda $, the opening angle $\theta ^{\ast
}$ and the excess length $\Delta L\,$\ as functions of the
\textquotedblright contact
parameter\textquotedblright\ $\sigma =L/2\Lambda $ and the shape parameter $%
m $, i.e., $\Lambda =\Lambda \left( \sigma ,m\right) $, $\theta ^{\ast
}=\theta ^{\ast }\left( \sigma ,m\right) $ and $\Delta L=\Delta L\left(
\sigma ,m\right) $. Inserting $\Lambda \left( \sigma ,m\right) $ and $\theta
^{\ast }\left( \sigma ,m\right) $ into Eq.~\ref{deltau} leads to the final
expression for the loop formation energy
\begin{eqnarray}
\Delta U\left( \sigma ,m\right) &=&\frac{4A}{R_{0}}\left\vert \frac{\left(
E\left( \sigma \mid m\right) +\left( m-1\right) \sigma \right) \left(
2E\left( \sigma \mid m\right) -\sigma \right) }{{\rm sn}\left( \sigma \mid
m\right) {\rm dn}\left( \sigma \mid m\right) }\right\vert  \nonumber \\
&&-2R_{0}\left( \frac{A}{2R_{0}^{2}}-k_{B}T\lambda \right) \arccos \left[
\pm \left( 2{\rm dn}^{2}\left( \sigma \mid m\right) -1\right) \right]
\label{EtotFinal}
\end{eqnarray}
with $\pm =sign\left( 2E\left( \sigma \mid m\right) -\sigma \right) $.

\begin{figure}
\begin{center}
\includegraphics*[width=10cm]{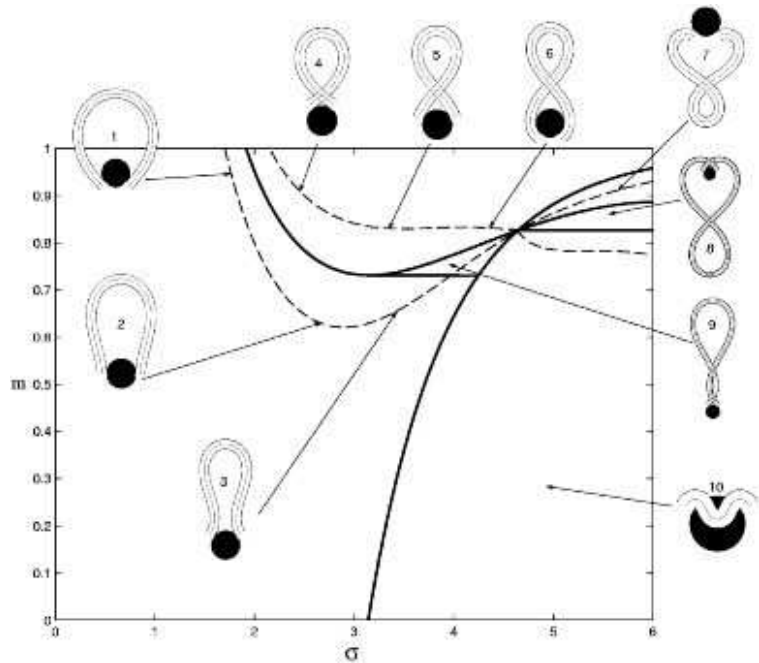}
\end{center}
\caption{Diagram of solutions of the Euler elastica providing an
overview of the possible loop shapes as a function of the shape
parameter $m$ and the contact parameter $\sigma$. Loops of
constant excess length $\Delta L=10$ bp are located on the dashed
lines. The solid lines separate regions with different geometrical
characteristics: simple loops ("1" to "3"), crossed loops ("4" to
"6") and more exotic shapes ("7" to "10").}
\end{figure}

Now the problem of finding the ground state loop for given excess length $%
\Delta L$ reduces to a two variable ($\sigma ,m$) minimization of Eq.~\ref%
{EtotFinal} under the constraint $\Delta L\left( \sigma ,m\right) \stackrel{!%
}{=}\Delta L$. This final step was performed numerically
\cite{kulic02}. An overview over the different solutions can be
obtained by inspecting some loop geometries in the resulting
$\left( \sigma ,m\right) $ parameter plane. Both parameter values
$\sigma $ and $m$ vary between $0$ and $\infty $, though the loops
of practical importance are all found within the range $0<m<1$ and
$0<\sigma<5$. In Fig.~14 this relevant section of the parameter
space is depicted together with a few example loops. The dashed
lines indicate parameter values which lead to constant excess
length $\Delta L=10\times 3.4$ nm (corresponding to 100 bp). On
these lines are located the loop shapes ''1'' to ''7'' that are
examples of such 100 bp-loops. The whole parameter plane is
subdivided into regions of structurally different
solutions that are separated by solid lines. The large region starting at $%
\sigma =0$ contains exclusively simple loops (like ''1'', ''2''
and ''3'') without self-intersections and nucleosome penetration.
Above that simple-loop region there is a region that contains
loops with a single self-intersection; it includes the branch of
100 bp-loops with the example configurations ''4'', ''5'' and
''6''. To the right there are nonphysical cases where the loops
penetrate the nucleosome, like example ''10''. There are also
three other regions with single and double crossing points (''7'',
''8'', ''9'') where the loop can be founds on the ''wrong'' side
of the nucleosome like in ''7'' and ''8''.

\begin{figure}
\begin{center}
\includegraphics*[width=10cm]{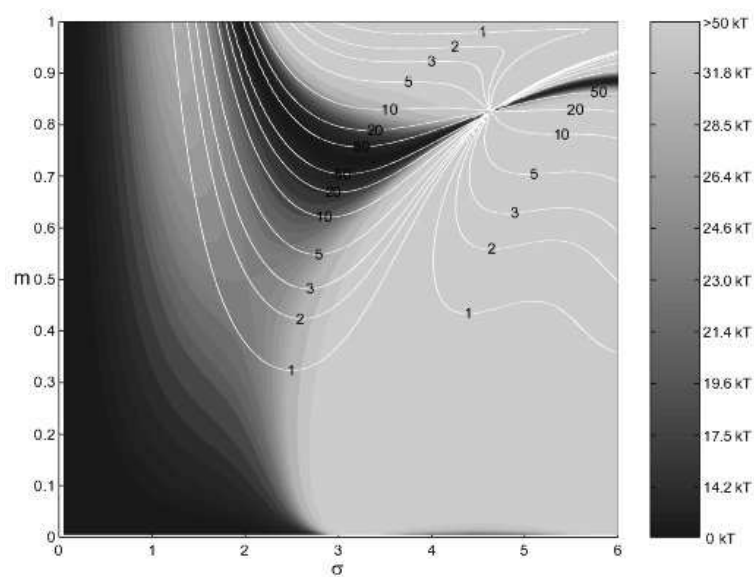}
\end{center}
\caption{Density plot of the total loop energy,
Eq.~\ref{EtotFinal}, as a function of $m$ and $\sigma$ (same
parameter range as in Fig. 14). The white lines denote lines of
contant excess length (in multiples of 10 bp).}
\end{figure}

In Ref.~\cite{kulic02} we determined the energy minimizing loop
for each value of excess length $\Delta L$. Fig.~15 shows a
contour plot of the loop energies, Eq.~\ref{EtotFinal}, in the
$\left( \sigma ,m\right) $ plane for the same range of parameters
as in Fig.~14. The nucleosome parameters chosen
in Fig.~15 are the same as in Ref.~\cite{kulic02}, namely $R_{0}=40$ \AA , $%
\lambda =0.23$ \AA $^{-1}$, $l_{P}=500$ \AA ; these values are
close (but not identical) to the one used in the previous section.
Shown in that picture are also the corresponding lines of constant
$\Delta L$ (with $\Delta L=1,2,..,50\times 3.4$ nm). As already
observed in Fig.~14 there are, for any given $\Delta L$, different
branches of $(\sigma ,m)$ values corresponding to uncrossed,
simply crossed and other, more exotic, structures. For short
excess lengths one finds that the loops with the smallest formation energy $%
\Delta U$, Eq.~\ref{EtotFinal}, belong to the simple, uncrossed
kind. For example, the optimal loop geometry for a 100bp-loop is
structure ''2'' in Fig.~14 which is located at the point where the
lines of the 100bp-loops in Fig.~15 encounter the total energy
minimum. Interestingly, the optimal loop shape switches from the
simple to the crossed type when an excess length of $\sim 500$
\AA\ is reached.

The ground state energy as a function of the excess length $\Delta
L$ is given in Fig.~16. Consider first the {\it simple loops} that
are energetically preferable for $\Delta L<50$ nm. Inspecting
Fig.~16 one finds the remarkable fact that the loop formation
energy is non-monotonous in that range. First it increases sharply
(namely as $\Delta U\propto \Delta L^{1/3}$, cf.
Ref.~\cite{kulic02}, in accordance with the
small loop behavior Eq.~\ref{deltau3}). Then at some critical excess length $%
\Delta L=\Delta L_{crit}$ (which is approximately $\Delta
L_{crit}\approx 2.2\times 3.4$ nm for the above given parameters)
the loop energy reaches a maximum $\Delta U(\Delta
L_{crit})\approx 26k_{B}T$. Beyond that the energy decreases with
increasing $\Delta L$.

In order to explain this behavior one might naively argue as
follows: For excess lengths shorter than the DNA persistence
length it is energetically unfavorable to store additional length
into the loop because it requires increasing deformation of the
loop DNA. On the other hand, for loops longer than $l_{P}$ the
bending energy contribution becomes very small; to add more length
should even decrease this energy since the loop can lower its
curvature. However, the occurrence of the maximum of $\Delta U$ at
$\sim 22$ bp excess DNA lengths, a value that is {\it
considerably} smaller than the persistence length, is surprising
at first sight.

\begin{figure}
\begin{center}
\includegraphics*[width=8cm]{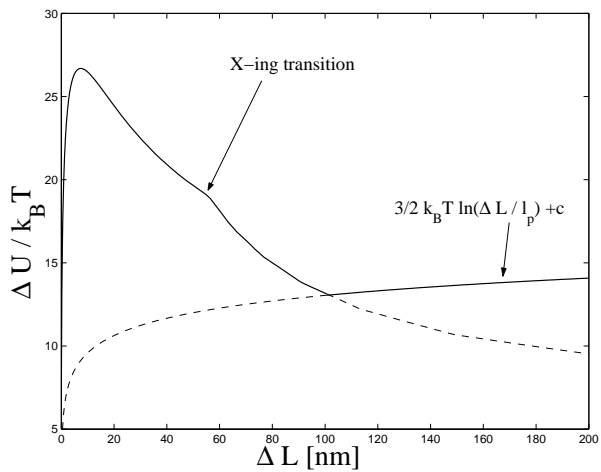}
\end{center}
\caption{The energy of the optimal loops for given excess length
$\Delta L$. The kink in the curve reflects the switch from simple
to crossed loops. The dashed line gives the free energy for
entropic loops that are much longer than their persistence
length.}
\end{figure}

The explanation for this small value is given in
Ref.~\cite{kulic02}. There it is shown that the condition for the
critical excess length $\Delta L_{crit}$ is given by a simple
geometric distinction between two loop shapes: the subcritical
loop (Fig.~17(a)) that has none of its tangents
parallel to the $X$ axis (i.e. $\theta \left( s\right) \neq \pi /2$ for all $%
s$) and the supercritical loop (Fig.~17(b)) that has two or more
tangents parallel to the $X$ axis where $\theta \left( s\right)
=\pi /2$. Now it can be easily envisaged that adding some extra
length $dL$ to a subcritical loop increases its energy $\Delta U$
(cf.~\cite{kulic02}) whereas for supercritical loops additional
length decreases its energy. In the latter case one might just cut
the loop at the two points $P_{L}$ and $P_{R}$ in Fig.~17(b) and
introduce there the additional length (this operation does not
change the energy) and {\it then} relax the shape by letting it
evolve to the new equilibrium while keeping $\theta ^{\ast }$
constant. It was demonstrated in Ref.~\cite{kulic02} that this
condition of the parallel tangents indeed leads to the above given
small value of $\Delta L_{crit}$.

\begin{figure}
\begin{center}
\includegraphics*[width=6cm]{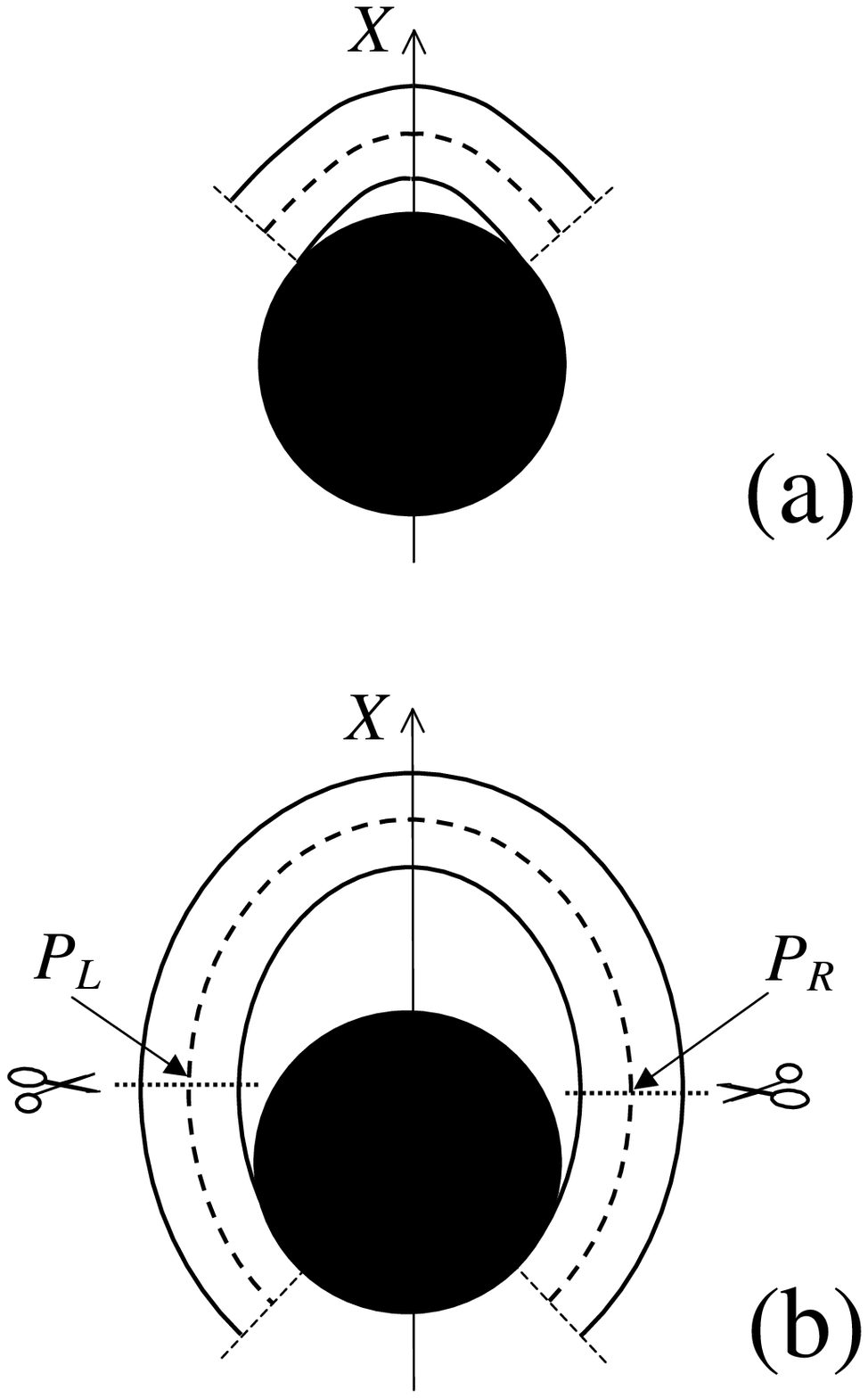}
\end{center}
\caption{Subcritical (a) and supercritical loops show a different
behavior when additional length is added. In the former case the
energy goes up, for the latter it goes down. This allows to
understand why the maximum of $\Delta U$ is reached already for a
very short excess length, cf. Fig.~16.}
\end{figure}

The ground state of loops switches from simple uncrossed loops to
{\it crossed loops} when one reaches an excess length of $\sim
500$ \AA\ . Here, however, arises an additional complication: As
can be seen by inspecting loops ''4'', ''5'' and ''6'' in Fig.~14
these structures contain a self-intersection at the crossing
point. Therefore in principle a planar theory cannot capture the
geometry of crossed loops. One might thus leave the plane and
describe the self-contacts of the rod with corresponding
point-forces in 3d as done by Coleman et al.~\cite{coleman00} in a
general theory of rod self-contacts. However, since the
self-avoiding crossed loops stay close to a plane in all cases of
practical interest (namely loops for which the self-contact point
is not too close to the nucleosome), it is here sufficient to
treat the self-interaction as a perturbation (cf.
Ref.~\cite{kulic02} for details). One finds an additional small
contribution for crossed loops due to the out-of-plane bending
caused by the self-contact. In Fig.~16 this contribution (a few
$k_{B}T$) has already been added; as can be seen from this figure
crossed loops are still favored for sufficiently long excess
lengths $\Delta L>\Delta L_{cross}$ (here $\sim 600$\AA ). This
can be rationalized by the fact that for long enough loops the
adsorption energy (proportional to $\theta ^{\ast }$) starts to
dominate over the bending energy so that loops with smaller
$\theta ^{\ast }$, namely crossed ones, become favorable.

Increasing the length even further one leaves the energy-dominated
regime in which entropic effects can be neglected due to short
loop length $L<l_{P}$. For larger lengths entropic effects become
important and one enters the {\it entropic loop} regime (cf. the
discussion of large-leafed rosettes, Eq.~\ref{flf}, and
Ref.~\cite{yamakawa72}). In the large loop limit where the loop is
longer than several $l_{P}$ the chain looses its ''orientational
memory'' exponentially and behaves as a random walk which starts
from and returns to the same point. The entropic cost for gluing
the ends of this random walk together is approximately given by
\begin{equation}
\Delta U\simeq 3/2k_{B}T\ln (\Delta L/l_{P})+E_{0}  \label{Uenropic}
\end{equation}
Here $E_{0}\approx 6.5$ $k_{B}T$ denotes the bending plus adsorption energy
contributions of the overcrossing DNA segments that enter and leave the
nucleosome in the large-loop limit $\Delta L\rightarrow \infty $. As already
discussed after Eq.~\ref{flf} the free energy minimum occurs at the
crossover between the elastic ($\Delta L<l_{P}$) and entropic ($%
\Delta L\gg l_{P}$) region where the decreasing elastic energy is
overtaken by the increasing entropic contribution -- as can also
be seen in Fig.~16.

The free energy, Eq.~\ref{Uenropic}, leads to an algebraically decaying
probability $w\left( \Delta L\right) $ for the jump lengths scaling as $%
w\propto \exp \left( -\Delta U/k_{B}T\right) \propto \left( \Delta
L\right) ^{-3/2}$. In general, power law distributions of the form
$w\propto \left( \Delta L\right) ^{-\gamma }$ with $\gamma >1$
lead to superdiffusive behavior of the random walker (here the
nucleosome displacement {\it along} the DNA). According to Levy's
limit theorem the probability distribution of the random walker
(more precisely, the distribution of the sums of independent
random variable drawn out from the same probability distribution
$w\propto \left( \Delta L\right) ^{-\gamma }$) converges to a
stable Levy distribution of index $\gamma -1$
\cite{bouchaud90,klafter93,sokolov97}. This so-called {\it
Levy-flight} \cite{mandelbrot82} differs in many respects from the
usual diffusion process as for short time intervals big jumps are
still available with significant probability. Moreover, all
moments (besides possibly the first few ones) diverge. For the
present case $\gamma =3/2$ even the first moment does not exist.
Note that the value $3/2$ is based on the assumption of an ideal
chain (no excluded volume); in general the excluded volume leads
to self-avoiding-walk statistics with a slightly larger
value of $\gamma $ around $2.2$ \cite{sokolov97} (cf. also Ref.~\cite%
{cloizeaux90}). In that case one has a finite value of the first moment,
i.e., a finite average jump length.

We presented in Ref.~\cite{kulic02} some numerical estimates of
the dynamics of the nucleosome repositioning on DNA fragments of
different lengths. The basic idea is that the transition rate
$w\left( \Delta L\right) $ for a jump of length $\Delta L$ is
proportional to $C_{A}\exp \left( -\Delta U\left( \Delta L\right)
\right) $ with $\Delta U$ being the loop formation energy. The
Arrhenius constant $C_{A}^{-1}$ involved in the loop formation has
in
principle to be determined experimentally. A theoretical estimate \cite%
{schiessel01c} was reported in the previous section, cf.
Eq.~\ref{rate}, where it was shown that $C_{A}$ corresponds
roughly to the inverse life time of the loop. Hence
$C_{A}^{-1}\approx l^{2}\eta R/k_{B}T\approx 10^{-5}-10^{-6}s$.
In Ref.~\cite{kulic02} we considered two DNA lengths: $\left( 147+90\right) $%
bp (short segment) and $\left( 147+300\right) $bp (intermediate
length). For the short piece the octamer repositioning occurs on
the time scale of hours (in accordance with the previous section);
on the intermediate segment the repositioning times is of the
order of seconds to minutes. Important in both cases is where the
nucleosome initially starts. If the start position is at an end of
the DNA fragment then the nucleosome jumps preferentially to the
other end since large jumps are energetically favored, cf.
Fig.~16. This leads to a fast relaxation of the initial position.
Smaller jumps take also place but less frequent; these jumps,
however, lead on long time scales to an equal distribution (in
accordance to Boltzmann's law) of the octamer along the DNA
fragment. On the other hand, if the initial position is chosen in
the middle of the DNA piece then the relaxation process is slower
since a smaller loop is initially required; this first jump is
then preferentially to an end position.

As mentioned in Section 2.4.1 the repositioning is often followed
via a gel electrophoresis (cf. Fig.~10). It is therefore helpful
to ask how the resulting band structure evolves with time
\cite{kulic02}. Let us start again with an end positioned
nucleosome. In that case -- as just mentioned -- the octamer
initially shows mainly jumps back and forth between the two ends.
Since the mobility is symmetric with respect to the middle
position this leads to the interesting conclusion that in a
standard gel electrophoresis these jumps would not be detected at
all! Only slowly shorter jumps will allow the nucleosome to
inhabit positions away from the ends. But this slower process
cannot be distinguished easily from a short-ranged diffusion (away
from the initial end), a process like the one discussed in the
previous section. On the other hand, when starting from the middle
position the preference to jump to the end positions will lead to
an initial population gap in the band structure between the fast
band (end positions) and the slow one (middle position). This gap
would not occur for short-ranged repositioning. For more
details on the expected band structures, the reader is referred to Ref.~\cite%
{kulic02}.

\subsubsection{Twist diffusion}

Let me now discuss twist diffusion which might be another possible
mechanism for nucleosome repositioning. I will give here some
first theoretical quantitative estimates for this mechanism
(without any free parameters) based on recent calculations by
Kuli\'{c} and myself; the full presentation will be given
elsewhere \cite{kulic02b}. If a 1 bp twist defect (one missing or
one extra bp) forms through thermal activation at one end and
manages to get through to the other end, this results in a 1 bp
step of the
nucleosome along the DNA and at the same time in a rotational motion by $%
\sim 36^{\circ}$, i.e., the nucleosome performs a short fraction
of a cork screw motion.

The possibility of twist defects was demonstrated as soon as the high
resolution crystal structure of the core particle was resolved \cite{luger97}%
. In that study the core particles were prepared from a
palindromic 146 bp DNA and core histones assuming that the
resulting complex would show a perfect two-fold symmetry. However,
it turned out that one base pair is localized directly on the dyad
axis so that one half of the nucleosomal DNA is of length 73 bp
whereas the other is only 72 bp long. The missing base pair of the
shorter half is, however, not localized at its terminus but
instead at a 10 bp stretch close to the dyad axis (cf. Fig. 4d in
\cite{luger97} that shows a superimposition of the two DNA
halves). The reason is presumably the attraction between the DNA
termini of adjacent particles in the crystal (cf. Fig. 4c in that
paper) that try to come close to mimic a base pair step at the
cost of forming a twist defect far inside the wrapped chain
portion. In fact, crystals of core particles with 147bp DNA do not
show this defect \cite{davey02}.

To proceed further we describe the DNA chain within a Frenkel-Kontorova
model, i.e., we view it as a chain of particles connected by harmonic
springs in a spatially periodic potential. The original Frenkel-Kontorova
model was introduced more than sixty years ago in order to describe the
motion of a dislocation in a crystal \cite{frenkel38}. In the meantime
variants of this model were applied to many different problems including
charge density waves \cite{floria96}, sliding friction \cite%
{braun97,strunz98}, ionic conductors \cite{pietronero81,aubry83},
chains of coupled Josephson junctions \cite{watanabe96} and
adsorbed atomic monolayers \cite{uhler88,schiessel03}. Here, in
the context of DNA adsorbed on the protein octamer, the beads
represent the basepairs. The springs connecting them have an
equilibrium distance of $\overline{b}=0.6$ nm (which is here taken
to be the distance along the DNA sugar-phosphate backbone, {\it
not} the distance $b=0.34$ nm along the fiber axis) and the spring
constant is chosen such that it reflect the DNA elasticity.
Specifically
\begin{equation}
E_{elastic}=\sum_{n}\frac{K\overline{b}^{2}}{2}\left( \frac{x_{n+1}-x_{n}}{%
\overline{b}}-1\right) ^{2}  \label{eelastic}
\end{equation}
with $x_{n}$ being the position of the the $n$th basepair measured
along the helical backbone and $K\overline{b}^{2}/2\simeq
70-100k_{B}T$ accounts for the coupled twist-stretch elasticity
\cite{marko98,marko97c,kamien97}. Finally, the external potential
comes from the contact points to the octamer. The distance between
neighboring contact points is 10 bp which corresponds to 60 nm
along the arclength of the minor groove. A contact point at
position $x_{0}$ is here modelled by the following function
\begin{equation}
E_{ads}=-U_{0}\left( \left( \frac{x-x_{0}}{a}\right) ^{2}-1\right)
^{2}\theta \left( a-\left| x-x_{0}\right| \right)  \label{eads}
\end{equation}
with $\theta \left(x\right)=1$ for $x \ge 0$ and zero otherwise.
The two parameters, the depth $U_{0}$ of the potential and its
width $a$, can be estimated as follows. $U_{0}$ represents the
pure adsorption energy per point contact which can be estimated
from competitive protein binding \cite{polach95,polach96} to be of
order $6k_{B}T$ (cf. Section 2.1). The other parameter, $a$, can
then be estimated from the fluctuations of the
DNA in the crystal (measured by the B-factor, cf. Fig. 1b in \cite{luger97}%
). The fluctuations of the DNA at the binding sites are much smaller than in
the middle in between. Using a quadratic expansion of Eq. \ref{eads} one
finds from a straightforward normal mode analysis \cite{kulic02b} that $%
a\approx \overline{b}/2$, i.e. the adsorption regions lead to a strong
localization of the DNA.

Now having all the numbers at hand we can answer the question
whether the twist defects are localized between two contact
points. The deformation energy of the defect localized along a
10bp stretch is of order $7k_{B}T$. On the other hand, by
distributing the defect homogenously over 20bp the elastic energy
goes down by $\sim 7/2k_{B}T$ at the cost of releasing the
adsorption point in the middle (roughly $6k_{B}T$). The smearing
out of the defect costs therefore $\sim 3k_{B}T$. This already
shows that the kink is not so strongly localized which points
towards a high mobility of the twist defect.

Let us assume now that a kink with one missing basepair is located between
two binding sites. When this kink jumps to the neighboring 10bp (say to the
right) it has to cross a barrier of height $\sim 3k_{B}T$. Using Eqs. \ref%
{eelastic} and \ref{eads} we obtain an explicit form for the
barrier and together with the single bead friction $\zeta \approx
10^{-9}k_{B}Ts/nm^{2}$ \cite{levinthal56,nelson99} are able to
calculate the Kramer's escape rate for the kink from the given
localized 10 bp stretch to a neighboring one. This leads us to a
typical time $t_{step}\approx 10ns$ for going from one stretch to
the next one (cf. Ref.~\cite{kulic02b} for details).

To determine the rate at which twist defects are formed at the
entry/exit points of the DNA one can now use an argument similar
to the one presented in Ref.~\cite{schiessel01c} (cf. also Section
2.4.2): the ratio of the life time $t_{life}$ of a kink to the
time interval $t_{inj}$ between two kink injection events at the
end of the wrapped DNA portion equals the probability to find a
defect on the nucleosome, i.e. $t_{life}/t_{inj}\simeq
N_{site}e^{-\Delta U/k_{B}T}\approx 10^{-2}$. Here $N_{site}=13$
denotes the number of possible positions of the defect between the
14 binding sites.

How is the average life time $t_{life}$ of a defect related to
$t_{step}$, the typical time needed for one step? It is possible
to calculate the mean first passage time for a defect that starts
at one end (say the left one) and leaves the nucleosome at the
same end, $\tau _{left}$, or at the other end, $\tau _{right}$.
From Ref.~\cite{vankampen} one finds $\tau _{left}=\left(
25/6\right) t_{step}$ and $\tau _{right}=28t_{step}$. Furthermore,
the probability to leave at the left end is $p_{left}=12/13$ and
at the right end $p_{right}=1/13$ \cite{vankampen} which gives the
life time as the weighted average $t_{life}=6 t_{step}$. Only a
fraction $p_{right}$ of the defects reaches the other end and will
lead to a repositioning step, i.e., the time of one diffusion step
of the nucleosome along the DNA is of order
$T=t_{inj}/p_{right}\approx 10^{-4}s$ where use was made of the
above presented relations between the time scales. From this
follows then directly the diffusion constant $D$ of the nucleosome
along the
DNA: $D=b^{2}/\left( 2T\right) \approx 7\times 10^{-12}cm^{2}/s$ (with $%
b=0.34$nm).

Therefore we find a diffusion constant that is {\it much} larger
than the one expected for the repositioning via bulges ($D\approx
10^{-16}cm^{2}/s$, cf. Eq. \ref{diff}). Most importantly, it is
also orders of magnitude larger than the diffusion constant
observed in the experiments. How can this apparent inconsistency
be resolved?

Most likely, the diffusion is considerably slowed down due to the
quenched disorder stored in the base pair sequence of the DNA. In
fact, the bulk of the repositioning experiments has been made on
DNA with rather strong positioning sequences leading to a strong
rotational positioning of the nucleosome. Starting from a
preferred position the nucleosome would arrive already after 5
steps to the left (or to the right) on the ''wrong'' site of the
nucleosome -- forcing the DNA to be bent in an unfavorable
direction. This means that the nucleosome needs to cross a barrier
in order to reach a position 10 bp apart. For instance, in the
case of the 5S rDNA sequence theoretical estimates indicate a
barrier height of the
order $10k_{B}T$ \cite%
{mattei02,anselmi00}. This leads to a strong reduction of the
effective diffusion constant, namely $D_{{\it eff}}\approx
De^{-10}\approx 10^{-16}cm^{2}/s$ -- a value comparable to the one
found for bulge diffusion. In Ref.~\cite{kulic02b} we formally
include these effects into the Frenkel-Kontorova framework by
introducing an octamer-fixed bending field and by attributing
''bending charges'' to the beads. This allows us to give a rough
quantitative treatment of the nucleosome mobility as a function of
the underlying base pair sequence.

\subsubsection{Discussion: Bulge versus twist diffusion}

Comparing the different repositioning mechanisms presented in the
previous sections one has to conclude that they lead to very
different ''sliding'' scenarios. On {\it short} DNA fragments
repositioning could in principle work via bulge (Section 2.4.2) or
twist diffusion (Section 2.4.4); large loops cannot occur because
there is not enough free DNA length available. Bulge diffusion is
rather slow (timescale of hours) since the formation of a small
loop is costly -- mainly because the opening angle of a bulge is
rather large leading to several open binding sites. The
repositioning rates should show a strong temperature dependence as
well as a strong dependence on the adsorption strength (i.e., a
strong dependence on the ionic conditions). The preferred
repositioning steps are multiples of ten base pairs.

On the other hand repositioning via twist defects should be much
faster (timescale of seconds). The nucleosome should slide in a
cork screw motion along the DNA and should forget its initial
position rather quickly. However, if the underlying DNA sequence
induces a strong rotational positioning signal the timescale
becomes comparable to that of small loop repositioning. Even more,
due to the underlying base pair sequence one should expect a 10 bp
spacing between the dominant positions that is, however, here not
the result of 10 bp jumps but just reflects the relative Boltzmann
weights of favorable and unfavorable positions. The estimates of
the diffusion constants of these two mechanisms are too unreliable
(activation energies appear in the exponent!) to allow one to
predict which of the mechanisms should be favored. If on the other
hand a rather homogenous DNA sequence is used, our prediction is
that cork screw motion is the much faster and therefore
predominant mechanism. The experiment by Flaus and Richmond
\cite{flaus98} goes already in this direction; comparing their
experimental results (cf. Section 2.4.1) with the theoretical
pictures seems to point towards sliding motion for one of the
positioning sequences (the one that has a homonucleotide tract)
whereas it is not clear whether the nucleosome escapes from the
rotational positioning trap via bulge or via twist diffusion.

On {\it long} DNA fragments single nucleosomes could also be repositioned
via large loops (Section 2.4.3). Our theoretical model suggests that large
loop repositioning would be much faster than bulge diffusion. Also it should
be expected that a similar mechanism allows the nucleosome to be transferred
to competing naked DNA chains. As discussed in the experimental section
above there is, however, not much evidence for such processes. Only Watkins
and Smerdon \cite{watkins85} report such a nucleosome transfer to free DNA
at higher ionic strength. This again allows to speculate that each of the
different mechanisms might play a dominant role in a certain parameter
range. To come to more definite conclusions more systematic experiments have
to be made on short as well as long DNA molecules with and without
positioning sequences under varying ionic conditions.

Repositioning {\it in vivo} might be actively facilitated by chromatin
remodelling complexes whose action is currently studied {\it in vitro }%
(reviewed in Refs.~\cite%
{kornberg99,varga-weisz98,guschin99,peterson00,flaus01}). There
are two major families: SWI/SNF and ISWI. They both burn ATP to
enhance nucleosome dynamics but their underlying modes of action
seem to be fundamentally different. The SWI/SNF class disrupts
many of the DNA-nucleosome contacts making the nucleosomal DNA
vulnerable to DNA digestion. It has been even observed that some
of these complexes are capable to transfer the octamer to another
DNA chain \cite{lorch99}. This might indicate that their mode of
action is the creation of large loops (similar to the one
discussed in Section 2.4.3) that could lead to large repositioning
steps. In fact, Bazett-Jones et al.~\cite{bazzettjones99} observed
that the SWI/SNF complex creates loops on naked DNA as well as on
bead-on-string nucleosome fibers (cf. the electron spectroscopic
images, Fig.~1 and 3, in that paper). On the other hand, the mode
of action of the ISWI family seems not to interrupt the
nucleosome-DNA contact on an appreciable level. Since these
complexes induce nucleosome repositioning it has been speculated
that they might work via twist or bulge diffusion. In the meantime
the latter mechanism seems to be more likely since ISWI induced
nucleosome sliding appears even if the DNA is nicked and hence a
torsion cannot be transmitted between the complex and the
nucleosome to be shifted \cite{laengst01}.

Another interesting and very prominent system known to mediate nucleosome
repositioning is unexpectedly the RNA polymerase. It is found to be able to
transcribe DNA through nucleosomes without disrupting their structure, yet
moving them {\it upstream} the DNA template, i.e., in the opposite direction
of transcription \cite%
{studitsky94,studitsky95,felsenfeld96,studitsky97,felsenfeld00}.
To rationalize this seemingly paradoxical finding Felsenfeld et
al.~\cite{felsenfeld00} introduced a model which assumes that the
polymerase crosses the nucleosome in a loop. This would indeed
explain the backwards directionality of repositioning. Note that
such a loop would have a different shape than the ones discussed
above since polymerases induce a kink at the DNA with a
preferential angle of $\sim 100^{\circ }$ \cite{rees93,schulz98,rivetti99}%
. This means, however, that an RNA polymerase sitting in an
intranucleosomal loop would soon get stuck since it transcribes
the DNA in a
cork screw fashion; this would complicate this mechanism \cite{felsenfeld00}%
. It might well be that this effect only occurs on short DNA
fragments as used in the experiments. If so, it would be an
artefact that would not work {\it in vivo}. In that case another
mechanism, namely induced cork screw motion of the bound DNA
towards the polymerase and subsequently the recapturing of the
nucleosome at its exposed binding sites by the other end of the
DNA fragment might also be a possible scenario \cite{kulicpc}.

\section{30-nm fiber}

\subsection{Solenoid versus crossed-linker model}

Whereas the structure of the core particle has been resolved up to
atomic resolution~\cite{luger97}, there is still considerable
controversy about the nature of the higher-order structures to
which they give rise. When stretched, the string of DNA/histone
complexes has the appearance of \textquotedblright
beads-on-a-string\textquotedblright . This basic structure can be
seen clearly when chromatin is exposed to very low salt
concentrations, and is sometimes referred to as the "10-nm
fiber"~\cite{thoma79}. When the ionic strength is increased
towards physiological values (100 mM),
the fiber appears to thicken, attaining a diameter of 30 nm \cite{widom86}%
. Linker histones (H1 or H5) play an important role in this compaction mechanism:
In their absence fibers form more open structures~\cite%
{thoma79}. These strongly cationic proteins act close to the
entry-exit point of the DNA. They carry an overall positive charge
and seem to bind the two strands together leading to a stem
structure \cite{bednar98}; in fact, this stem is missing in the
absence of linker histones.

There is a longstanding controversial discussion concerning the
structure of the 30-nm fiber
\cite{vanholde89,widom89,vanholde95,vanholde96}. There are
mainly two competing classes of models: the solenoid models \cite%
{thoma79,finch76,widom85}; and the zig-zag or crossed-linker models \cite%
{woodcock93,horowitz94,leuba94,bednar98,schiessel01b,woodcock01}.
In the solenoid model (depicted in Fig.~18(a)) it is assumed that
the chain of nucleosomes forms a helical structure with the axis
of the core particles being perpendicular to the solenoid axis.
The DNA entry-exit side faces inward towards the axis of the
solenoid. The linker DNA (shown as a thick lines at the bottom of
Fig.~18(a)) is required to bent in order to connect neighboring
nucleosomes in the solenoid which in turn requires strong
nucleosome-nucleosome interactions to hold this structure
together. The other class of models posits straight linkers that
connect nucleosomes located on opposite sides of the fiber. This
results in a three-dimensional zig-zag-like pattern of the linker
(cf. Fig.~18(b)).

\begin{figure}
\begin{center}
\includegraphics*[width=8cm]{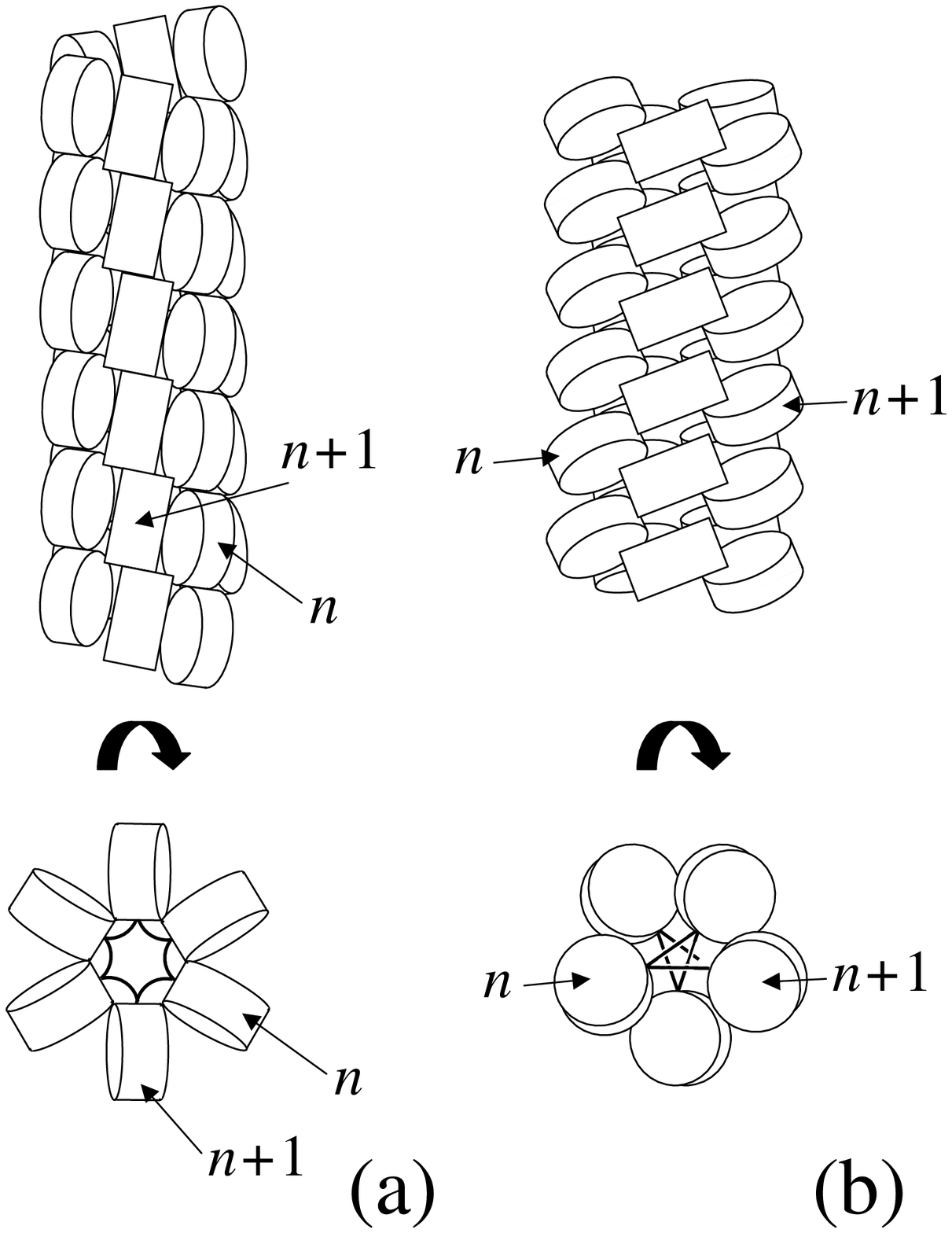}
\end{center}
\caption{The two competing models for the 30-nm fiber: (a) the
solenoid model and (b) the crossed linker model. Both are shown
from the side and from the top; the latter view allows to
distinguish the different linker geometries. Note that these are
idealized models; real fibers are believed to be less regular
\cite{vanholde95}.}
\end{figure}

Images obtained by electron cryomicroscopy \cite{bednar98} should
in principle be able to distinguish between the structural
features predicted by the two different models. The micrographs
show a zig-zag motif at lower salt concentrations and they
indicate that the chromatin fiber becomes more and more compact
when the ionic strength is raised towards the physiological value.
A similar picture emerges also from atomic force microscopy
\cite{leuba94,zlatanova98}. However, neither method allows to
identify the linker geometry at physiological ionic conditions so
that one still cannot exclude the possibility that the fiber folds
close to physiological conditions into a solenoid-like structure
by a bending of its linkers. This is in fact the structure that is
depicted in most of the standard textbooks on cell biology (e.g.
Ref.~\cite{alberts94}). Also X-ray diffraction data that
constituted the basis for many models lead to controversial
interpretations, cf. Ref.~\cite{vanholde95} for a critical
discussion.

In view of this fact it is an important recent experimental
achievement that single
chromatin fibers can be stretched via micromanipulation techniques \cite%
{cui00,bennink01,brower02}. The force-extension curves allow in principle to
discern between the different structures. So far computer simulations \cite%
{katritch00} as well as analytical approaches \cite{schiessel01b,ben-haim01}
to chromatin fiber stretching seem -- when comparing their predictions to
the experimental data -- to support the crossed linker models.

Another intriguing way that might allow to discriminate experimentally
between these two types of structures is to measure the fiber orientation in
strong magnetic fields -- as it has been done already long ago \cite{maret76}%
. Such a method has been used successfully to determine the
persistence length of naked DNA \cite{maret75,maret83}. One makes
use of the anisotropic magnetic susceptibility of the base pairs
that cause the DNA double helix to orient its axis perpendicular
to the field. As a consequence, single core particles orient their
DNA superhelix axis parallel to the field \cite{maret76}. Since
the nucleosome axes in the two fiber models are oriented in
different directions with respect to the fiber axis (cf. Fig. 18),
an external field would induce orientations of the two fiber
models in different directions.

In the following I will focus on analytical models and computer
modeling of the chromatin fiber which all belong to the class of
crossed-linker models. In the next section the possible
geometrical structures that follow from regular two-angle fibers
(a ''generalization'' of crossed-linker models) are presented.
Section 3.3 gives a speculation about the ''optimal'' fiber design
from a biological point of view. Then in Section 3.4 I will give a
detailed account on the mechanical properties of the fiber
comparing analytical results and computer models to recent
stretching experiments. Finally, in Section 3.5 I report on a
recent model that relates the degree of fiber swelling to the
ionic strength.

\subsection{Structure diagram of the two-angle fiber}

To address the folding problem of DNA at the level of the 30-nm fiber
myself, Gelbart and Bruinsma \cite{schiessel01b} introduced a mathematical
description for the different possible folding pathways which was based on
Woodcock's crossed-linker model \cite{woodcock93} (cf. also a related study
on closed minichromosomes \cite{martino99}). At the simplest level, we
assumed that the geometric structure of the 30-nm fiber can be obtained from
the intrinsic, single-nucleosome structure. The specific roles of linker
elastic energy, nucleosome-nucleosome interaction, preferred binding sites,
H1 involvement, etc. was then treated afterwards as ''corrections''\ to this
basic model \cite{schiessel01b,schiessel02}. To see how single-nucleosome
properties can control the fiber geometry, consider the fact that DNA is
wrapped a non-integral number of turns around the nucleosome, e.g., 1-and-$%
3/4$ times (147 bp) in the case of no H1. This implies that the
incoming and outgoing linker chains make an angle $\theta $ with
respect to each other -- the entry-exit angle $\pi -\theta $ is
nonzero. In the presence of the histone H1 (or H5) the in- and
outcoming DNA are glued together along a short section resulting
in a stem-like structure \cite{bednar98}. While the precise value
of the resulting exit-angle depends on salt concentration,
presence or absence of linker histones, degree of acetylation of
the histones, etc. (discussed in Section 3.4) one may nevertheless
assume $\theta $ to be a quantity {\it that is determined purely
at the single-nucleosome level}.

\begin{figure}
\begin{center}
\includegraphics*[width=4cm]{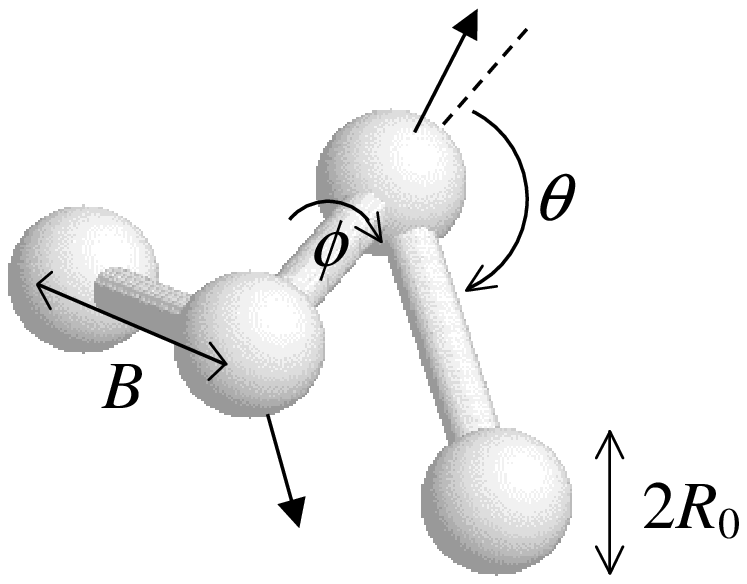}
\end{center}
\caption{Fraction of a two-angle fiber containing four
nucleosomes. The two angles, the defection angle $\theta$ and the
dihedral angle $\phi$ are depicted together with the nucleosome
diameter $2R_0$ and the "linker length" $B$. The arrows denote the
nucleosomal superhelix axis, cf. Fig.~2.}
\end{figure}

Next, there is a rotational (dihedral) angle $\phi $ between the
axis of neighboring histone octamers along the necklace (see
Fig.~19). Because nucleosomes are rotationally positioned along
the DNA, i.e., adsorption of DNA always begins with the minor
groove turned in towards the first histone binding site, the angle
$\phi $ is a {\it periodic} function of the linker length $B$,
with the 10 bp repeat length of the helical twist of DNA as the
period. There is experimental evidence that the linker length
shows a preferential quantization involving a set of values that
are related by integral multiples of this helical twist
\cite{widom92}, i.e., there is a preferred value of $\phi $.

Treating the pair of angles $\left( \theta ,\phi \right) $,
together with the linker length $B$, as {\it given} physical
properties (even though {\it in vivo} they are likely under
biochemical control), the geometrical
structure of the necklace is determined entirely by $\theta $, $\phi $ and $%
B $. The model only describes linker geometry and does not account
for excluded volume effects and other forms of
nucleosome-nucleosome interaction; it assumes that the core
particles are pointlike ($R_{0}=0$) and that they are located at
the joints of the linkers. The model also assumes that the linkers
are {\it straight}. The $\left( \theta ,\phi \right) $-model is
similar to the freely rotating chain model encountered in polymer
physics literature (see, for instance, Ref.~\cite{doi86}). The
main difference is that in the present case there is no free
rotation around the linker and so torsion is transmitted (see also
Ref.~\cite{plewa00}).

\begin{figure}
\begin{center}
\includegraphics*[width=10cm]{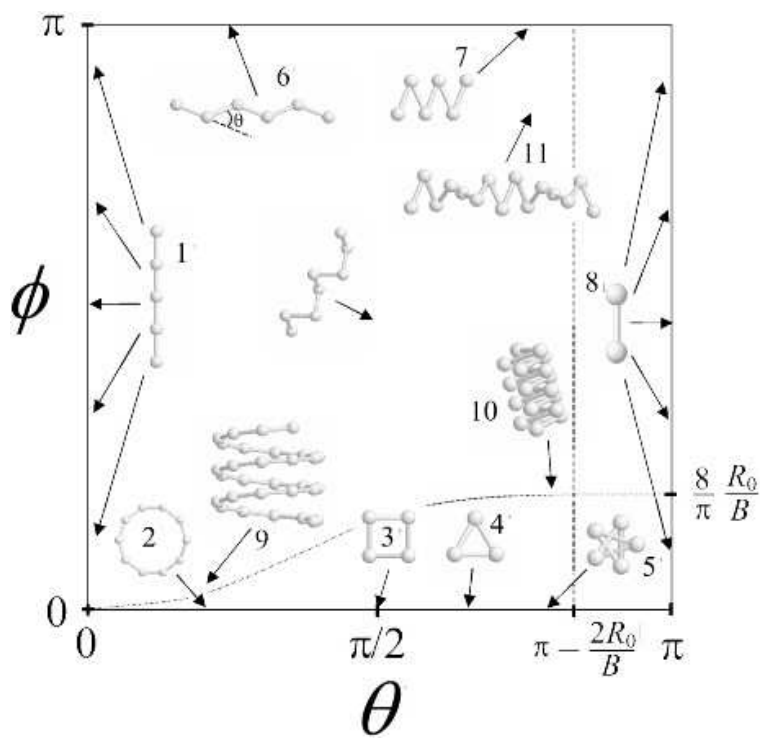}
\end{center}
\caption{Diagram of geometries of two-angle fibers in the
$(\theta,\phi)$-plane. Shown are some example configurations with
the arrows denoting their position in the plane. The lines give
the boundaries to the forbidden structures due to short-range
excluded volume (large $\theta$ values) and long-range excluded
volume (small $\phi$-values).}
\end{figure}

Before giving a detailed discussion of two-angle geometries let me
provide a short overview of the possible structures in the $\left(
\theta ,\phi \right) $-space that is shown in Fig.~20. Both angles
$\theta $ and $\phi $ can each vary over the range $0$ to $\pi $.
At the edges of the diagram where one of the angles assumes an
extremal value, the configurations are always planar. On the line
$\phi =0$ are located circles (see structure ''2'' in Fig.~20) and
star-type polygons (that are closed for specific values of $\theta
$ like ''5''). The planar zig-zag-structures are found on the line
$\phi =\pi $ (cf. ''6'' and ''7''); for $\theta =0$ one has
straight configurations (''1'') and for $\phi =\pi $ ''dimer''
structures (''8''). If one moves from the line $\phi =0$ towards
larger values of $\phi $ the circles and star-like polygons
stretch out into the direction perpendicular to their plane,
forming solenoids (''9'') and fibers with crossed linkers
(''11''), respectively. On the other hand, if one starts at the top of the diagram ($%
\phi =\pi $) and decreases the value of $\phi $ the planar zig-zag
structure extends into the third dimension by becoming twisted
(''11''). Various examples of two-angle fibers were displayed by
Woodcock et al.~\cite{woodcock93} in their Fig.~2, namely fibers
with $\theta =150{{}^{\circ }}$ and many different values of $\phi
$, corresponding to a vertical trajectory on the right-hand side
of Fig.~20. Three different configurations with a fixed value of
$\phi $ and different values of $\theta $ are displayed in
Fig.~3(c) in another paper by these authors \cite{bednar98}.

An analytical decription of the structures can be achieved as
follows (cf. Ref.~\cite{schiessel01b}): It is possible to
construct a spiral of radius $R$ and pitch angle $\gamma $ such
that the nucleosomes -- but not necessarily the linker chain --
are located on this spiral. The nucleosomes are placed along the
spiral in such a way that successive nucleosomes have a fixed
(Euclidean) distance $B$ from one another. From straightforward
geometrical considerations we derived in Ref.~\cite{schiessel01b}
analytical expressions that relate pitch angle $\gamma $ and
radius $R$ of the solenoid as well as $s_{0}$ (defined as the
vertical distance between successive ''nucleosomes'' along the
helical axis) to the pair of angles
$\theta $, $\phi $ and linker length $B$. The corresponding relations $%
B=B\left( \gamma ,R,s_{0}\right) $, $\theta =\theta \left( \gamma
,R,s_{0}\right) $ and $\phi =\phi \left( \gamma ,R,s_{0}\right) $
are Eqs.~(32) to (34) in Ref.~\cite{schiessel01b}. I present here
the reverse relations that have the advantage that they allow the
direct calculation of the overall fiber geometry from the local
geometry. Specifically, the radius $R$ of the master solenoid is
given by
\begin{equation}
R=\frac{B\sin \left( \theta /2\right) }{2\allowbreak -2\cos ^{2}\left(
\theta /2\right) \cos ^{2}\left( \phi /2\right) }  \label{r}
\end{equation}
and its pitch angle $\psi $ by
\begin{equation}
\cot \psi =\frac{\tan \left( \theta /2\right) \arccos \left( 2\cos
^{2}\left( \theta /2\right) \cos ^{2}\left( \phi /2\right) -1\right) }{2\sin
\left( \phi /2\right) \sqrt{1\allowbreak -\cos ^{2}\left( \theta /2\right)
\cos ^{2}\left( \phi /2\right) }}  \label{alp}
\end{equation}
Finally, the distance $s_{0}$ of neighboring nucleosomes along the fiber
axis is obtained from
\begin{equation}
s_{0}=\frac{B\sin \left( \phi /2\right) }{\sqrt{\sec ^{2}\left( \theta
/2\right) -\cos ^{2}\left( \phi /2\right) }}  \label{s0}
\end{equation}
Using these relations, it is straightforward to construct a
catalog of structures.

If either one of the angles $\theta $ or $\phi $ assumes the value $0$ or $%
\pi $, then the resulting structure is {\it planar}. Consider
first the line $\phi =0$. If one also has $\theta =0$ the fiber
forms a straight line (''1'' in Fig.~20). For small
non-vanishing $\theta $ the structure forms a {\it circle} of radius $%
R\simeq B/\theta $ (as follows directly from Eq.~\ref{r}). For the
special case $\theta =2\pi /n$, with $n$ an integer, the ring
contains $n$ monomers before it repeats itself and one obtains a
{\it regular polygon} (''2''). The special case $\theta =\pi /2$
corresponds to the square (''3''). With increasing $\theta $ the
radius of the circle shrinks and approaches
asymptotically the value $B/2$. For $\theta =\pi \left( n-1\right) /n$ with $%
n$ being an odd integer one encounters a series of closed {\it
star-like polygons} with $n$ tips. In particular, $n=3$
corresponds to the equilateral triangle (''4'') and $n=5$ to the
regular pentagram (''5'').

Next consider the case $\phi =\pi $ and $\theta $ arbitrary. This
case corresponds to 2D {\it zig-zag-like structures}, as shown by
''6'' and ''7'' at the top of Fig.~20. The length of a fiber
consisting of $N$ monomers is given by $L=s_{0}N=B\cos \left(
\theta /2\right) N$ (cf. Eq.~\ref{s0}) and the diameter is given
by $2R=B\sin \left( \theta /2\right) $ (cf. Eq.~\ref{r}). Note
that the length of the fiber increases with decreasing $\theta $.
Finally, there are two remaining cases of planar structures:
$\theta =0$ with an arbitrary value of $\phi $ leads to the
straight line mentioned earlier (''1''); $%
\theta =\pi $ and arbitrary $\phi $ corresponds to linkers that go
back and forth between two positions (''8'').

Structures with $\theta \neq 0$ {\it and} $\phi \neq 0$ form {\it %
three-dimensional fibers}. For small angles, $\theta \ll 1$ and $\phi \ll 1$%
, structures resemble {\it solenoids} (see ''9'') where the
linkers themselves follow closely a helical path corresponding to
that of the master solenoid. For these structures one finds from
Eqs.~\ref{r} and \ref{s0} the following limiting behavior of the
fiber radius and length (for $N$ monomers):
\begin{equation}
R\simeq \frac{B\theta }{\phi ^{2}+\theta ^{2}},\,\,\,\,\,L\simeq \frac{%
BN\phi }{\sqrt{\phi ^{2}+\theta ^{2}}}  \label{sol1}
\end{equation}
Furthermore, the pitch angle $\psi $ is given by
\begin{equation}
\cot \psi \simeq \frac{\theta }{\phi }  \label{alpha2}
\end{equation}
This suggests a classification of solenoids into dense helices with small
pitch angle $\psi \simeq \phi /\theta $ for $\phi \ll \theta $ and open
helices with large pitch angle $\psi \simeq \pi /2-\theta /\phi $ for $\phi
\gg \theta $. Other geometrical information can be obtained easily. For
instance, the vertical distance $d$ between two turns follows from $d=2\pi
R/\cot \psi $ to be
\begin{equation}
d\simeq \frac{2\pi \phi B}{\phi ^{2}+\theta ^{2}}  \label{d}
\end{equation}
Dense helices, $\phi \ll \theta $, are characterized by $d\ll R$ and open
ones by $d\gg R$.

Structures where $\phi $ is still small but where the entry-exit angle $%
\theta $ is large, i.e. $\pi -\theta \ll \pi $, form {\it fibers with
crossed linkers}. As discussed above for $\phi =0$ one encounters star-shape
polygons that are closed for $\theta =\pi \left( n-1\right) /n$ with $n$
odd. For {\it non}-vanishing $\phi \ll 1$ the star-shaped polygons open up
in an accordion-like manner into a three-dimensional fiber with the
following radius and length (for $N$ monomers):
\begin{equation}
R\simeq \frac{B}{2\sin \left( \theta /2\right) },\,\,\,\,\,L\simeq \frac{%
BN\phi }{2}\cot \left( \theta /2\right)  \label{r2}
\end{equation}
Assume now that $\theta _{n}=\pi \left( n-1\right) /n$ so that the
projection of the fiber is a closed polygon (this is only strictly true for $%
\phi =0$ but it is still a good approximation for $\phi \ll 1$). For this
set of angles monomer $i$ and $i+n$ come very close in space; their distance
$d$ follows from the master solenoid that has $n-1$ turns in between these
two monomers:
\begin{equation}
d\simeq \frac{2\pi \left( n-1\right) R}{\cot \psi }\simeq \frac{\pi \phi B}{4%
}  \label{d2}
\end{equation}

Finally structures with a rotational angle $\phi $ close to $\pi $, say $%
\phi =\pi -\delta $ with $\delta \ll 1$, lead to {\it twisted zig-zag
structures\ }-- see ''11''. In this case monomer $i+1$ is located nearly
opposite to the $i$th monomer, but slightly twisted by an angle $\delta $.
Monomer $i+2$ is then on the same side as monomer $i$ but slightly twisted
by an angle $2\delta $ and so on. The geometrical properties of the
resulting fiber are the following
\begin{equation}
R\simeq \frac{B}{2}\sin \left( \theta /2\right) ,\,\,\,\,\,L\simeq BN\cos
\left( \theta /2\right)  \label{r3}
\end{equation}
and show only a higher order dependence on $\phi $ that we gave
explicitly in Ref.~\cite{schiessel01b}. For $\phi =\pi $ one
recovers the planar zig-zag structure for which Eq.~\ref{r3}
becomes exact.

If one takes into account the excluded volume of the core
particles, then certain areas in that phase diagram are forbidden
-- reminiscent of the familiar Ramachandran plots used in the
study of protein folding \cite{stryer95}. For simplicity we assume
in the following that the core particles are spherical with a
radius $R_{0}$ and that their centers are located at the joints of
two linkers, cf. Fig.~19. There are two different types of
interactions. One is between monomers at position $i$ and $i\pm 2$
(short range interaction), and leads to the requirement that the
entry-angle must be sufficiently small:
\begin{equation}
\theta <2\arccos \left( R_{0}/B\right)  \label{ev1}
\end{equation}
This condition excludes a vertical strip at the right side of the diagram,
as indicated in Fig.~20 by a dashed line.

There is {\it also} a long-range excluded volume interaction that comes into
play for small values of $\phi $. This is apparent for the case $\phi =0$
where one finds planar structures that run into themselves. Starting with a
circular structure one has to increase $\phi $ above some critical value so
that the pitch angle of the resulting solenoid is large enough so that
neighboring turns do not interact. This leads to the requirement $d>2R_{0}$
with $d$ given by Eq.~\ref{d} (using $\phi \ll \theta $), i.e.,
\begin{equation}
\phi >\frac{1}{\pi }\frac{R_{0}\theta ^{2}}{B}  \label{ev2}
\end{equation}
For the large $\theta $-case (fibers with crossed linkers) one finds from
Eq.~\ref{d2} the condition
\begin{equation}
\phi >\frac{8}{\pi }\frac{R_{0}}{B}  \label{ev3}
\end{equation}
The two conditions, Eqs.~\ref{ev2} and \ref{ev3}, shown schematically as a
dotted curve in Fig.~20, lead to a forbidden strip in the structure diagram
for small values of $\phi $.

Figure 20 does not show the interesting ''fine-structure'' of the
boundary of that forbidden strip that is due to
commensurate-incommensurate effects. I already noted that there
are special $\theta $-values for which the projection of the
linkers forms a regular polygonal star ($\theta _{n}=\pi \left(
n-1\right) /n$) or a regular polygon ($\theta _{n}^{\prime }=2\pi
/n$) (for small values of $\phi $). In these cases the nucleosomes
$i$ and $i+n$ ''sit'' on top of each other. On the other hand, for
other values of $\theta $, monomers of neighboring loops will be
displaced with respect to each other. In this case monomers of one
loop might be able to fill in gaps of neighboring loops so that
the minimal allowed value of $\phi $ is smaller than estimated
above. We are currently exploring the interesting mathematical
problem of the exact boundary line that is also sensitive to the
exact nucleosome shape \cite{mergell03}. The dotted line in
Fig.~20 only represents the upper envelope of the actual curve.

The above given discussion of the two-angle model was based of the
assumption of a perfectly homogeneous fiber where $B$, $\theta $
and $\phi $ are constant throughout the fiber. For a discussion of
the effect of randomness in these values on the fiber geometry I
refer the reader to Ref.~\cite{schiessel01b}.

\subsection{Chromation fiber: optimization of design?}

If one assumes that the chromatin fiber has a relatively regular
structure and that the linker DNA is straight, then the two-angle
model might be a good description of the fiber geometry. In that
case the question arises where in the structure diagram, Fig.~20
the 30-nm fiber is actually located. The diagram, by itself, does
not favor any structure over another. However,
the diagram plus the formulae given above allowed use in Ref.~\cite%
{schiessel01b} to invoke the following two criteria to optimize
the structure of the 30-nm fiber and to check {\it a posteriori}
their usefulness. The two suggested criteria are:
\begin{eqnarray*}
&&({\it i})\,\,{\it maximum\ compaction} \\
&&({\it ii})\,\,{\it maximum\ accessibility}
\end{eqnarray*}
The first criterion is obvious: inactive chromatin should be
packed as dense as possible because of the very large ratio of DNA
length to nucleus size (cf. also Ref.~\cite{daban00}\ to see how
severe this packing problem actually is). By the second criterion
we meant that a {\it local accessibility }mechanism is required
for gene transcription and that this mechanism should somehow be
optimized (see below).

\smallskip In order to attain maximum compaction one needs structures that
lead to high bulk densities $\rho =1/\left(
2\sqrt{3}R^{2}s_{0}\right) $ (assuming that the 30-nm fibers are
packed in parallel forming a hexagonal lattice). A comparison of
the 3d densities of all possible structures shows that fibers with
internal linkers have highest densities $\rho $, namely (cf.
Eq.~\ref{r2})
\begin{equation}
\rho \simeq \frac{16}{2\sqrt{3}\phi \left( \pi -\theta \right) B^{3}}
\label{rho}
\end{equation}
In particular, the highest density is achieved for the largest possible
value of $\theta $ and the smallest possible value of $\phi $ that is still
in accordance with the excluded volume condition. This set of angles is
located at the point where the dotted curve and the dashed line in Fig.~20
cross each other. Apparently this {\it also} represents the only region in
the phase diagram where excluded volume effects are operative on a
short-range and a long-range scale at the {\it same} time, i.e., nucleosome $%
i$ is in close contact with nucleosome $i-2$ and $i+2$ {\it as
well as} with nucleosomes father apart along the contour length of
the necklace. This unique set of angles is given by $\theta _{\max
}=2\arccos \left( R_{0}/B\right) $, cf. Eq.~\ref{ev1}, and $\phi
_{\min }\simeq \left( 8/\pi \right) \left( R_{0}/B\right) $, cf.
Eq.~\ref{ev3}.

In order to achieve maximum accessibility we looked in Ref.~\cite%
{schiessel01b} for structures that, for a given entry-exit angle $\pi
-\theta $ of a highly compacted structure, achieve the maximum reduction in
nucleosome line density $\rho _{L}=s_{0}^{-1}$ for a given small change $%
\Delta \theta $ of the angle $\theta $. In other words, we looked
for a maximum of $d\rho _{L}/d\theta \,$which we called the
\textquotedblright accessibility\textquotedblright . {\it
Interestingly, the accessibility is maximized at the same unique
pair of angles} $\left( \theta _{\max },\phi _{\min }\right) $.
This can be seen from its angle dependence for fibers with crossed
linkers
\begin{equation}
\frac{d\rho _{L}}{d\theta }\simeq \frac{4}{\phi \left( \pi -\theta
\right) ^{2}B}  \label{acc}
\end{equation}
Note that this change in $\rho _{L}$ with $\theta $ is achieved by changing
the number of monomers per vertical repeat length $d$. The length $d$ itself
is only weakly dependent on $n$ according to Eq.~\ref{d2}.

The above given formulas are now compared with experimental
results. For chicken erythrocyte chromatin one has $B\approx 20$
nm (center-to-center distance of nucleosomes, \cite{vanholde96}).
Together with
$R_{0}\approx 5$ nm this leads to $\theta _{\max }\simeq 151^{\circ }$, $%
\phi _{\min }\simeq 36^{\circ }$ and $\rho _{L}\simeq 6.9$
nucleosomes per 11 nm (using Eqs.~\ref{s0}, \ref{ev1} and
\ref{ev3}; the approximate formula, Eq.~\ref{r2} gives $\rho
_{L}\simeq 6.8$). The theoretically derived values can now be
compared with the experimental ones reported by Bednar et
al.~\cite{bednar98} for chicken erythrocyte chromatin fibers. From
their table 1 one finds that for an ionic strength of 80 mM (which
is close to the physiological value) $\theta \approx 145^{\circ }$
and $\rho _{L}=6.0$ nucleosomes per 11 nm. Furthermore, electron
cryotomography-constructed
stereo pair images of an oligonucleosome (cf. Fig.~3(b) in Ref.~\cite%
{bednar98}) indicate that the chromatin fiber might indeed have
the structure of a fiber with crossed linkers, with $n\approx 5$;
this would correspond to $\theta =\pi \left( n-1\right) /n\approx
144^{\circ }$.

Information concerning the preferred value for $\phi $ can be
obtained from the measured statistical distribution of the
nucleosome repeat lengths.
This distribution shows statistically preferred linker lengths equal to $%
10k+1$bp with $k$ a positive integer \cite{widom92}, which, in
turn, indicates that the rotation angle $\phi $ corresponds to a
change in helical pitch associated with 1 bp, i.e. $360^{\circ
}/10=36^{\circ }$. This value coincides with $\phi _{\min }$, the
value that we estimated for maximum compaction. However, the
statistical uncertainty around the expectation values for the
nucleosome repeat length is sufficiently large to make this
estimate for $\phi $ less reliable.

The second feature, the local accessibility, can be monitored {\it in vitro}
by changing the salt concentration. Bednar et al. report, for example, that $%
\theta $ decreases with decreasing ionic strength, namely $\theta
\approx 145^{\circ }$ at 80 mM, $\theta \approx 135^{\circ }$ at
15 mM and $\theta \approx 95^{\circ }$ at 5 mM \cite{bednar98}. In
the biochemical context the change of $\theta $ is accomplished by
other mechanisms, especially by the depletion of linker histones
and the acetylation of core histone tails (cf. my discussion in
Section 3.5), both of which are operative in transcriptionally
active regions of chromatin. These mechanisms lead effectively to
a decrease of $\theta $.

As pointed out below Eq.~\ref{acc}, the decrease of $\theta $ is
accompanied by a decrease of the line-density $\rho _{L}=n/d$ of
nucleosomes at an essentially fixed value of $d$. In other words,
the number of vertices of the projected polygon decreases
significantly with decreasing $\theta $ because $\theta _{n}=\pi
\left( 1-1/n\right) $. In that respect the effect of reducing
$\theta $ below the optimal packing value might be best viewed as
an ''untwisting'' of the 30-nm fiber. Using the experimentally
determined values of $\theta $ one finds that the line density
(the number of nucleosomes per 11 nm) is given by $\rho
_{L}\approx 5.7$ for $\theta \approx 145^{\circ }$, $\rho
_{L}\approx
4.3 $ for $\theta \approx 135^{\circ }$ and $\rho _{L}\approx 2.0$ for $%
\theta \approx 95^{\circ }$, which is in reasonable agreement with
the experimental values $\rho _{L}\approx 6.0$, $\rho _{L}\approx
3.2$ and $\rho _{L}\approx 1.5$ \cite{bednar98}. Furthermore, the
number of polygonal
vertices $n=\pi /\left( \pi -\theta \right) $ decreases as follows: $%
n\approx 5.1$ for $\theta \approx 145^{\circ }$, $n\approx 4.0$ for $%
\theta \approx 135^{\circ }$ and $n\approx 2.1$ for $\theta
\approx 95^{\circ }$, consistent with the stereo pair images by
Bednar et al., suggesting $n\approx 5$ at an ionic strength of 80
mM and $n\approx 3$ at 5 mM (cf. Figs. 3(a) and (b) in
\cite{bednar98}).

Let me close this section with a cautionary remark
\cite{schiessel01b}. The 3d density and the line density of the
fiber can not only be changed by changing $\theta $ or $\phi $ but
also by changing the linker length (in multiples of 10 bp). A
variation in $B$ changes the location of the point $\left( \theta
_{\max },\phi _{\min }\right) $ in the diagram of geometrical
states, and thus the values of the maximum 3d and line densities
that can be achieved, namely
\begin{equation}
\rho ^{\left( \max \right) }\simeq \frac{16}{2\sqrt{3}\phi _{\min }\left(
\pi -\theta _{\max }\right) B^{3}}\simeq \frac{\pi }{2\sqrt{3}R_{0}^{2}B}
\label{rhomax}
\end{equation}
and
\begin{equation}
\rho _{L}^{\left( \max \right) }\simeq \frac{4}{B\phi _{\min }\left( \pi
-\theta _{\max }\right) }\simeq \frac{\pi }{4}\frac{B}{R_{0}^{2}}
\label{lamdamax}
\end{equation}
This shows that fibers with smaller values of $B$ can achieve
higher 3d densities but have a smaller maximal line density (and
accessibility $d\rho _{L}/d\theta \propto B^{2}$). From this one
might infer that active cells should have larger nucleosome repeat
lengths in order to maximize the accessibility to their genetic
material. An overview on nucleosome repeat lengths in different
organisms and tissues is given in table 7-1 of van Holde's book
\cite{vanholde89}. The data shown there do not follow this rule,
unfortunately. In fact, very active cells like yeast cells and
neuronal cells have in general short nucleosome repeat lengths
while inactive ones like sperm cells have large ones. This shows
that the optimization principle of high density and accessibility
has to be used with caution.

\subsection{Mechanical properties of the two-angle model}

The two angle model -- as discussed in the previous sections -- is purely
geometrical. Could it be useful as well for predicting {\it physical
properties} of the 30-nm fiber? The response of the 30-nm fiber to {\it %
elastic stress} was indeed one of the major issues in our paper on
the two-angle model \cite{schiessel01b}. In an independent study
on the two-angle model by Ben-Ha\"{\i}m, Lesne and Victor
\cite{ben-haim01} this question has been the major focus. By
combining in this section results from both papers will allow for
the first time to give analytical expression for the elastic
properties of the two-angle model as a function of the underlying
pair of angles $\theta $ and $\phi $.

Before doing so let me remark that the elastic stress can either
be of external or of internal origin. External stresses are
exerted on the chromatin during the cell cycle when the mitotic
spindle separates chromosome pairs \cite{janninck96}. The 30-nm
fiber should be both highly flexible and extensible to survive
these stresses. The {\it in vitro} experiments by Cui and
Bustamante demonstrated that the 30-nm fiber is indeed very
''soft'' \cite{cui00}. The 30-nm fiber is also exposed to {\it
internal stresses}. Attractive or repulsive forces between the
nucleosomes will deform the linkers connecting the nucleosomes.
For instance, electrostatic interactions, either repulsive (due to
the net charge of the nucleosome core particles) or attractive
(bridging via the lysine-rich core histone tails \cite{luger97})
could lead to considerable structural adjustments of the model.

Before considering the elastic properties of the two-angle model,
it is helpful to briefly recall some results concerning the
large-scale elasticity of the DNA itself \cite{marko98,cluzel96}.
The measured force-extension curve of naked DNA breaks up into two
highly distinct regimes: the ''entropic'' and ''enthalpic''
elastic regimes. For very low tension $F$ ($\le$ pN), the
restoring force is provided by ''entropic elasticity''
\cite{degennes79}. In the absence of any force applied to its
ends, the DNA's rms end-to-end distance (chain length, $L$) is
small compared to its contour length ($L_{0}$) and the chain
enjoys a large degree of conformational disorder. Stretching DNA
reduces its entropy and increases the free energy. The
corresponding force $f$ increases linearly with the extension $L$:
\begin{equation}
F\simeq \frac{3k_{B}T}{l_{P}}\frac{L}{L_{0}},\,\,\,\,\,\,L\ll
L_{0} \label{dna1}
\end{equation}
with $l_{P}\approx 500$ \AA\ being the thermal persistence length
of DNA \cite{hagerman88}.

For higher forces ($F>10$ pN), the end-to-end distance $L$ is
close to $L_{0}$ and the elastic restoring force is due to
distortion of the internal structure of DNA. In this regime, the
force extension curve can be approximated by
\begin{equation}
F\simeq k_{B}T\gamma \frac{L-L_{0}}{L_{0}},\,\,\,\,\,\,L> L_{0}
\label{dna2}
\end{equation}
The stretching modulus $\gamma =\left( \partial f/\partial
L\right) L_{0}/k_{B}T$ of DNA is about 300 nm$^{-1}$
\cite{cluzel96,smith96}, i.e., almost four orders of magnitude
larger than the corresponding value $3/l_{P}$ obtained from
Eq.~\ref{dna1}.

In the following I shall first discuss how the mechanical
properties of the linker backbone (modelled as a two-angle fiber)
can be derived analytically from its geometry. Then, in Section
3.4.2 the influence of nucleosome-nucleosome interaction is
considered before I compare in Section 3.4.3 the theoretical
results with that of stretching experiments on chromatin fibers \cite%
{cui00,bennink01,brower02}.

\subsubsection{The elasticity of the linker backbone}

That the chromatin fiber is highly flexible due to the large
amount of twistable and bendable linker DNA has been pointed out
by myself, Gelbart and Bruinsma \cite{schiessel01b}. For a few
special cases we were also able to calculate the stretching
modulus of the two-linker model. A complete analysis of the
elastic properties of the two-angle model has been given by
Ben-Ha\"{\i}m, Lesne and Victor \cite{ben-haim01}. In that paper
the authors managed to relate the macroscopic mechanical
properties of the fiber to the
geometrical properties of the master solenoid (i.e. to quantities like $R$, $%
s_{0}$ and $\psi $). Their underlying microscopic geometrical
model was more complicated since it was assumed that the linker
DNA leaves the octamer as a straight line so that entering and
exiting strand are displaced with respect to each other. A similar
arrangement has also been assumed in the original study by
Woodcock et al.~\cite{woodcock93}. From cryo-EM pictures it is
known, however, that in the presence of linker histones the
entering and exiting strand are glued together in a stem
\cite{bednar98} and this is also the situation encountered in the
mechanical stretching experiments by Cui and Bustamante
\cite{cui00}. Therefore it might be more appropriate to model the
influence of the nucleosome on the linker DNA just as inducing a
kink on the DNA -- as modelled in the above discussed variant of
the two-angle model. Since for this case we have the exact
relations between the geometrical parameters of the master
solenoid ($R$, $s_{0}$ and $\psi $) and the underlying two-angle
geometry, Eqs.~\ref{r} to \ref{s0}, the problem of calculating the
mechanical properties of the two-angle fiber is now completely
analytically solved.

Let me sketch in the following the elegant line of arguments used
by Ben-Ha\"{\i}m, Lesne and Victor \cite{ben-haim01} to determine
the mechanical parameters. The basic
idea is that the two-angle fiber can be described as an extensible WLC \cite%
{marko98,marko97c,kamien97} as already suggested in Refs.~\cite%
{cui00,katritch00}. In the linear response regime the relation is
\[
\left(
\begin{array}{l}
F \\
M_{t} \\
M_{b}%
\end{array}
\right) =\left(
\begin{array}{lll}
k_{B}T\widetilde{\gamma } & k_{B}T\widetilde{g} & 0 \\
k_{B}T\widetilde{g} & \widetilde{C} & 0 \\
0 & 0 & \widetilde{A}%
\end{array}
\right) \left(
\begin{array}{l}
u \\
\Omega \\
R^{-1}%
\end{array}
\right)
\]
Here $F$, $M_{t}$ and $M_{b}$ denote the external force and torque
components: $F$ is the force along the fiber axis, $M_{t}$ the
torsional torque (the torque component of ${\bf M}$ parallel to
the fiber axis) and $M_{b}$ denotes the flexural torque which is
the torque component perpendicular to the fiber axis. These
stresses are linearly related to the strains: $u$ is the relative
extension, $\Omega $ the twist rate and $R^{-1}$ is the curvature
of the fiber. The components of the stress-strain tensor give the
mechanical properties of the fiber: the stretching modulus $\widetilde{%
\gamma }$, the bending stiffness $\widetilde{A}$, the torsional stiffness $%
\widetilde{C}$ and the twist-stretch coupling constant
$\widetilde{g}$. These quantities follow from the underlying
properties of the linker DNA that is modelled as a {\it
non}-extensible WLC (like in Eq.~\ref{WLC}). In
Ref.~\cite{ben-haim01} the authors wrote down the energy density
as a function of the stresses (and not of the strains as usual).
Then they compared the resulting energy per linker with the energy
that follows from a microscopic calculation of the fiber elastic
energy (again as function of the applied stresses). The
microscopic calculation was based on the equilibrium condition for
Kirchhoff rods (WLCs) \cite{nizette99}, applied to the linker DNA:
The
force ${\bf f}\left( s\right) $ on the linker at any given point ${\bf r}%
\left( s\right) $ ($s$: arclength) equals the external tension
\begin{equation}
{\bf f}\left( s\right) ={\bf F}  \label{fsf}
\end{equation}
and the local torque obeys
\begin{equation}
{\bf m}\left( s\right) ={\bf M}-{\bf v}\wedge {\bf F}  \label{msmv}
\end{equation}
where ${\bf v}$ is the vector pointing from the fiber axis to the point $%
{\bf r}\left( s\right) $. With these assumptions it was possible
to obtain analytical expressions for the mechanical fiber
properties \cite{ben-haim01}. Specifically:
\begin{equation}
\widetilde{\gamma }=\frac{s_{0}}{k_{B}TB}\frac{C+\Delta S\cos ^{2}z}{%
R^{2}\cos ^{2}\left( \eta /2\right) }f\left( \eta ,z\right)  \label{gammaw}
\end{equation}
\begin{equation}
\widetilde{A}=\frac{As_{0}}{B}\frac{2C}{A+C-\Delta S\cos ^{2}z}  \label{lpw}
\end{equation}
\begin{equation}
\widetilde{C}=\frac{s_{0}}{B}\left( \frac{C}{3}\tan ^{2}\left( \eta
/2\right) +A-\Delta S\cos ^{2}z\right) f\left( \eta ,z\right)  \label{cw}
\end{equation}
and
\begin{equation}
\widetilde{g}=-\frac{s_{0}}{k_{B}TB}\frac{\Delta S\cos z\sin z}{R\cos \left(
\eta /2\right) }f\left( \eta ,z\right)  \label{gw}
\end{equation}
where\footnote{In the original work \cite{ben-haim01} this
function is called $K\left( \eta ,3\right) $. Note that there is a
printing error in the denominator (last line of Eq.~(16) in that
paper). The factor $1/3A$ has to be removed.}
\begin{equation}
f\left( \eta ,z\right) =\frac{3A}{3A+\tan ^{2}\left( \eta /2\right) \left(
C+\Delta S\cos ^{2}z\right) }  \label{fez}
\end{equation}
Note that all the parameters occurring in Eqs.~\ref{gammaw} to
\ref{fez} can be deduced analytically from the two-angle geometry.
Specifically $R$ is the fiber radius, Eq.~\ref{r}, and $z$ denotes
the angle between the fiber axis
and the linker, $z=\arccos \left( s_{0}/B\right) $ with $s_{0}$ given by Eq.~%
\ref{s0}. Furthermore $\eta =\cot \left( \psi \right) s_{0}/R$ is the angle
between neighboring nucleosomes as seen when viewed down the fiber axis,
i.e. $\eta /s_{0}$ is the twist rate of the unperturbed fiber. From Eqs.~\ref%
{r} to \ref{s0} follows:
\begin{equation}
\eta =\arccos \left( 2\cos ^{2}\left( \theta /2\right) \cos ^{2}\left( \phi
/2\right) -1\right)  \label{etaarc}
\end{equation}
The other parameters describe the mechanical properties of the
DNA: the bending stiffness $A$ and the torsional stiffness $C$ as
well as their difference $\Delta S=A-C$. Therefore we know now the
macroscopic mechanical properties of the two-angle fiber as
explicit functions of
the microscopic parameters.

These functions, $%
\widetilde{\gamma }=\widetilde{\gamma }\left( \theta ,\phi \right) $, $%
\widetilde{A}=\widetilde{A}\left( \theta ,\phi \right) $, $\widetilde{C}=%
\widetilde{C}\left( \theta ,\phi \right) $ and $\widetilde{g}=\widetilde{g}%
\left( \theta ,\phi \right) $, are, however, rather unwieldy. To
get an idea of the overall behavior one might resort to numerical
calculations as done in Ref \cite{ben-haim01} where the mechanical
moduli of the fiber were calculated as a function of $\phi $ for
two values of $\theta $, cf. Fig.~11 in that paper. It was found
that the moduli vary strongly with $\phi $ (and thus with the
linker length) and it was argued that this strong dependence might
be used in the biological context as a regulatory factor.

\begin{figure}
\begin{center}
\includegraphics*[width=7cm]{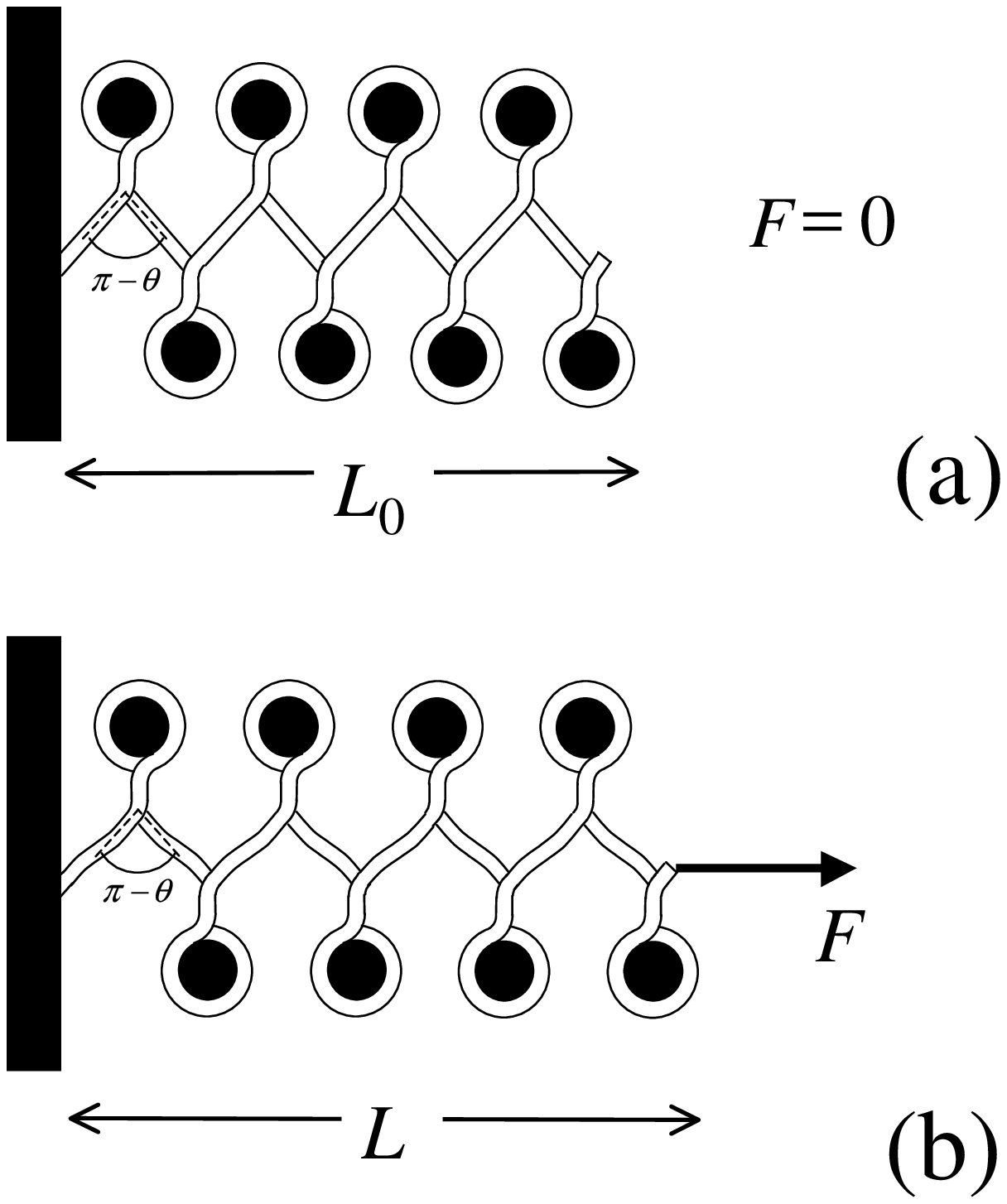}
\end{center}
\caption{Stretching of a zig-zag chain. (a) The unperturbed chain,
$F=0$, with a total length $L_0$ has straight linker DNA. (b) The
same fiber under tension $F>0$ stretches to an end-to-end distance
$L>L_0$ via the bending of its linkers.}
\end{figure}

Having the analytical relations at hand, another approach has now
become available, namely to look at limiting cases (solenoids,
fibers with crossed linkers and zig-zag structures) which show
simple dependences on the underlying geometry (i.e., on the angles
$\theta $ and $\phi $) --
as discussed in Section 3.2. Not surprisingly $\widetilde{\gamma }$, $%
\widetilde{A}$, $\widetilde{C}$ and $\widetilde{g}$ are also
simple functions of these underlying angles in all the limiting
cases. We will give a complete overview in a forthcoming
publication \cite{mergell03}. Here I will restrict myself to two
limits only.

Let me start with the {\it planar zig-zag fiber}. Such a chain can
be stretched via the bending of its linkers -- maintaining the
deflection angle $\theta $ at each kink, cf. Fig.~21. It is also
clear that the linker will not be twisted in this planar geometry.
In Ref.~\cite{schiessel01b} we calculated the stretching modulus
$\widetilde{\gamma }$ for this special arrangement. In order to do
so we wrote down the elastic energy of the linker (similar to
Eq.~\ref{WLC}) and determined the deformed shape (for small
perturbations) from the corresponding Euler-Lagrange equation
taking the boundary conditions into account. We found
\begin{equation}
\widetilde{\gamma }=\frac{12A\cos \left( \theta /2\right) }{k_{B}TB^{2}\sin
^{2}\left( \theta /2\right) }  \label{gammaw2}
\end{equation}
This result can now also be obtained directly from the general formula, Eq.~%
\ref{gammaw}, by simply setting $\phi =\pi $. It is evident from Eq.~\ref%
{gammaw2} that the stretching occurs via linker bending since
$\widetilde{\gamma }$ depends on $A$ only whereas the DNA
torsional stiffness $C$ does not enter the expression. I also note
that the planar zig-zag fiber shows other interesting features,
especially two different persistence lengths for
bending in plane and bending out of plane as discussed in Refs.~\cite%
{schiessel01b,ben-haim01,mergell02}. General features of such
polymers with highly anisotropic bending rigidities have been
considered by Nyrkova et al.~\cite{nyrkova96}.

Now I consider the case that might be of importance for 30-nm
fibers: the chains with crossed linkers ($\phi \ll 1$, $\pi -\theta \ll \pi $%
). Starting from the general expressions it is straightforward to
show that the linker geometry leads in this case to the following
overall mechanical properties:
\begin{equation}
\widetilde{\gamma }\simeq \frac{3A}{k_{B}TB^{2}}\phi \left( \pi -\theta
\right)  \label{gammaw3}
\end{equation}
\begin{equation}
\widetilde{A}\simeq \frac{AC}{A+C}\frac{\phi \left( \pi -\theta \right) }{2}
\label{aw2}
\end{equation}
\begin{equation}
\widetilde{C}\simeq \frac{A\phi \left( \pi -\theta \right) }{4}  \label{cw2}
\end{equation}
and
\begin{equation}
\widetilde{g}\simeq -\frac{3A\Delta S}{16k_{B}TCB}\phi ^{2}\left( \pi
-\theta \right) ^{3}  \label{gw2}
\end{equation}
It can be seen from Eq.~\ref{gammaw3} that stretching occurs via
linker bending (as in the case of zig-zag fibers) and from
Eq.~\ref{cw2} that also the {\it twisting} of the overall fiber is
achieved via the {\it bending} of the linkers. The dependence of
$\widetilde{A}$ on the DNA parameters, Eq.~\ref{aw2}, shows that
fiber bending involves both bending and twisting of the linkers, a
fact that is due to the different orientations of individual
linkers with respect to the bending direction. Finally, the
twist-stretch coupling is very small (cf. the angle dependence in
Eq.~\ref{gw2}).

The elasticity of the linker backbone is predicted to be very
soft: For instance, the stretching modulus $\widetilde{\gamma }$
scales for fibers with crossed linkers and zig-zag chains as
$A/\left( k_{B}TB^{2}\right) $. This is of the order one (per nm)
for an effective linker length of 20 bp as compared to a $\gamma $
of $\sim 300$ nm$^{-1}$ for free DNA (see above). Of course,
depending on the values of $\theta $ and $\phi $ this value varies
in a wide range. Also the other mechanical parameters of the
two-angle fiber indicate an extremely soft structure. Because of
this it is evident that the presence of the nucleosomes play a
crucial role in determining the mechanical properties of the 30-nm
fiber. The excluded volume will not allow a strong bending that
would lead to overlapping nucleosomes and the
nucleosome-nucleosome attraction counteracts the stretching of the
fiber under external tension. Therefore before I compare in
Section 3.4.3 the theoretical expressions and the results from
fiber stretching experiments it is indispensable to discuss first
how the nucleosomes modify the mechanical properties of the
two-angle fiber.

\subsubsection{Role of the nucleosome interaction}

The effect of attractive interaction between nucleosomes is to cause a {\it %
compression} of the 30-nm fiber. Phase behavior studies of linker-free
nucleosome solutions, i.e.,{\it \ }solutions of disconnected nucleosomes \cite%
{livolant00} (cf. also Ref.~\cite{fraden00}) indicate that
nucleosome core particles spontaneously form fiber-like {\it
columnar structures}, presumably due to attractive
nucleosome-nucleosome interaction. Attractive nucleosome
interaction could be mediated for instance by the lysine-rich core
histone tails \cite{luger97}, as mentioned above.

Let me first discuss the role of this internucleosomal attraction
on the stretching elasticity of a fiber. Following
Ref.~\cite{schiessel01b} the special case of a planar zig-zag
structure with elastic linkers is considered where a short-range
interaction between nucleosomes is assumed in addition. This
interaction, denoted by $U_{inter}$, is taken to be a short range
attraction, of strength $-U_{min}$, that acts only when the
nucleosomes are in close contact, i.e., at a distance $x\approx
2R_{0}$ of the order of the hardcore diameter. For a given
nucleosome, say the $i$th, the closest nucleosomes in space are
number $i+2$ and $i-2$ (cf. Fig.~21). The interaction between
other pairs is disregarded. The elastic interaction $U_{elastic}$
follows directly from Eq.~\ref{gammaw2} applied to a trinucleosome
($N=2$):
\begin{equation}
U_{elastic}\left( x\right) =\frac{3}{\sin ^{2}\left( \theta /2\right) }\frac{A}{%
B^{3}}\left( x-x_{0}\right) ^{2}=\frac{\widetilde{K}}{2}\left(
x-x_{0}\right) ^{2} \label{ubend}
\end{equation}
where $x_{0}=2B\cos \left( \theta /2\right) $ denotes the distance
between nucleosome $i$ and $i+2$ for straight linkers (cf.
Eq.~\ref{r3}). The total internucleosomal $U\left( x\right) $
equals $U_{inter}\left( x\right) +U_{elastic}\left( x\right) $.

\begin{figure}
\begin{center}
\includegraphics*[width=6cm]{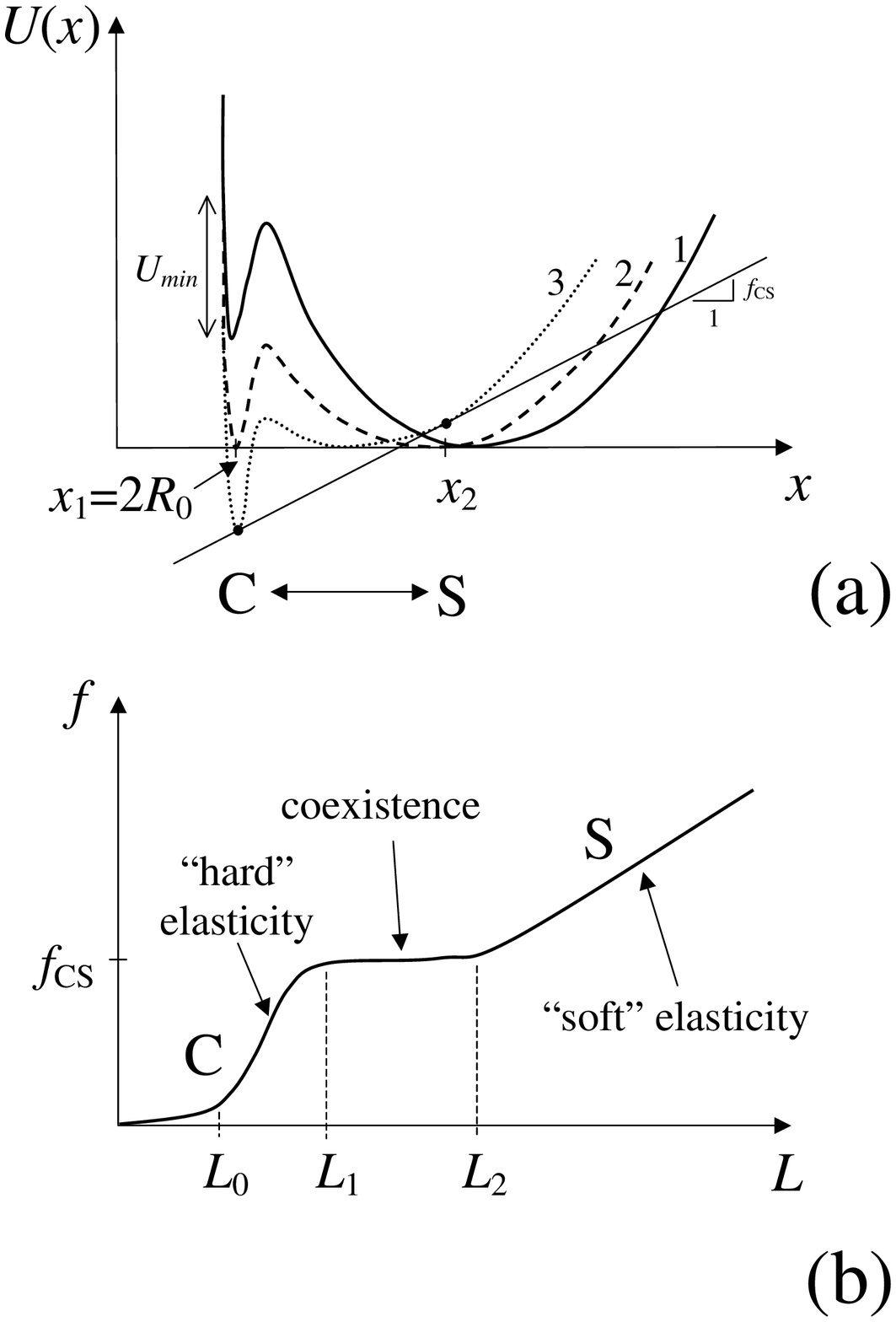}
\end{center}
\caption{(a) Internucleosomal interaction potential $U$ between
nucleosome $i$ and $i+2$ as a function of distance $x$. In
addition to the elastic contribution there is a short range
attraction for nucleosome at close contact, $x=2R_0$. The
different curves correspond to different values of the angle
$\theta $. Curve ''1'' has the global minimum at large $x$
(swollen state ''S'') whereas curve ''3'' has the minimum for
nucleosomes in close contact (condensed state ''C''). Curve ''2''
corresponds to the transition point. Also depicted is the common
tangent for curve ''3''. Its slope corresponds to the critical
stretching force $f_{CS}$ at which nucleosomes are transferred
from the state C to state S. (b) Force-extension curve of a
condensed fiber. For extensions $L$ with $L_{1}<L<L_{2}$ one finds
a coexistence plateau with the restoring force $f_{CS}$.}
\end{figure}

Fig.~22(a) shows $U\left( x\right) $ for different values of
$\theta $. Let me assume for simplicity that the interaction
energy $U_{inter}$ remains unchanged. Curve ''1'' in Fig.~22(a)
shows $U\left( x\right) $ for a small value of $\theta $ where the
global minimum of $U\left( x\right) $ is located at $x=x_{0}$
denoted by ''S'' (swollen state). Curve ''2'' corresponds to an
intermediate value of $\theta $ at which the minima at ''S'' and
''C'' have the same value. For this value of $\theta $, $\theta
=\theta _{c}$, the energy minimum shifts from ''S'' to a new
minimum, representing the
condensed state ''C''. The change in $%
\theta $ produced a {\it structural transition} from a swollen
state to a condensed state. Finally, curve ''3'' depicts $U\left(
x\right) $ for a deflection angle $\theta >\theta _{c}$ with the
minimum at ''C''. The critical angle for the ''S'' to ''C''
transition can be determined by comparing the bending
energy at close contact, $U_{elastic}\left( 2R_{0}\right) $, and the strength $%
U_{min}$ of the short range attraction. Equating both leads to the following
condition for $\theta _{c}$:
\begin{equation}
\cos \left( \theta _{c}/2\right) -\sqrt{\frac{BU_{min}}{12A}}\sin \left(
\theta _{c}/2\right) =\frac{R_{0}}{B}  \label{thetac}
\end{equation}
In the swollen state the elastic properties should be in principle the one
that were discussed in the previous section. In the condensed state, the
elastic properties are determined by the detailed form of the nucleosome
interaction potential.

If the condensed state has a lower free energy, i.e. if $\theta >\theta _{c}$%
, then an external stretching force $f$ can induce a transition
from the condensed to the swollen state. The transition point
$f_{CS}$ follows from a ''common-tangent'' construction. The
conditions are $U^{\prime }\left( x_{1}\right) =U^{\prime }\left(
x_{2}\right) =f_{CS}$ and $\left( U\left( x_{2}\right) -U\left(
x_{1}\right) \right) /\left( x_{2}-x_{1}\right) =f_{CS}$ (cf.
Fig.~22(a)) leading to \cite{schiessel01b}
\begin{equation}
f_{CS}=\sqrt{2KU_{min}}-\widetilde{K}\left( x_{0}-2R_{0}\right)
\label{fcs2}
\end{equation}
The corresponding force-extension curve has a ''coexistence
plateau'', cf. Fig.~22(b). If the imposed end-to-end distance is
smaller than $L_{0}$ (the contour length of the condensed fiber)
then the restoring force is entropic. For $L_{0}<L<L_{1}$ the
force rises sharply with increasing $L$. This ''hard elasticity''
is governed by the nucleosomal interaction potential $U_{inter}$.
Then at $L=L_{1}$ the coexistence plateau is reached. Between
$L=L_{1}$ and $L=L_{2}$ parts of the fiber are in the ''S'' state
and parts are in the
''C'' state. For larger extensions, $%
L>L_{2}$, the fiber shows soft elasticity due to the bending and twisting of
the linkers as discussed in the previous section.

Katritch, Bustamante and Olson \cite{katritch00} presented a
Monte-Carlo simulation of the chromatin fiber that was based on a
model very similar to the two-angle model. The nucleosomes were
modelled as spheres and attached to the kinks in the linker
backbone via a short stem. The only difference from the above
discussed two-angle model was that the rotational angle $\phi $
between each pair of nucleosomes was chosen randomly from the
interval $-\pi $ to $\pi $. These fibers were then stretched as in
a micromanipulation experiment \cite{cui00} and their
force-extension relationships were measured. The values for
$\widetilde{\gamma }$ were in good agreement with what is expected
on theoretical
grounds (a detailed discussion is given in Appendix D of Ref.~\cite%
{schiessel01b}). What is of special interest here is that they
also studied the effect of a short-ranged nucleosome-nucleosome
attraction. Using a value $U_{min}$ of order $2k_{B}T$ (or larger)
they observed very clearly the occurrence of a pseudoplateau in
the force-extension curve similar to Fig.~22(b).

The nucleosomes have also a large effect on the persistence length $%
\widetilde{l}_P$ of the fiber. This has been demonstrated most
clearly in a computer simulation by Wedemann and Langowski
\cite{wedemann02} (cf. also an earlier preliminary study of this
group \cite{ehrlich97}). Their model is again very closely related
to the two-angle model discussed above. Differences are that the
entering and exiting DNA at the nucleosome are slightly displaced
in the direction of the nucleosome axis and that the screened
electrostatic interaction between linkers was taken explicitly
into account. Nevertheless, Eq.~\ref{lpw} should be expected to
give a good estimation of the contribution of the linker DNA to
the fiber persistence
length. Using the values of that simulation ($\theta \simeq 143^{\circ }$%
, $\phi \simeq 80^{\circ }$, $B=10$ bp) gives
$\widetilde{l}_{P}\simeq 13$ nm. However, the persistence length
observed in the simulation is 265 nm, i.e. 20 times larger! This
is clearly an effect of the nucleosomes. The role of the linkers
is to bring the nucleosomes into contact. The nucleosomes
(modelled here as ellipsoids) experience then in addition an
attractive force where $U_{min}$ has been chosen to be of order
$k_{B}T$. This leads to a very dense structure with the
nucleosomes in contact so that there is hardly any space for fiber
bending. Most clearly this is seen in Fig.~9 of that paper that
shows a contraction of a fiber that has been stretched out first.
As long as the nucleosomes are not in contact the fiber shows
sharp bends and strong shape fluctuations. The fibers stiffens
very strongly as soon as the dense state is reached.

This all shows that the nucleosome interaction is a crucial
element determining the mechanical fiber properties. Therefore a
more microscopic model taking details of the nucleosome structure
into account might be important for a theoretical prediction of
the properties of chromatin fibers. A first step in this direction
has been done by Beard and Schlick \cite{beard97}. They performed
a molecular dynamics simulation where the nucleosomes were
represented by disks made of several hundred charges that were
chosen to match the crystal structure \cite{luger97}. Di- and
trinucleosomes as well as whole fibers have been studied. The
authors demonstrated also that a fiber with a crossed linker
geometry unfolds into an open zig-zag fiber as a result of
changing ionic conditions. However, in their nucleosome model they
neglected most of the histone tails that constitute very likely
the crucial ingredient for the nucleosome-nucleosome attraction.

\subsubsection{Stretching chromatin}

It is now possible to measure the mechanical behavior of single
chromatin fibers via micromanipulation techniques as it has been demonstrated in three studies \cite%
{cui00,bennink01,brower02}. Each of the studies focused on a
different variant of the 30-nm fiber. Cui and Bustamante
\cite{cui00} stretched native chicken erythrocyte chromatin fibers
containing linker histones and contrasted the cases of low and
high ionic strength. For low ionic strength (5 mM NaCl) it was
found that fibers are very soft. By fitting their data to that of
an
extensible WLC they found a stretching modulus of $k_{B}T%
\widetilde{\gamma }\approx 5$ pN and a persistence length of
$\widetilde{l}_{P}\approx 30$ nm. The theoretical values are
$k_{B}T\widetilde{\gamma }\simeq
6.3$ pN (from Eq.~\ref{gammaw}) and $\widetilde{l}_{P}\simeq 16$ nm (from Eq.~%
\ref{lpw}) for $\phi =36^{\circ }$ and $\theta =95^{\circ }$ (cf.
Section 3.3) and for a linker length of 20 bp (chosen on the basis
of the 210 bp repeat length of chicken erythrocyte chromatin
\cite{vanholde89} minus roughly 190 bp associated with the core
and linker histones). The theoretical and experimental values are
close which indicates that the mechanical properties of a swollen
fiber at low salt concentrations are mainly determined by the
elasticity of its linker backbone. The nucleosomes are less
important since they are not close enough in such a swollen fiber.
For high stretching forces around 20 pN there is an irreversible
change in the overall length of the fiber due to histone
''evaporation'' which has been seen more clearly in the other two
stretching experiments (see below).

Stress-strain curves for fibers at higher ionic strength (40 mM
NaCl) are also reported in Ref.~\cite{cui00}. In this case the
fiber is much denser and nucleosomes approach each other closely.
Attractive short-range forces and the increase of $\theta $
associated with higher ionic strength should favor the condensed
phase. Indeed a plateau appears at 5 pN in the force-extension
curve (cf. Fig.~4 of Ref.~\cite{cui00}). From the extend of the
plateau,  0.6 $\mu m$, its height, 5 pN, and the number of
nucleosomes in the stretched fiber, $\approx 280$, it was
estimated that there is an attractive interaction energy of $\sim
3k_{B}T$ per nucleosome \cite{cui00}. In Ref.~\cite{schiessel01b}
we used Eq.~\ref{fcs2} to independently estimate the strength of
the internucleosomal attraction from the value of the
critical force alone. Neglecting the second term in that equation one finds $%
U_{min}\approx f_{CS}^{2}/\left( 2\widetilde{K}\right) \approx
3k_{B}T$ assuming $\theta =140^{\circ }$ and again $B=20$ bp. Note
that one finds with these values that the stretching modulus of
the linker backbone, Eq.~\ref{gammaw}, is of order
$k_{B}T\widetilde{\gamma }\simeq 4.7$ pN, i.e., even lower than
the value 6.3 pN predicted above for low ionic strength. The fact
that these fibers appear much stiffer with respect to (small)
deformations indicates that its mechanical properties are mainly
determined by the nucleosome-nucleosome attraction and {\it not}
by the backbone elasticity. This is also in accordance with the
estimated large value of the persistence length of condensed
chromatin fibers ($\sim 260$ nm, cf.~\cite{munkel98}).

Bennink et al.~\cite{bennink01} assembled chromatin fibers by exposing $%
\lambda $-DNA to {\it Xenopus} egg extract before stretching
them.\footnote{A detailed discussion of the kinetics of the
chromatin assembly in this kind
of experiment is given in Ref.~\cite{ladoux00}; cf. also Ref.~\cite%
{bennink01b}.} This extract contains core histones but no linker
histones. There is, however, an abundance of non-histone proteins
some of which act close to the DNA entry-exit point similar to
linker histones. For small forces a stretch modulus $\sim 150$ pN
was extracted from the data. This indicates again that chromatin
fibers are quite stiff at high salt concentrations (here 150 mM
NaCl) compared to the pure linker backbone elasticity but still
quite soft compared to naked DNA ($k_{B}T\gamma \simeq 1200$ pN).
For stretching forces of order 20 pN irregular sawtooth-like
fluctuations were observed, each being a result of a sudden fiber
lengthening by multiples of $\sim 65$ nm. This was attributed to
the unraveling of single or multiple nucleosomes.

Finally, Brower-Toland et al.~\cite{brower02} used well
characterized fibers that were prepared from tandem repeats of the
5S rDNA positioning sequence and core histones (no linker
histones). As shown in Ref.~\cite{pennings91} (cf. my discussion
of that paper in Section 2.4.1) most of the nucleosomes are
localized at the preferred positions. Around 20 pN the
force-extension curve showed a regular sawtooth pattern
reminiscent of the one observed
during the unravelling of tandem repeat domains in the protein titin \cite%
{rief97,oberhauser01}. The spacing of the peaks, $\sim 27$ nm, is
indicative of the unravelling of only one turn of the nucleosomal
DNA. The outer sections of the DNA were detected to be released at
much smaller forces. Brower-Toland et al.~\cite{brower02} gave
also a theoretical explanation of this observation: They speculate
that the first 76 bp are unwrapped much more easily due to weaker
binding between DNA and the octamer whereas strong binding sites
occur as soon as these first 76 bp are unwound. Only when a
sufficiently large force is applied these binding sites are broken
on the time scale of the experiment.

However, this explanation is questionable since it does not take into
account the actual unwinding geometry. Already Cui and Bustamante \cite%
{cui00} pointed out that it requires a twisting of the core
particle to unwrap the inner part of the nucleosomal DNA. The free
DNA has to be bent strongly close the nucleosome which leads to a
considerable barrier that has to be crossed during unwrapping. We
are currently calculating this barrier analytically in the WLC\
framework \cite{kulic02c}. This high barrier might explain why the
nucleosomes are dissociated only at surprisingly high tensions --
even in the absence of linker histones. In fact, as Marko and
Siggia pointed out in Ref.~\cite{marko97}, one would find
nucleosome release (and a corresponding plateau in the
force-distance curve) around 2 pN -- if such a barrier could be
neglected. This value follows from the comparison
of adsorption energy ($\sim 30k_{B}T$, cf.~Section 2.1) to wrapping length ($%
\sim 50$ nm): $f\approx 30k_{B}T/50$ nm$\approx 2$ pN.

\subsection{Fiber swelling}

In this section I discuss how the entry-exit angle $\alpha =\pi
-\theta $ of the DNA at the nucleosomes is controlled via
electrostatics. A more detailed account on this subject is
provided in Ref.~\cite{schiessel02}. As mentioned above it can be
seen in the cryo EM studies~\cite{bednar98} that the fibers open
up and therefore become more accessible when the ionic strength is
reduced and that this opening is directly linked to an increase in $\alpha $%
. It was suggested that via other mechanisms (for instance, the
acetylation of the lysine-rich histone tails~\cite{vanholde96}, as
explained in more detail below) the angle $\alpha $ and therefore
the degree of swelling can be changed for a given section of the
fiber and that this constitutes a biochemical means to control the
transcriptional activity of genes.

Whereas the X-ray studies of the core particle~\cite{luger97}
allow a detailed knowledge of the wrapped part of the DNA it does
not give insight into the conformational properties of the
entering and exiting strands. One has therefore to refer to the
electron cryomicrographs. In these micrographs it can be seen
clearly that 10 nm stretches of the entering and exiting DNA
strands are glued together forming a unique ''stem motif''
\cite{bednar98} (cf. also Fig.~23(a)). The gluing of the two
equally charged chains is accomplished -- amongst other things --
via the linker histone H1 as shown schematically in Fig.~23(b).

\begin{figure}
\begin{center}
\includegraphics*[width=9cm]{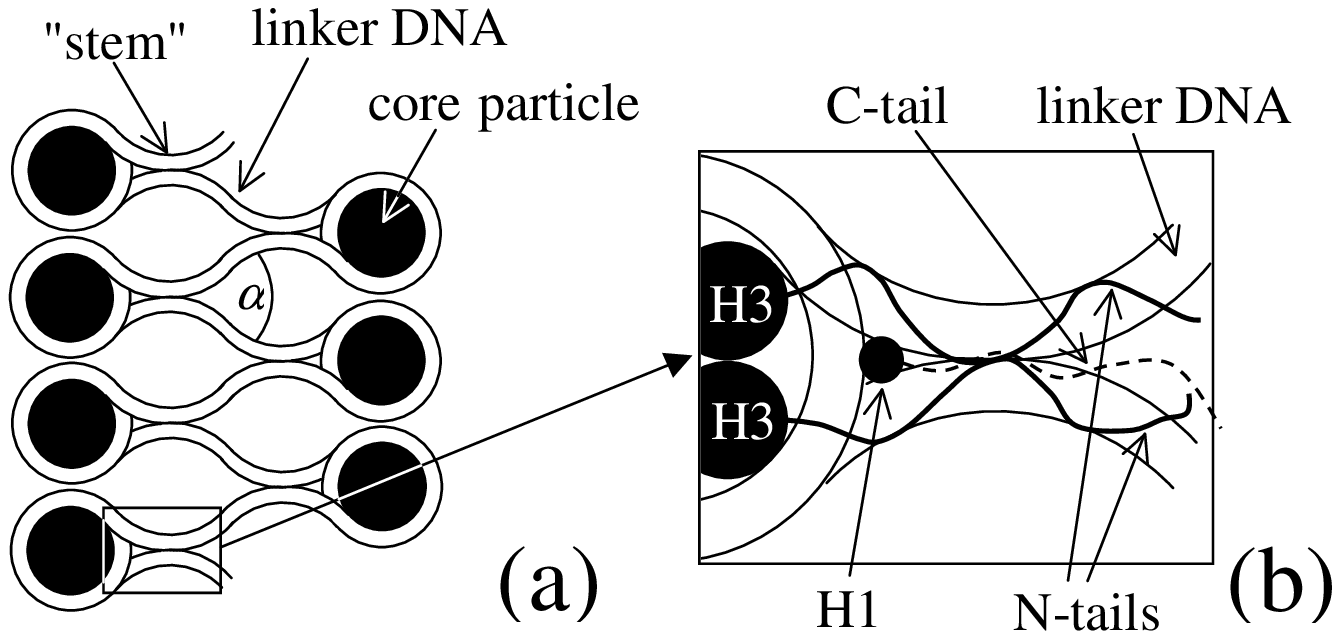}
\end{center}
\caption{(a) Schematic view of a section of the 30-nm fiber (for
simplicity shown here as a two-dimensional zig-zag). (b) Enlarged
view of the stem region showing a speculative model of the role of
the H1 histone and some N-tails from the core histones, cf. also
Fig.~7 of Ref.~\cite{zlatanova98}.}
\end{figure}

At physiological concentrations the electrostatics is essentially
short-ranged ($\kappa ^{-1}\simeq 10$ \AA\ for 100 mM salt). It
seems therefore reasonable to assume that $\alpha $ is set within
the small region where the two linker DNA are in close contact,
i.e., within the stem region. This value of $\alpha $ in turn
controls the large-scale secondary structure of chromatin, the
30-nm fiber, as discussed in Section 3.2. To mimic this situation
I assumed in Ref.~\cite{schiessel02} a geometrical arrangement
where two parallel DNA strands are hold together tightly at $y=0$
for $x\leq 0$ and are free for $x>0$, cf. Fig.~24. Because of
their mutual electrostatic repulsion the two strands bend away
from each other. When the two strands are far enough from each
other their interaction is screened so that they asymptotically
approach straight lines defining the opening angle $\alpha $ as
indicated in Fig.~24.

\begin{figure}
\begin{center}
\includegraphics*[width=5cm]{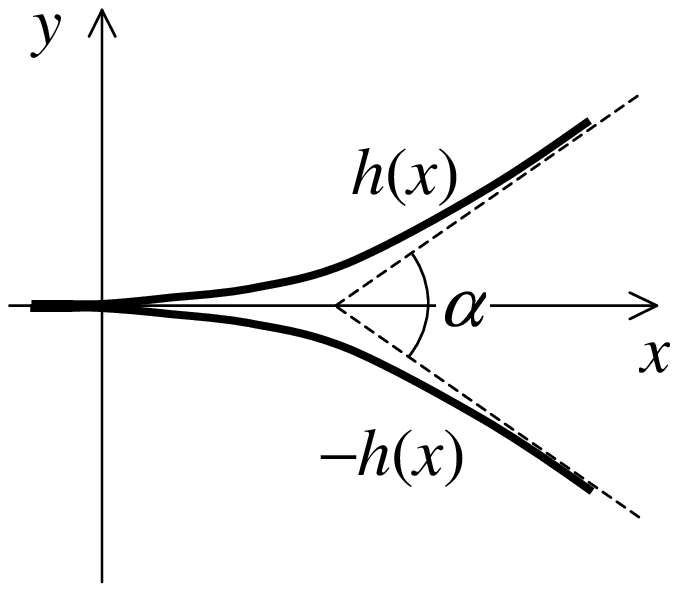}
\end{center}
\caption{Idealized model for the entry-exit region of the DNA at
the nucleosome. The thick curved lines represent the entering and
exiting DNA that enclose a well-defined angle $\alpha$ that in
turn determines the overall geometry of the 30-nm fiber.}
\end{figure}

The conformation of the upper DNA chain can be described by the height
function $h\left( x\right) $. By symmetry the position of the lower strand
is then given by $-h\left( x\right) $. To mimic the stem two boundary
conditions are imposed at $x=0$, namely $h\left( 0\right) =h^{\prime }\left(
0\right) =0$. The entry-exit angle $\alpha $ is related to the slope of $%
h\left( x\right) $ at infinity via $\tan \left( \alpha /2\right)
=h^{\prime }\left( \infty \right) $. The two DNA chains are
modelled as semiflexible polymers with persistence length $l_{P}$
and line-charge density $-e/b$ that interact via a screened
electrostatic potential. The free energy of the system is then
given by
\begin{equation}
\frac{{\cal F}\left\{ h\left( x\right) \right\} }{k_{B}T}\simeq
\int_{0}^{\infty }dx\left[ l_{P}\left( \frac{d^{2}h}{dx^{2}}\right) ^{2}+%
\frac{2l_{B}}{b^{2}}K_{0}\left( 2\kappa h\left( x\right) \right) \right]
\label{freee}
\end{equation}
The first term in the integral accounts for the bending of the
{\it two} DNA strands and the second term describes the
interaction between the two chains ($K_{0}\left( x\right) $ being
the $0th$ order modified Bessel function). Here the interaction of
a given charge on one chain with all the charges on the other
chain is approximated by the interaction of this charge with a
{\it straight} chain at the distance $2h$.\footnote{This can be shown \cite{schiessel02} to be a good approximation as long as $%
l_{P}\gg l_{OSF}=l_{B}/4b^{2}\kappa ^{2}$ (the Odijk-Skolnick-Fixman length
that describes the electrostatic stiffening of the chain, cf. Ref.~\cite%
{barrat96}). In fact, it is also precisely that limit at which the {\it intra%
}molecular interaction can be neglected (as done here). One can
reformulate the above condition in the simple form $\alpha \ll 1$
(cf. Eq.~\ref{tan} below); for instance it is found that for
$\alpha =45^{\circ}$ the approximation is still excellent and for
$\alpha =90^{\circ}$ the chain-chain repulsion is overestimated by
$\sim $20\%.} The conformation of the upper chain, $h\left(
x\right) $, is then the solution of the corresponding
Euler-Lagrange equation
\begin{equation}
l_{P}\frac{d^{4}h}{dx^{4}}-\frac{2l_{B}\kappa }{b^{2}}K_{1}\left( 2\kappa
h\right) =0  \label{euler1}
\end{equation}
together with four boundary conditions: Two are given at the origin (see
above) and two follow from the condition of straight ''linkers'' at
infinity: $h^{\prime \prime }\left( \infty \right) =h^{\prime \prime \prime
}\left( \infty \right) =0$. Defining $\tilde{h}=2\kappa h$ and introducing
the dimensionless quantity $\tilde{x}=\left( 4l_{B}\kappa
^{2}/b^{2}l_{P}\right) ^{1/4}x$, Eq.~\ref{euler1} can be rewritten as $d^{4}%
\tilde{h}/d\tilde{x}^{4}=K_{1}\left( \tilde{h}\right) $ with a
solution
showing the dimensionless asymptotic slope $c_{0}=\left. d\tilde{h}/d\tilde{x}%
\right| _{\tilde{x}=\infty }$. It follows immediately that $\tan \left(
\alpha /2\right) $ is given by
\begin{equation}
\tan \left( \alpha /2\right) =h^{\prime }\left( \infty \right) =\frac{c_{0}}{%
\sqrt{2}}\left( \frac{l_{B}}{l_{P}}\right) ^{1/4}\left( \frac{1}{\kappa b}%
\right) ^{1/2}  \label{tan}
\end{equation}
where $c_{0}$ is of order one (cf. Ref.~\cite{schiessel02} for
details).

This result might help to understand better how the local
electrostatics controls the geometry of the chromatin fiber {\it
in vitro} and, on a more tentative level, {\it in vivo}. The {\it
in vitro}-experiments show that chromatin fibers ''open up'' with
decreasing salt concentrations. As already mentioned in Section
3.3 it was estimated from
electron cryomicrographs that $\alpha _{exp}\approx 85^{\circ }$ for $%
c_{s}=5$ mM and $\alpha _{exp}\approx 45^{\circ }$ for $c_{s}=15$
mM and from electron cryotomography that $\alpha _{exp}\approx
35^{\circ }$ for $c_{s}=80$ mM~\cite{bednar98} (cf.
Refs.~\cite{vanholde96} and \cite{toth01} for other approaches to
determine $\alpha$). One expects from Eq.~\ref{tan} that $\alpha
\simeq 2\arctan \left( Cc_{s}^{-1/4}\right) $ with $C$ being a
constant. Let me take the angle at the highest salt concentration,
$c_{s}=80$ mM, as the reference value. From this follows $C=0.94$.
With this value of $C$ the prediction is $\alpha \approx 51^{\circ
}$ for $c_{s}=15$ mM and $\alpha \approx 64^{\circ }$ for
$c_{s}=5$ mM. Whereas the theoretical value $\alpha \approx
51^{\circ }$ at intermediate ionic strength is close to $\alpha
_{exp}\approx 45^{\circ }$, the value $\alpha \approx 64^{\circ }$
for low salt concentrations is noticeably too low ($\alpha
_{exp}\approx 85^{\circ }$).
However, as mentioned above, for such a large value of $%
\alpha $ the chain-chain repulsion is underestimated by $\sim
$20\%.

How can the degree of swelling of the chromatin fiber be
controlled {\it in vivo}? Under the assumption that the above
mentioned geometry is valid the only parameter that might be under
biochemical control is the linear charge density $b^{-1}$. It is
known that the formation of a dense chromatin fiber is dependent
on the presence of several components, especially of the cationic
linker histones and of some of the lysine-rich (i.e., cationic)
N-tails of the core histones that appear to be long, flexible
polyelectrolyte chains~\cite{luger97}. In Fig.~23(b) I give a
tentative picture of the conformation of two N-tails that protrude
from the histone core. It is known that if either of these
components is missing the fiber does not fold properly (cf.
Ref.~\cite{vanholde96} and references therein). As indicated in
the Figure the tails might form a complex with the entering and
exiting linker DNA in such a way that they effectively reduce its
linear charge density $b^{-1}$. It is known that transcriptionally
active regions in chromatin show an acetylation of the core
histone tails (i.e., the cationic groups of the lysines are
neutralized). In that tentative picture this acetylation mechanism
would increase $b^{-1}$ and according to Eq.~\ref{tan} this would
lead to an opening of the entry-exit angle $\alpha $. The
acetylation might therefore be the first step in the
decondensation of a stretch of the chromatin fiber that needs to
be accessed for transcription.

Let me note that processes that are involved in the acetylation
and deacetylation are quite specific and involved as, for instance, discussed in Ref.~\cite%
{grunstein97}. The histone tail modifications might serve specific
functions via the modification of their secondary structure that
in turn modifies their interaction with certain proteins
\cite{hansen98}. Recently there are even attempts to decipher a
specific \textquotedblright
language\textquotedblright\ of covalent histone modifications \cite{strahl00}%
. It might be that such specific processes act in concert with the
more basic charge neutralization principle discussed here.

\section{Conclusions and outlook}

Chromatin is of fundamental importance for a host of biological
processes ranging from gene expression to cell division.
Consequently there is a huge research activity among biologists in
this area. For physicists chromatin becomes now also of interest
since there are more and more experiments available that work
under quite well-defined conditions. Such experiments typically
involve only a few components (DNA, histone proteins...) but no
active protein "machines." These experiments either focus on
elucidating properties of single nucleosomes or of
beads-on-a-string complexes ("chromatin fibers"). They study the
behavior of these systems under changing ionic conditions and/or
under an externally applied tension. Also the dynamics of these
systems can be investigated that is solely driven by thermal
fluctuations. Theoretical treatments and computer simulations that
capture the essential features of the chromatin system are now
possible and thus allow to estimate the energy- and time-scales
occurring in chromatin. Other approaches look at simplified model
systems and try to identify general physical principles that
govern complexes of charged chains and macroions. With the better
understanding of the mechanical and dynamical properties of
nucleosomes and chromatin fibers one hopefully gains also a deeper
insight into more complicated questions like the working of
chromatin remodeling complexes, the interaction between RNA
polymerase and nucleosomes etc.

To proceed in this direction it is crucial to obtain reliable
numbers from experiments. One energy scale that dominates many
processes is the adsorption energy of DNA on the octamer that has
now been measured quite directly through stretching experiments.
Another important feature -- especially in 30-nm fibers -- is the
nucleosome-nucleosome interaction energy. Again detailed
experimental studies have been performed and await a detailed
theoretical treatment.

Chromatin is a very active and exciting field in biology where
tremendous progress has been made in recent years. I hope that at
least a few of the ideas gained from the physical models will be
of help to biologists to develop a clearer picture of the working
of chromatin and to design appropriate experimental setups.

\section*{Acknowledgments} The author thanks R. Bruinsma, W. M.
Gelbart, I. Kuli\'{c} and J. Widom for many valuable discussions.
Useful conversations with R. Everaers, K. Kremer, J. Langowski, F.
Livolant, S. Mangenot, G. Maret and B. Mergell are also
acknowledged.

\appendix

\setcounter{equation}{0} \renewcommand{\thesection}{\Alph{section}} %
\renewcommand{\theequation}{\Alph{section}\arabic{equation}}

\section{Rosette in d dimensions}

The heterogeneity found for open loops (cf. Eq.~\ref{lsize}) is
reminiscent of a phase coexistence. To make clearer why the loop
sizes are so sensitive to small details, especially why there is
no "phase separation" for closed chains (cf. Eq.~\ref{ll}; even
though $d^{2}f/dl^{2}<0$ at larger separations) I will present
here some unpublished results in which the free energy is formally
recomputed for arbitrary space dimensions $d$. In order to do so
one has to replace the entropy term in Eq.~\ref{fli} by $\left(
d/2\right) \ln \left( l/l_{P}\right) $. Also in that case one
obtains an analytical expression for the partition function,
namely (for an open chain)
\begin{equation}
Z_{M}=\left[ \frac{2l_{P}^{\frac{d+2}{4}}}{r}\left( \frac{2\chi k_{B}T}{P}%
\right) ^{\frac{2-d}{4}}K_{\frac{d}{2}-1}\left( \sqrt{\frac{8\chi l_{P}P}{%
k_{B}T}}\right) \right] ^{M}\left( \frac{k_{B}T}{rP}\right) ^{2}  \label{znd}
\end{equation}
with $K_{\nu }\left( x\right) $ denoting the modified Bessel function of $%
\nu $th order. In the case of large ''pressure'' $P$, $l_{P}P/k_{B}T\gg 1$,
it follows from the asymptotic form of $K_{d/2-1}\left( x\right) $ for large
$x$, $K_{d/2-1}\left( x\right) \simeq \sqrt{\pi /2x}e^{-x}$, that the
leading term of the resulting free energy $G\left( P\right) $ is independent
of $d$. Therefore one recovers the 3D case, i.e., Eqs.~\ref{rp}-\ref{deltal}
for $L/M\ll Ml_{P}$. For the low ''pressure'' regime, $l_{P}P/k_{B}T\ll 1$,
I use the asymptotics $K_{d/2-1}\left( x\right) \simeq 2^{-1}\Gamma \left(
\left| d/2-1\right| \right) \left( 2/x\right) ^{\left| d/2-1\right| }$ for $%
x\ll 1$ and $d\neq 2$. This leads to the following asymptotic behavior of
the partition function
\begin{equation}
Z_{M}\simeq \left\{
\begin{array}{ll}
\left[ \Gamma \left( \frac{2-d}{2}\right) \frac{l_{P}^{d/2}}{r}\left( \frac{%
k_{B}T}{P}\right) ^{\frac{2-d}{2}}\right] ^{M}\left( \frac{k_{B}T}{rP}%
\right) ^{2} & \mbox{for}\;d<2 \\
\left[ \Gamma \left( \frac{2-d}{2}\right) \frac{l_{P}}{r}\left( 2\chi
\right) ^{\frac{2-d}{2}}\right] ^{M}\left( \frac{k_{B}T}{rP}\right) ^{2} & %
\mbox{for}\;d>2%
\end{array}
\right.  \label{znd2}
\end{equation}
It follows then from $L=\partial G/\partial P$ that $P$ is given in leading
order by
\begin{equation}
P\simeq \left\{
\begin{array}{ll}
\left( \frac{2-d}{2}M+2\right) \frac{k_{B}T}{L} & \mbox{for}\;d<2 \\
\frac{2k_{B}T}{L} & \mbox{for}\;d>2%
\end{array}
\right.  \label{pd1}
\end{equation}
The average leaf size can in principle be calculated, as before,
from $Z_{1}$. Here, however, it turns out to be more convenient to calculate $%
\left\langle l_{leaf}\right\rangle $ directly:
\begin{eqnarray}
\left\langle l_{leaf}\right\rangle &=&\frac{\int_{0}^{\infty
}dl\,l^{1-d/2}\exp \left( -\frac{2\chi l_{P}}{l}-\frac{Pl}{k_{B}T}\right) }{%
\int_{0}^{\infty }dl\,l^{-d/2}\exp \left( -\frac{2\chi l_{P}}{l}-\frac{Pl}{%
k_{B}T}\right) } \nonumber \\
&&=\sqrt{\frac{2\chi l_{P}k_{B}T}{P}}\frac{K_{2-d/2}\left(
\sqrt{8\chi l_{P}P/k_{B}T}\right) }{K_{1-d/2}\left( \sqrt{8\chi l_{P}P/k_{B}T%
}\right) }  \label{leafd}
\end{eqnarray}
Now using the above given power law behavior of the Bessel
function together with Eq.~\ref{pd1} the average leaf size
follows:
\begin{equation}
\left\langle l_{leaf}\right\rangle \simeq \left\{
\begin{array}{ll}
\frac{L}{M+\frac{4}{2-d}} & \mbox{for}\,\,\,d<2 \\
\frac{\Gamma \left( 2-d/2\right) }{\Gamma \left( d/2-1\right) }\frac{\chi
^{d/2-1}}{2^{3-d}}\left( \frac{l_{P}}{L}\right) ^{\frac{d-2}{2}}L & %
\mbox{for}\,\,\,2<d<4 \\
\frac{4\chi l_{P}}{d-4} & \mbox{for}\,\,\,d>4%
\end{array}
\right.  \label{leafd2}
\end{equation}
First note that for $d<2$ the leaf size is set by the overall length of the
chain but does not depend on $l_{P}$; on the other hand, for $d>4$ $%
\left\langle l_{leaf}\right\rangle $ is solely determined by
$l_{P}$. Speaking in the picture of interacting particles on a
track of length $L$ one can explain these two extreme cases as
follows. For $d<2$ the increase of the ''nearest neighbor pair
potential'' beyond a distance $l_{P}$ (given by $\left( d/2\right)
\ln \left( l/l_{P}\right) $) is too small to keep the particles
together; instead they explore all available space. For higher
space dimensions than 4 the prefactor of the log-term is large
enough to keep neighboring particles close to the ideal distance
$\sim l_{P}$ given by the shallow minimum of $f\left( l\right) $,
Eq.~\ref{fli}. The case $d=3$
which I already gave in Eq.~\ref{lsize} (a result recovered in Eq.~\ref%
{leafd2}) is an intermediate case where $\left\langle
l_{leaf}\right\rangle $ reflects the overall chain length $L$ as
well as the position $\sim l_{P}$ of the shallow minimum. Note
further that in the limit $L\rightarrow \infty $ the average size
per leaf goes to infinity for $d<4$ (but the ''particles'' will
only be spread out over the whole volume, $M\left\langle
l_{leaf}\right\rangle \approx L$, for $d<2$).

Concerning the role of dimensionality one also gains some insight by the
following simple argument (similar to the famous Onsager/Manning argument
for the condensation of counterions on an infinitely long charged rod \cite%
{manning78}). Consider a pair of two particles at distance $l$ in one
dimension that attract each other via $f\left( l\right) =k_{B}T\left(
d/2\right) \ln \left( l/l_{P}\right) $. Now assume that the particles move
further apart from the distance $l_{1}$ to the distance $l_{2}>l_{1}$. This
leads to an increase in energy by $\Delta E=k_{B}T\left( d/2\right) \ln
\left( l_{2}/l_{1}\right) $. On the other hand the particles gain entropy
since they are now less confined: $-k_{B}T\Delta S=k_{B}T\ln \left(
l_{1}/l_{2}\right) $. Hence for $d<2$ the particles will ''loose'' each
other since their attraction to the nearest neighbors is overruled by the
gain in entropy -- as derived rigorously in Eq.~\ref{leafd2}.\smallskip

Finally, I mention that the same extension to arbitrary dimensions
$d$ can be performed for closed chains. One finds then phase
separation for molten rosettes if $d>4$. More specifically, loops
have a preferred spacing $L/N$ for $d<4$ and $4\chi l_{P}/\left(
d-4\right) $ for $d>4$. This is different from the results on open
chains, Eq.~\ref{leafd2}, in the interval $2<d<4$; hence in $d=3$
molten rosettes respond strongly to a cutting of the chain.

\setcounter{equation}{0}

\section{Formation energy for small intranucleosomal loops}

Since the configurations of small loops are essentially planar, it is
convenient to describe them in terms of the function $r\left( \theta \right)
$, cf. Fig.~12(b), where $r$ and $\theta $ are the polar coordinates of an
arbitrary point on the loop (with the origin chosen on the cylinder axis and
the $X$ axis running through the center of the loop). In these terms the
line element $ds\,\ $takes the form $ds=d\theta \sqrt{r^{2}+\left(
dr/d\theta \right) ^{2}}$ and the loop excess length is given by

\begin{equation}
\Delta L=\int_{-\theta ^{*}}^{\theta ^{*}}d\theta \sqrt{r^{2}+\left(
dr/d\theta \right) ^{2}}-2\theta ^{*}R_{0}  \label{deltall}
\end{equation}
with $2\theta ^{*}$ being the aperture angle of the bulge, $2\theta
^{*}R_{0}=L^{*}$. The local curvature $1/R=\left| d^{2}{\bf r}\left(
s\right) /ds^{2}\right| $ of the loop takes the form $1/R=\left| r^{\prime
\prime }r-2r^{\prime 2}-r^{2}\right| /\left( r^{2}+r^{\prime 2}\right)
^{3/2} $, with primes and double primes denoting the first and second
derivative, respectively, with respect to $\theta $. Restricting ourselves
here to {\it small} loops, one can write $r\left( \theta \right)
=R_{0}+u\left( \theta \right) $ with $u\ll R_{0}$ everywhere. Keeping only
quadratic terms in $u$ and its derivatives one obtains $ds\simeq
R_{0}d\theta \left( 1+u/R_{0}+u^{\prime 2}/\left( 2R_{0}^{2}\right) \right) $
and
\begin{equation}
ds\frac{1}{R^{2}\left( s\right) }\simeq d\theta \left[ \frac{1}{R_{0}^{3}}%
\left( \left( u^{\prime \prime 2}+\frac{3}{2}u^{\prime 2}+u^{2}\right)
+4uu^{\prime \prime }\right) -\frac{2u^{\prime \prime }}{R_{0}^{2}}-\frac{u}{%
R_{0}^{2}}+\frac{1}{R_{0}}\right]  \label{ds2}
\end{equation}
The bending energy of the loop is then
\begin{equation}
E_{elastic}\simeq \frac{1}{2}A\int_{-\theta ^{*}}^{\theta ^{*}}d\theta \left[
\frac{1}{R_{0}^{3}}\left( u^{\prime \prime 2}-\frac{5}{2}u^{\prime
2}+u^{2}\right) -\frac{u}{R_{0}^{2}}+\frac{1}{R_{0}}\right]  \label{be}
\end{equation}
where the boundary conditions $u\left( \theta \right) =du\left( \theta
\right) /d\theta =0$ at $\theta =\pm \theta ^{*}$ have been used.

The variational energy of the loop $F=E_{elastic}-T\Delta L$ for a {\it given%
} aperture angle $\theta ^{*}$ and subject to the constraint of a
fixed loop contour length $L=L^{*}+\Delta L$ follows from
Eqs.~\ref{deltall} and \ref{be} to be:
\begin{equation}
{\cal F}\left\{ u\left( \theta \right) \right\} =\frac{1}{2}\frac{A}{%
R_{0}^{3}}\int_{-\theta ^{*}}^{\theta ^{*}}d\theta \left[ u^{\prime \prime
2}-\frac{5}{2}u^{\prime 2}+u^{2}\right] -\frac{T}{2R_{0}}\int_{-\theta
^{*}}^{\theta ^{*}}d\theta u^{\prime 2}  \label{f}
\end{equation}
Here $T$ is the Lagrange multiplier that constrains an extra length $\Delta
L $ to be adsorbed as the loop is formed ($T$ can be interpreted as a
''tension'' pulling in extra length). In Eq.~\ref{f} the constant terms as
well as the term linear in $u$ have been dropped since solutions to $F$ with
and without the linear $u$-term differ just by a constant. The optimal loop
shape (with given values of $\Delta L$ and $\theta ^{*}$) obeys the Euler
Lagrange equation $\delta F/\delta u=0$:
\begin{equation}
u^{\prime \prime \prime \prime }+\left( \frac{5}{2}+\frac{TR_{0}^{2}}{A}%
\right) u^{\prime \prime }+u=0  \label{euler}
\end{equation}
Solutions of Eq.~\ref{euler} are of the form $u\propto e^{i\lambda \theta }$
with four possible values of $\lambda =\pm \lambda _{\pm }$ where
\begin{equation}
\lambda _{\pm }^{2}=\frac{1}{2}\left( \frac{5}{2}+\frac{TR_{0}^{2}}{A}%
\right) \pm \frac{1}{2}\sqrt{\left( \frac{5}{2}+\frac{TR_{0}^{2}}{A}\right)
^{2}-4}  \label{lpm}
\end{equation}
that show the following asymptotics:
\begin{equation}
\lambda _{+}\simeq \frac{1}{\lambda _{-}}\simeq \left\{
\begin{array}{ll}
\sqrt{2} & \mbox{for}\;T\ll A/R_{0}^{2} \\
\sqrt{TR_{0}^{2}/A} & \mbox{for}\;T\gg A/R_{0}^{2}%
\end{array}
\right.  \label{lp}
\end{equation}
One expects symmetric solutions of the form $u\left( \theta \right)
=C_{1}\cos \left( \lambda _{+}\theta \right) +C_{2}\cos \left( \lambda
_{-}\theta \right) $. The boundary conditions $u=0$ and $u^{\prime }=0$ at $%
\theta =\theta ^{*}$ have then the form $C_{1}\cos \left( \lambda _{+}\theta
^{*}\right) +C_{2}\cos \left( \lambda _{-}\theta ^{*}\right) =0$ and $%
C_{1}\lambda _{+}\sin \left( \lambda _{+}\theta ^{*}\right) +C_{2}\lambda
_{-}\sin \left( \lambda _{-}\theta ^{*}\right) =0$. The solubility condition
leads to the transcendental equation
\begin{equation}
\frac{\lambda _{-}}{\lambda _{+}}=\frac{\tan \left( \lambda _{+}\theta
^{*}\right) }{\tan \left( \lambda _{-}\theta ^{*}\right) }  \label{trans}
\end{equation}
For vanishing ''tension'', $\smallskip T=0$, one finds from Eq.~\ref{lp}
that the condition \ref{trans} is of the form $2\tan \left( \sqrt{2}\theta
^{*}\right) =\tan \left( \theta ^{*}/\sqrt{2}\right) $ that has no
non-trivial solution. For large $T$, $T\gg A/R_{0}^{2}$, one obtains from
Eqs.~\ref{lp} and \ref{trans}:
\begin{equation}
\frac{1}{\left( TR_{0}^{2}/A\right) ^{3/2}}\simeq \frac{1}{\theta ^{*}}\tan
\left( \sqrt{TR_{0}^{2}/A}\theta ^{*}\right)  \label{trans2}
\end{equation}
The left hand side of Eq \ref{trans2} is small and hence solutions are
approximately given by $\sqrt{TR_{0}^{2}/A}\theta ^{*}\simeq \lambda
_{+}\theta ^{*}\simeq k\pi $ with $k=1,2,3,...$ In the following I consider
only the $k=1$ solution which is the solution that leads to the smallest
elastic energy.

Using partial integration and Eq.~\ref{euler} the loop formation energy, Eq.~%
\ref{deltau}, can be cast into the form
\begin{equation}
\Delta U=2k_{B}T\lambda R_{0}\theta ^{*}+\frac{T}{R_{0}}\int_{0}^{\theta
^{*}}d\theta u^{\prime 2}-A\int_{0}^{\theta ^{*}}d\theta \frac{u}{R_{0}^{2}}
\label{deltau5}
\end{equation}
To proceed further one makes use of the explicit solution given below Eq.~%
\ref{lp}. Assume the large tension case $T\gg A/R_{0}^{2}$ (the quality of
this approximation will be checked {\it a posteriori}); then $\lambda
_{-}\ll 1$. The condition $u\left( \theta ^{*}\right) =0$ takes then the
form $C_{2}\simeq -C_{1}\cos \left( \lambda _{+}\theta ^{*}\right) $ which
leads together with $\lambda _{+}\theta ^{*}\simeq \pi $ to $C_{2}=C_{1}$.
The loop shape is thus approximately given by $u\left( \theta \right) \simeq
C_{1}\left( 1+\cos \left( \pi \theta /\theta ^{*}\right) \right) $ from
which follows its formation energy
\begin{equation}
\Delta U=2k_{B}T\lambda R_{0}\theta ^{*}+\frac{\pi ^{2}TC_{1}^{2}}{%
2R_{0}\theta ^{*}}-\frac{AC_{1}\theta ^{*}}{R_{0}^{2}}  \label{deltau2}
\end{equation}
and excess length, Eq.~\ref{deltall}, $\Delta L\simeq 2C_{1}\theta ^{*}+\pi
^{2}C_{1}^{2}/\left( 2R_{0}\theta ^{*}\right) \simeq 2C_{1}\theta ^{*}$
where I used the fact that small loops have small amplitudes: $C_{1}\ll
R_{0}\left( \theta ^{*}\right) ^{2}$. For fixed $\Delta L$ and $\theta ^{*}$
follows $C_{1}\simeq \Delta L/\left( 2\theta ^{*}\right) $. Inserting this
into Eq.~\ref{deltau2} and using $T\simeq \pi ^{2}A/\left( R_{0}\theta
^{*}\right) ^{2}$ leads to
\begin{equation}
\frac{\Delta U}{k_{B}T}\simeq 2\lambda R_{0}\theta ^{*}+\frac{\pi ^{4}}{8}%
\frac{l_{P}\left( \Delta L\right) ^{2}}{R_{0}^{3}\left( \theta ^{*}\right)
^{5}}-\frac{l_{P}\Delta L}{2R_{0}^{2}}  \label{deltau4}
\end{equation}
Now minimizing $\Delta U$ with respect to $\theta ^{*}$ (for $\Delta L$
fixed) gives the optimal aperture angle
\begin{equation}
\theta ^{*}\simeq \left( \frac{5\pi ^{4}}{16}\frac{l_{P}}{\lambda }\right)
^{1/6}\frac{\Delta L^{1/3}}{R_{0}^{2/3}}  \label{theta}
\end{equation}
Combining Eqs.~\ref{deltau4} and \ref{theta} one arrives at the final
expression for the formation energy of a (small) loop of excess length $\Delta L$%
, Eq.~\ref{deltau3}.


\end{document}